\documentclass[notitlepage,superscriptaddress,preprintnumbers,aps,nofootinbib,tightenlines,floatfix]{revtex4-2}

\usepackage{graphicx}
\usepackage{dcolumn}
\usepackage{bm}
\usepackage{hyperref}

\usepackage{braket}
\usepackage{physics}
\usepackage{float}
\usepackage{graphicx}
\usepackage{placeins}
\usepackage{amsmath,amssymb}
\usepackage{xcolor}

\begin{document}
	\preprint{IQuS@UW-21-017}
	\title{Preparation of the SU(3) Lattice Yang-Mills Vacuum with Variational Quantum Methods}
	
	\author{Anthony N Ciavarella}
	\email{aciavare@uw.edu}
	\affiliation{InQubator for Quantum Simulation (IQuS), Department of Physics, University of Washington, Seattle, WA 98195}
	
	\author{Ivan A Chernyshev}
	\email{ivanc3@uw.edu}
	\affiliation{InQubator for Quantum Simulation (IQuS), Department of Physics, University of Washington, Seattle, WA 98195}
	
	\date{\today}
	
	\begin{abstract}
	Studying QCD and other gauge theories on quantum hardware requires the preparation of physically interesting states. The Variational Quantum Eigensolver (VQE) provides a way of performing vacuum state preparation on quantum hardware. In this work, VQE is applied to pure SU(3) lattice Yang-Mills on a single plaquette and one dimensional plaquette chains. Bayesian optimization and gradient descent were investigated for performing the classical optimization. Ansatz states for plaquette chains are constructed in a scalable manner from smaller systems using domain decomposition and a stitching procedure analogous to the Density Matrix Renormalization Group (DMRG). Small examples are performed on IBM's superconducting {\tt Manila} processor.
	\end{abstract}
	
	\maketitle
	
{
\small
\twocolumngrid
\tableofcontents
\onecolumngrid
}


\section{Introduction}
Quantum chromodynamics (QCD) plays an important role in a number of phenomena ranging from nuclear forces
holding together nuclei to inelastic hadron collisions to the behavior of matter under extreme conditions (such as in
supernovas and the early universe). A number of analytic and numerical tools have been developed to study QCD since its discovery in the 1970s. One of the most successful approaches to numerical calculations is lattice QCD \cite{PhysRevD.10.2445,PhysRevD.21.2308}. High precision calculations of hadronic spectra \cite{doi:10.1126/science.1257050,PhysRevLett.105.252002,2003resonance}; electroweak matrix elements \cite{PhysRevD.96.054505,PhysRevLett.115.132001,PhysRevLett.112.112001,PhysRevD.99.114509,gamiz2014,PhysRevD.98.074512,PhysRevD.90.074509,2018axial,PhysRevD.94.054508}; properties of high-temperature, low density systems and some multi-hadron systems \cite{PhysRevLett.119.062002,PhysRevD.77.094507,BEANE200762,PhysRevLett.124.032001} have been performed using lattice QCD (for recent reviews see Ref. \cite{DAVOUDI20211,Aoki_2020}). However, lattice QCD calculations of some important observables of interest are limited by sign problems present in the stochastic sampling used. For example, the simulation of QCD at high densities \cite{PhysRevD.66.074507,DEFORCRAND2002290,SEILER2013213,HASENFRATZ1992539,Density2020}, relevant to supernovas and the early universe, or with a $\theta$ term suffer from sign problems \cite{PhysRevD.86.105012} and are out of the reach of classical computers at scale. The limitations of classical computers to simulate quantum physics was recognized by Feynman \cite{Feynman1982} and Benioff \cite{Benioff:1980} in the 1980s, and they proposed the use of controlled quantum systems to perform simulations of quantum systems of interest.

The recent rapid improvements in the control of quantum systems in the laboratory have led to the creation of the first few generations of quantum computers. Many different platforms have been explored including, but not limited to, superconducting circuits, trapped ions, and photonic systems (for recent reviews see Ref. \cite{huang2020superconducting,2019Trap,2019Photon}). These experimental efforts have been accompanied by a corresponding growth in the theoretical understanding of how to use quantum computers to simulate quantum systems. Algorithms for aspects of quantum simulation such as state preparation and time evolution have been developed for application in the future regime of error-corrected quantum computers and for near term applications on noisy intermediate quantum (NISQ) computers. To apply these algorithms, the basis of the theory being studied must be mapped onto the basis of the quantum computer being used. The simulation of scalar field theories has been studied in the eigen-basis of the field operator \cite{preskill2018simulating,jordan2019quantum}, the basis of the local free-field eigenstates \cite{10.5555/3179430.3179434,PhysRevA.98.042312,Klco:2018zqz}, the momentum basis \cite{Yeter_Aydeniz_2019} and using single particle digitization \cite{PhysRevA.103.042410}. Relativistic fermionic field theories have been studied using both the Jordan-Wigner and Bravyi-Kitaev encodings \cite{jordan2014quantum,Lamm_2020,Mazza_2012}. Non-linear $\sigma$ models have been studied using fuzzy spheres, qubit regularizations, and clock approximations \cite{Alexandru_2019_2,Singh:2019uwd,Bhattacharya:2020gpm,Hostetler_2021}. Aspects of superstring theory have been mapped onto quantum computers by making use of matrix models \cite{Gharibyan_2021}. There have been many approaches made to the quantum simulation of lattice gauge theories \cite{Klco:2018kyo,Kokail_2019,Kharzeev_2020,Lu_2019,chakraborty2020digital,Shaw_2020,Bender_2018,Klco_2020,Zohar_2013,Zohar_2015,Zohar_2015_2,Lamm_2019,Alexandru_2019,Ba_uls_2020,Tagliacozzo_2013,Tagliacozzo_2013_2,PhysRevLett.105.190404,Tavernelli20_094501,Byrnes_2006,Ciavarella_2021,kan2021lattice,stryker2021shearing,Paulson_2021,davoudi2021simulating,Zohar:2012ay,Zohar:2012xf,Banerjee:2012xg,Banerjee:2012pg,Martinez2016,Muschik:2016tws,Zohar:2016iic,Banuls:2017ena,Kaplan:2018vnj,Zache:2018jbt,Stryker:2018efp,Raychowdhury:2018tfj,Davoudi:2019bhy,Haase:2020kaj,Paulson:2020zjd,Davoudi:2020yln,Buser:2020cvn,Raychowdhury:2018osk,Raychowdhury:2019iki,Tagliacozzo:2012df,Ji:2020kjk,Chandrasekharan:1996ih,Brower:2003vy,Wiese:2006kp,atas2021su2,klco2021standard,dejong2021quantum,meurice2021theoretical,zohar2021quantum,armon2021photonmediated,andrade2021engineering}, mostly by making use of the Kogut-Susskind Hamiltonian \cite{PhysRevD.11.395,RevModPhys.51.659,Robson:1980nt,LIGTERINK2000983c,PhysRevD.31.2020}. These different approaches to mapping theories onto quantum computers are important to explore, as the optimal choice of basis for a given quantum computer will likely depend on the details of the hardware.

To use quantum computers to study physical systems of interest, physical states, such as the vacuum of a QFT, need to be prepared. The Variational Quantum Eigensolver (VQE) is a NISQ algorithm that can be used to variationally prepare the lowest energy state of a quantum system \cite{Peruzzo_2014}. The application of VQE to quantum chemistry problems has been studied in great detail \cite{Peruzzo_2014,McClean_2016,O_Malley_2016,Kandala_2017,PhysRevX.8.031022,PhysRevX.8.011021,PhysRevA.100.022517,PhysRevA.100.010302,Grimsley_2019,Kandala_2019,2020,gyawali2021insights,yamamoto2019natural,2020QNG}.  Additionally, use of VQE in the preparation of the vacuum state for various quantum field theories, including the Abelian Higgs model with a topological $\theta$ term \cite{zhang2021simulating},  has recently been examined. The VQE algorithm has been previously applied to find the vacuum state of small lattices for the Schwinger model \cite{Klco:2018kyo,PhysRevLett.126.220501,Kokail_2019}. It has also been used to prepare hadron states in an SU(2) gauge theory in 1+1 dimensions \cite{atas2021su2} and to model the force between mesons in the Schwinger model \cite{Lu_2019}. VQE requires an ansatz circuit to prepare the system's state and a classical optimizer to determine the angles in the ansatz circuit. To scale these calculations to situations with a useful quantum advantage, it will be necessary to understand how to connect these small lattice calculations to a calculation on a larger lattice and how the optimization procedure performs as system size is increased.

In this work, the application of VQE to pure SU(3) lattice Yang-Mills gauge theory is studied. This provides a starting point for understanding the resources required to simulate lattice QCD on a quantum computer. We performed a VQE calculation of the vacuum state for one and two plaquette systems using superconducting quantum processors. We also examine how to apply ideas from domain decomposition in lattice QCD calculations on classical computers to the construction of ansatz states for VQE of large lattices from the vacuum state of smaller lattices.

\section{Electric Multiplet Basis}
\FloatBarrier
Quantum simulation of SU(3) Yang-Mills theory on a lattice can be performed with link variables connecting neighboring sites of the lattice. The Hamiltonian, first discussed by Kogut and Susskind \cite{PhysRevD.11.395}, is
\begin{equation}
	\hat{H} = \frac{g^2}{2a^{d-2}} \sum_{b,\text{links}} |\hat{E}^b|^2 + \frac{1}{2 a^{4-d} g^2} \sum_{\text{plaquettes}}\left(6 - \hat{\Box}(\mathbf{x})  - \hat{\Box}^\dagger(\mathbf{x}) \right) \ \ \ ,
\end{equation}
where $g$ is the coupling constant, $a$ is the lattice spacing and $d$ is the number of spacial dimensions. The plaquette operator $\hat{\Box}(\mathbf{x})$ is defined by
\begin{equation}
	\hat{\Box}(\mathbf{x}) = \text{Tr} \left(\hat{U}(\mathbf{x}, \mathbf{x} + a \mathbf{i}) \hat{U}(\mathbf{x} + a \mathbf{i}, \mathbf{x} + a \mathbf{i}  + a \mathbf{j}) \hat{U}(\mathbf{x} + a \mathbf{i} + a \mathbf{j}, \mathbf{x} + a \mathbf{j}) \hat{U}(\mathbf{x} + a \mathbf{j}, \mathbf{x}) \right) \ \ \ ,
\end{equation}
where $\hat{U}(\mathbf{x},\mathbf{y})$ is an SU(3) matrix on the link between sites $\mathbf{x}$ and $\mathbf{y}$ and $\mathbf{i}$ and $\mathbf{j}$ are unit vectors that define the orientation of the plaquette. This theory can be described in the electric field basis, where each link's Hilbert space is spanned by the state vectors $\ket{\mathbf{R}, m_L, m_R}$, where $\mathbf{R}$ is an irreducible representation of $SU(3)$, $m_L$ labels the component of the representation on the left side of the link, and $m_R$ labels the component of the representation on the right side of the link. Physical states in this Hilbert space are subject to a constraint from Gauss's law which requires the wavefunction of the links meeting at each vertex to form a singlet state.
\begin{figure}
	\includegraphics[scale=0.25]{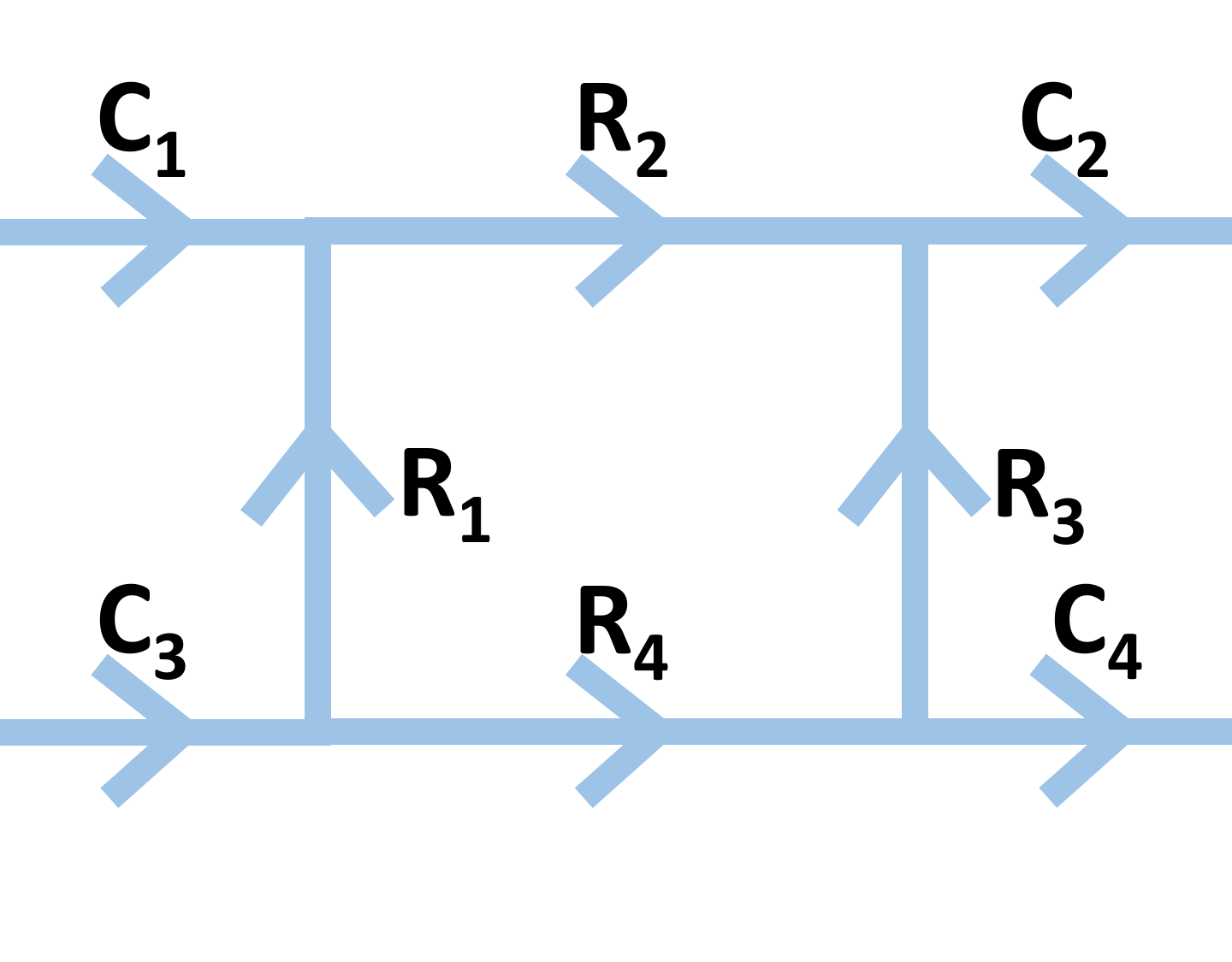}
	\caption{An SU(3) plaquette in a 1D chain of plaquettes. The electric multiplet basis states of the links in this figure are represented by $
		\Bigg | \chi\begin{pmatrix}\mathbf{C_1}, \mathbf{R_2},\mathbf{C_2} \\  \mathbf{R_1}, \mathbf{R_3} \\
			\mathbf{C_3}, \mathbf{R_4}, \mathbf{C_4}\end{pmatrix} \Bigg\rangle $ where each $\mathbf{C_i}$ and $\mathbf{R_i}$ labels the irrep on the corresponding link. }
	\label{fig:plaq_controls}
\end{figure}
In previous work, it was noted that for a lattice consisting of a chain of plaquettes, the Gauss's law constraint can be used to integrate out the irrep state labels $m_L$ and $m_R$ on each link \cite{PhysRevX.7.041046,Klco_2020,Ciavarella_2021}. Integrating out $m_L$ and $m_R$ allows basis states to be described by only specifying $\mathbf{R}$ on each link. Fig. \ref{fig:plaq_controls} shows an example of a plaquette in a chain with the basis labels necessary to specify its state.  For an SU(3) gauge theory, the representation on each link can be labeled by a pair of non-negative integers $p$ and $q$ that count the upper and lower tensor indices. These labels can be mapped onto a quantum computer in a local basis by using two registers of qubits on each link to represent $p$ and $q$ in binary. Alternatively, Gauss's law can be solved at each vertex on the lattice and the resulting physical states can be mapped onto the basis of a quantum computer. This global basis construction is not scalable to large lattices, but can be used to map small lattices onto near term devices. The number of states in the global basis that need to be considered can be reduced by making use of symmetries to study different sectors of the theory. For example, SU(3) lattice Yang-Mills theory has a color parity (CP) symmetry related to the invariance of the theory under reversal of the direction of the links. The global and CP invariant bases were studied in detail for one and two plaquettes in Ref. \cite{Ciavarella_2021}.

\FloatBarrier
	
\section{Single Plaquette}
\label{section:OnePlaq}
\FloatBarrier
A single plaquette is one of the simplest systems that can be considered in lattice gauge theory. In this work, the single plaquette system will be studied in the electric multiplet basis described in the previous section. In this formulation, Gauss's law guarantees that each link in the plaquette will have the same representation. Therefore, the basis states of the plaquette can be specified by $\ket{p,q}$, where $p$ and $q$ are specified earlier.  In units where the lattice spacing equals one, the Hamiltonian for a single plaquette is
\begin{figure}
	\includegraphics[scale=0.25,  bb = 750 0 200 500]{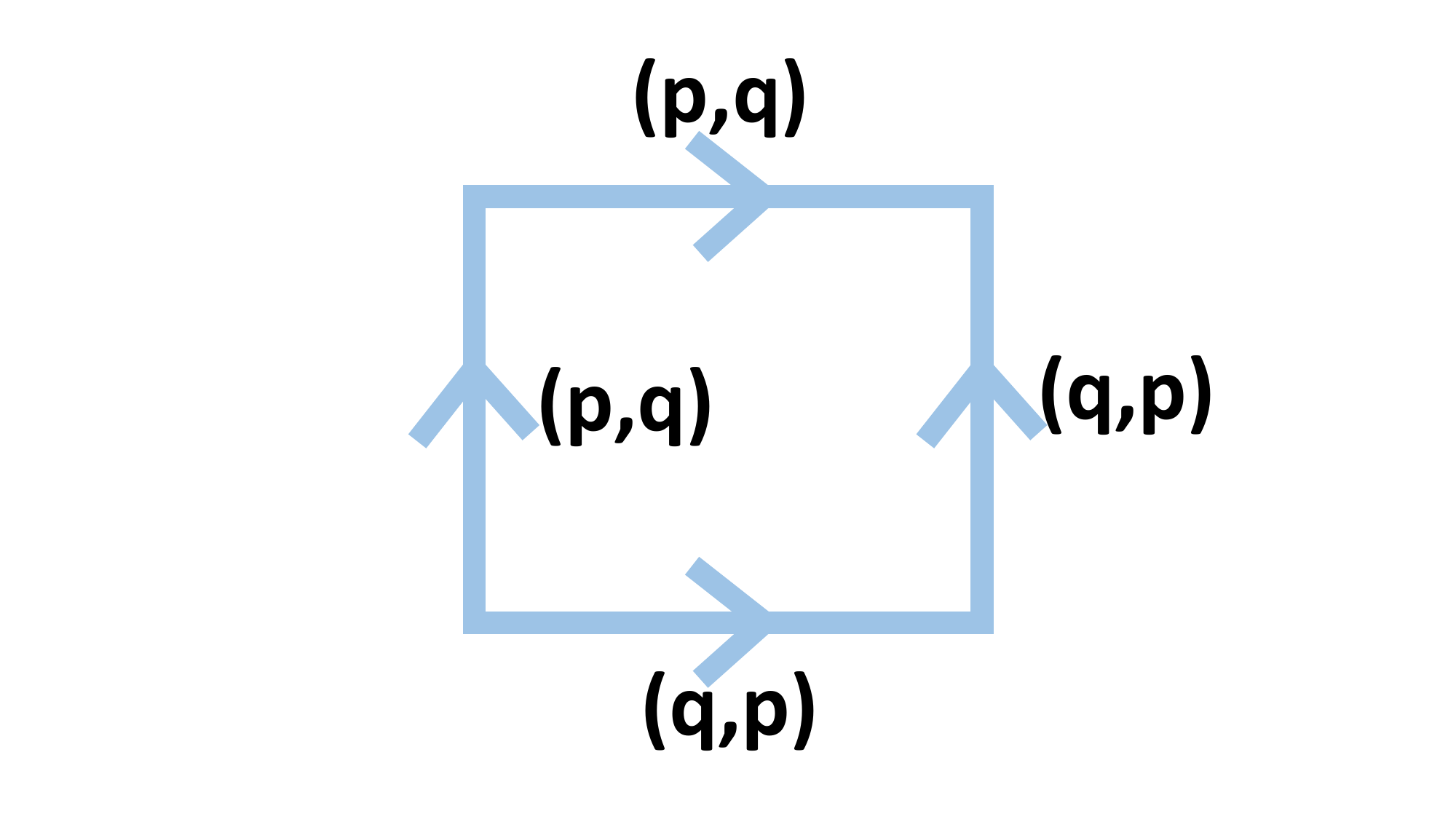}
	\caption{A single SU(3) plaquette. p and q label the chromo-electric flux on each link.}
\end{figure}
\begin{equation}
	\label{eq:HamOnePlaq}
	\hat{H} = 2g^2 \sum_b |\hat{E}^b|^2 + \frac{1}{2 g^2} \left(6 - \hat{\Box} - \hat{\Box}^\dagger \right) \ \ \ ,
\end{equation}
where $\sum_b |\hat{E}^b|^2$ is the Casimir for the chromo-electric field representation, given by
\begin{equation}
	\sum_b |\hat{E}^b|^2 \ket{p,q} = \frac{p^2 +q^2 + pq + 3p + 3q}{3} \ket{p,q} \ \ \ ,
\end{equation}
and the plaquette operator $\hat{\Box}$ acts on the basis states by
\begin{equation}
	\hat{\Box} \ket{p,q} = \ket{p+1,q} + \ket{p-1,q+1} + \ket{p,q-1} \ \ \ .
\end{equation}

While the exponential decay of correlations in gapped quantum systems is known to allow for state preparation using circuits localized in position space \cite{Klco_2020F}, the depth of the circuits needed to prepare the local color-space degrees of freedom has not been studied in as much detail. Due to Gauss's law guaranteeing every link in a single plaquette has the same chromo-electric flux, the single plaquette system can be used to study state preparation of the local color-space while avoiding the complications of spacial correlations.

\subsection{Vacuum Preparation}
\subsubsection{Initialization}
\label{section:VQEInitial}
\begin{figure}
	\includegraphics{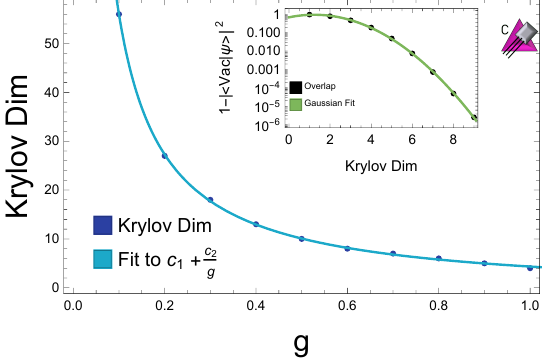}
	\caption{This figure shows the dimension of the Krylov subspace required for the overlap of the state prepared by the Lanczos algorithm, $\ket{\psi}$, with the true vacuum, $\ket{\text{Vac}}$, to satisfy $\abs{\bra{\psi} \ket{\text{Vac}}}^2 \geq 0.999999$. The inset panel shows the overlap with the true vacuum as a function of Krylov dimension for $g=0.5$.}
	\label{fig:one_plaq_krylov}
\end{figure}

VQE is a hybrid quantum algorithm that can improve the overlap of an initial state with the vacuum state. The performance of VQE has a strong dependence on the initial state used \cite{Peruzzo_2014,McClean_2016,Kandala_2017}. In applications of VQE to electronic structure problems, Hartree-Fock states and unitary coupled cluster states computed on classical computers have been used as initial starting points for VQE. However, lattice gauge theory does not have comparable classical calculations in the Hamiltonian formulation available. As an alternative, the Lanczos algorithm can used to initialize VQE for a single plaquette.\footnote{This application of Krylov subspaces to quantum simulation was developed in collaboration with other members of IQuS during the spring of 2020.} The Lanczos algorithm works by constructing the Krylov subspace spanned by $\{\ket{\psi}, \hat{H} \ket{\psi}, \dots, \hat{H}^n \ket{\psi}\}$ for some integer $n$ and initial state $\ket{\psi}$ and diagonalizing the Hamiltonian in this subspace \cite{Lanczos:1950zz}. Quantum variations of the Lanczos algorithm have also been proposed for use in the study of state preparation \cite{Motta_2019}. The result of applying the Lanczos algorithm to a single plaquette using the electric vacuum as the initial state is shown in Fig. \ref{fig:one_plaq_krylov}.\footnote{The icons in the corners of the plots in this text were introduced in Ref. \cite{icons2020} and are available at {\tt iqus.uw.edu/resources/icons/}. The pink icons indicate the calculations in the figure were performed on a classical computer and the blue icons indicate the calculations in the figure were performed on a quantum computer.}
For a fixed coupling, the overlap with the true vacuum is shown to scale asymptotically as a Gaussian with the Krylov dimension used in the Lanczos algorithm. The dimension of the Krylov subspace needed to reach a fixed accuracy scales as $\frac{1}{g}$. This behavior can be seen to follow from the structure of the single plaquette vacuum wavefunction. The vacuum wavefunction is asymptotically Gaussian in the chromo-electric field with a width inversely proportional to $g$. Each time $\hat{H}$ is applied to increase the dimension of the Krylov subspace, the maximum $p$ and $q$ included in the Krylov subspace is increased by 1. Therefore, the size of the vacuum wavefunction components added by increasing the Krylov dimension fall off asymptotically as a Gaussian, and the Krylov dimension needed to reach a desired accuracy $\epsilon$ scales as $\frac{\log(\frac{1}{\epsilon})}{g}$. It should be noted that an exponential convergence with field truncation has also been observed in the simulation of scalar field theories \cite{Klco:2018zqz} and $U(1)$ gauge theories \cite{zache2021achieving}, and has been proven to be a rather generic property of theories involving bosonic modes \cite{tong2021provably}.

The Lanczos algorithm provides approximate wavefunction components of the vacuum state that must be mapped into a quantum circuit to be useful for state preparation. The state prepared by using a $d$-dimensional Krylov subspace potentially spans all basis states with $p,q < d$. Therefore, a state with nontrivial support on $d^2$ basis states must be prepared, which can be done using a circuit of length $O(d^2)$ using standard state preparation procedures \cite{Nielsen:2011:QCQ:1972505}. Using the previous result on the Krylov dimension required to reach an accuracy $\epsilon$, a quantum circuit of size
\begin{equation}
	S = O \left(\left(\frac{\log(\frac{1}{\epsilon})}{g}\right)^2 \right) \ \ \ ,
\end{equation}
can be used to prepare the vacuum of a single plaquette with coupling $g$ on a quantum computer within an accuracy of $\epsilon$.

\subsubsection{Optimization}
\label{section:VQEOpt}
The VQE algorithm makes use of a classical optimizer to improve the overlap of the ansatz state with the actual vacuum. In previous work, Bayesian optimizers have been used in the VQE algorithm to prepare the ground state of the Schwinger model \cite{Klco:2018kyo} and to prepare hadron states in an SU(2) gauge theory \cite{atas2021su2} on small lattices. Bayesian optimization minimizes an objective function by iteratively constructing an interpolator, usually a Gaussian process, from existing data and optimizing the interpolator.  It is ideal for optimizations where the number of available evaluations of the objective function is limited (typically to a few hundred evaluations), the objective function is continuous, and the dimensionality of the domain is no more than 20 \cite{frazier2018tutorial}. On existing hardware that only has a handful of qubits available, circuits that can prepare a generic ansatz state can be implemented with fewer than 20 parameters. However, as quantum computers grow in qubit count and coherence time, this will no longer be true. To reach a quantum advantage, it will be important to understand when Bayesian optimization breaks down. To test the performance of a Bayesian optimizer for lattice gauge theory, VQE was simulated without noise on a classical computer for a single SU(3) plaquette with a truncation of $p,q \leq 3$. This system can be represented using 4 qubits on a quantum processor. The vacuum state of this system lies in a 10-dimensional CP-invariant subspace which can be parametrized in spherical coordinates with 9 degrees of freedom. The details of how the Bayesian optimization was performed are available in Appendix \ref{app:bayes}.

\begin{figure}
	\includegraphics{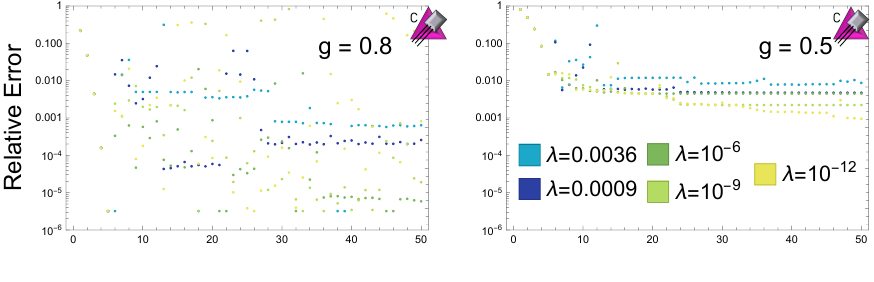} \\
	
	\includegraphics{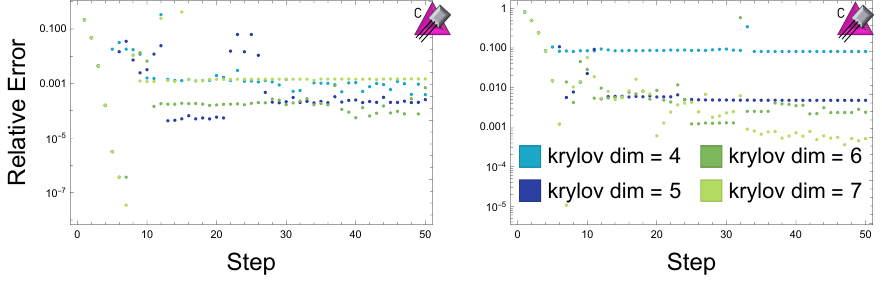}
	\caption{The relative error in the estimation of the vacuum energy obtained by performing a classical simulation of VQE using Bayesian optimization for a single plaquette with $p,q\leq 3$. The left panel is for $g=0.8$ and the right panel is for $g=0.5$. The top panel shows the results of Bayesian optimization as a function of the number of iterations of the optimization for different values of the regulator $\lambda$. Each of the calculations in the top panel was initialized with the vacuum states obtained from the Lanczos algorithm with subspace of Krylov dimension equal to 5. The bottom panel shows the result of Bayesian optimization using $\lambda=0.0009$ with different maximum Krylov dimensions.}
	\label{fig:BOpreccouplingcomparison}
\end{figure}

The results of the simulation of VQE with a Bayesian optimizer are shown in Fig. \ref{fig:BOpreccouplingcomparison}. In these calculations, the Gaussian process used to model the energy function being minimized suffered from multicollinearity. This was mitigated with Tikohonov regularization, which in this context is equivalent to adding a small constant term $\lambda$ to the covariance matrix of the energies \cite{mittikhonovnotes}. As this figure shows, the convergence of the Bayesian optimizer has a dependence on the regulator $\lambda$. The energy that the Bayesian optimizer converges to cannot be made arbitrarily close to the vacuum energy because at sufficiently small values of $\lambda$, multicollinearity returns and the covariance matrix cannot be inverted, causing the Bayesian optimizer to fail. The lower panels in Fig. \ref{fig:BOpreccouplingcomparison} show the dependence of the Bayesian optimizer's convergence on the dimension of the Krylov subspace used to initialize the calculation. For certain initializations, the Bayesian optimizer is not able to improve upon the initial state's overlap with the actual vacuum. Even for this modest system size, Bayesian optimization has limitations in how close it can get to the vacuum state.

Gradient descent is an alternative classical optimizer that can be used in VQE. Gradient descent evaluates the gradient of the energy, $\grad{f(\bf{x})}$, at the current step's ansatz parametrization ${\bf{x}}_i$, then selects the next step's ansatz parametrization ${\bf{x}}_{i+1}$ according to
\begin{equation}
	{\bf{x}}_{i+1} = {\bf{x}}_{i} - \eta \grad{f({\bf{x}}_i)} \ \ \ ,
\end{equation} 
where $\eta$ is a learning rate that controls the convergence of the gradient descent. Convergence to a local minimum can be guaranteed by the use of backtracking, where $\eta$ is steadily decreased during the course of the calculation \cite{truong2019convergence}. Alternatively, the step size can be selected by using Bayesian optimization to perform a line search \cite{tamiya2021stochastic}. In applications to VQE, the gradient can be computed on a quantum processor by making use of parameter shift formulas which give the gradient without discretization errors due to large shift size \cite{PhysRevA.101.032308}. The use of gradient descent as the classical optimizer in VQE will require the energy of the state to be calculated on the quantum processor a number of times equal to two times the number of parameters in the circuit ansatz per step in the optimization. For comparison, Bayesian optimization only requires the energy to be computed once per step. The increase in quantum resources per step in the optimization may be offset by a faster rate of convergence and ability to converge to the actual vacuum state. As an optimizer, gradient descent also requires fewer classical resources per step than Bayesian optimization. This is because with gradient descent, the classical computer only needs to perform subtraction during gradient descent. Bayesian optimization, on the other hand, requires the computations of determinants and inverses of a matrix whose dimension is equal to the number of times the energy was previously evaluated.

\begin{figure}
	\includegraphics{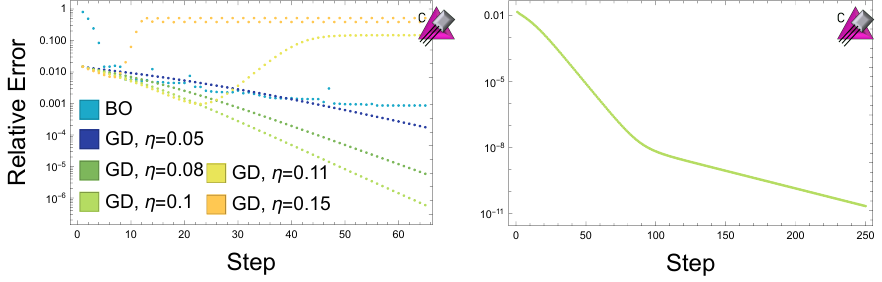}
	\caption{The relative error in the estimation of the vacuum energy obtained by performing a classical simulation of VQE for a single plaquette with $p,q\leq3$. The coupling is $g = 0.5$ and the initial state was obtained from the Lanczos algorithm using a Krylov dimension of 5.  The left panel shows a comparison of the results obtained by performing VQE using a Bayesian optimizer to those obtained by performing VQE using a numerical gradient descent for different learning rates $\eta$.  The right panel shows the results of 250 iterations of gradient descent with $\eta=0.1$.  }
	\label{fig:compareBOvsGD}
\end{figure}

\begin{figure}
	\includegraphics{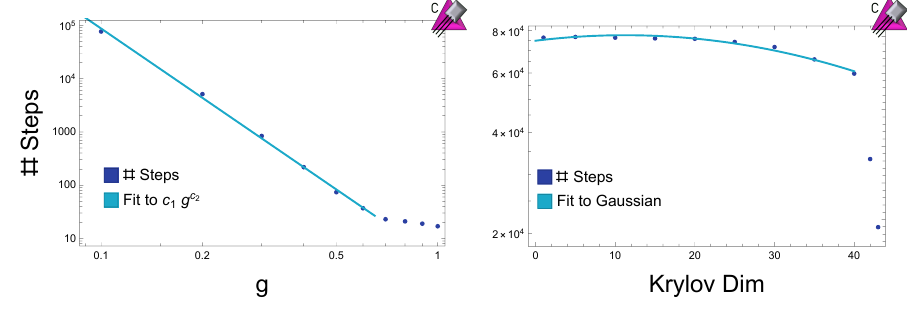}
	\caption{The left panel shows the number of steps needed for VQE using a backtracking gradient descent to converge to the true vacuum with an accuracy of 0.999 as a function of coupling for a single plaquette with $p,q \leq 31$. The right panel shows the number of steps needed for a backtracking gradient descent to converge to the true vacuum with an accuracy of 0.999  for $g=0.1$ as a function of the dimension of the Krylov subspace used to obtain the initial state.}
	\label{fig:grad_scaling}
\end{figure}

Fig. \ref{fig:compareBOvsGD} compares,  for a single plaquette truncated at $p,q\leq3$ and with $g=0.5$, the results of using Bayesian optimization for the classical optimizer to those of using numerically-computed gradient descent. The Bayesian optimizer shown in this plot was run with $\lambda=10^{-12}$. Both optimizers were initialized with the vacuum obtained using the Lanczos algorithm with a Krylov dimension of $5$. As this plot shows, the Bayesian optimizer converges above the vacuum energy, while VQE using gradient descent with a sufficiently small $\eta$ is limited only by the number of steps performed in the optimization. To understand if VQE can offer a quantum advantage, it is helpful to know how many steps in the optimizer must be performed to reach a certain level of accuracy. Fig. \ref{fig:grad_scaling} shows the number of steps needed for a backtracking gradient descent to converge for a single plaquette with a truncation of $p,q\leq 31$. This truncation was chosen so that the relative error in the mass gap and the vacuum expectation of the plaquette operator due to field truncation was $\leq 1 \%$ for each coupling studied. The left panel shows that as $g$ is decreased, the number of steps needed by the gradient descent algorithm to start from the electric vacuum and reach a state $\ket{\psi}$ with $\abs{\bra{\text{Vacuum}} \ket{\psi}}^2 \geq 0.999$ scales as $O(g^{-4})$. The number of steps needed to reach this level of accuracy can be decreased by beginning the optimization at a state closer to the vacuum, such as a state obtained from the Lanczos algorithm. The right panel in Fig. \ref{fig:grad_scaling} shows the number of steps needed by a backtracking gradient descent to converge to $\abs{\bra{\text{Vacuum}} \ket{\psi}}^2 \geq 0.999$ for a coupling $g=0.1$ as a function of the dimension of the Krylov subspace used in the Lanczos algorithm to initialize the starting state. From the fit in the right panel, it appears that the number of steps required for the gradient descent to converge scales asymptotically as a Gaussian as a function of the Krylov dimension used. This is expected, as the discussion in the previous section showed that the error in the state obtained from the Lanczos algorithm falls off asymptotically as a Gaussian as a function of the Krylov dimension. By beginning in a state obtained from the Lanczos algorithm and performing the optimization step using gradient descent, classical simulations of the VQE algorithm are able to reach the vacuum state of a single plaquette at weak couplings that are beyond the reach of Bayesian optimization. Based on these results, Bayesian optimization will not be a practical optimizer for VQE calculations at scale, while gradient based methods have a chance of reaching the vacuum state at scale.

\FloatBarrier
\subsection{Hardware Implementation}
\label{section:hardwareimp}
The discussion in the previous section suggests that VQE should be capable of preparing the vacuum state for a single plaquette. However, existing quantum hardware suffers from the effects of noise and imperfect gate implementations. This will have an impact on how VQE performs in practice. To understand how near-term hardware will perform in the simulation of SU(3) lattice Yang-Mills theory, IBM's {\tt Manila} superconducting quantum processor was used to perform a VQE calculation for a single plaquette \cite{ibmManila}.
\begin{figure}
	\includegraphics{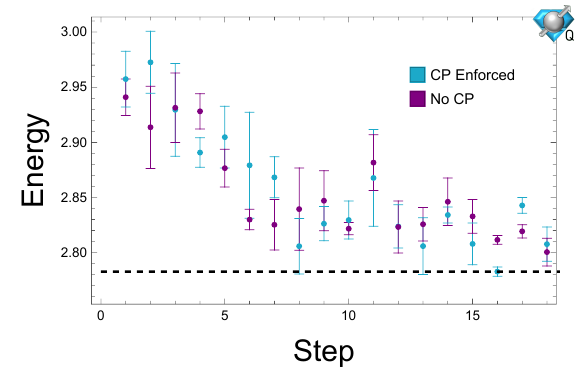}
	\caption{Variational state preparation of the vacuum state for a single plaquette truncated at $\mathbf{8}$ with $g=1$ run on the {\tt Manila} quantum processor. The blue points show the results of gradient descent with CP symmetry enforced in the rotation angles in the ansatz circuit and the purple points show the result of not explicitly enforcing CP symmetry in the state. The data in this figure is available in Table \ref{tab:OnePlaq8Data}.}
	\label{fig:VQE_one_plaq_8}
\end{figure}

The SU(3) lattice Yang-Mills Hamiltonian possesses a CP symmetry that guarantees that the amplitude of a given representation in the vacuum wavefunction will be the same as the amplitude of the conjugate representation. In principle, this symmetry can be used to restrict the state preparation circuit used in VQE which will reduce the number of free parameters. However, in the presence of noise and imperfect gate implementations, attempting to explicitly enforce the symmetry may prevent the actual state prepared on the quantum processor from respecting the symmetry. This would be the case if, hypothetically, the rotations in the circuit suffered from a constant offset error that was not corrected for. To understand if this is an issue on existing hardware, a single plaquette was simulated in the global basis truncated at a representation of $\mathbf{8}$. The Hamiltonian is given by Eq. (14) of Ref. \cite{Ciavarella_2021}. A VQE state preparation procedure described in Appendix \ref{appendix:hardware} was used to prepare the vacuum state starting from the electric vacuum and to optimize the angles using gradient descent. VQE was performed both by enforcing CP symmetry in the rotation angles in the circuit ansatz and by allowing all three of the angles to vary freely. The results of both calculations are displayed in Fig. \ref{fig:VQE_one_plaq_8}. As this figure shows, explicitly enforcing the CP symmetry in the VQE calculation does not break the symmetry in the vacuum state prepared using VQE on this hardware. The ability to explicitly enforce CP symmetry in the ansatz circuit will be helpful when performing VQE calculations on larger systems where the number of free parameters is much greater.

\begin{figure}
	\includegraphics{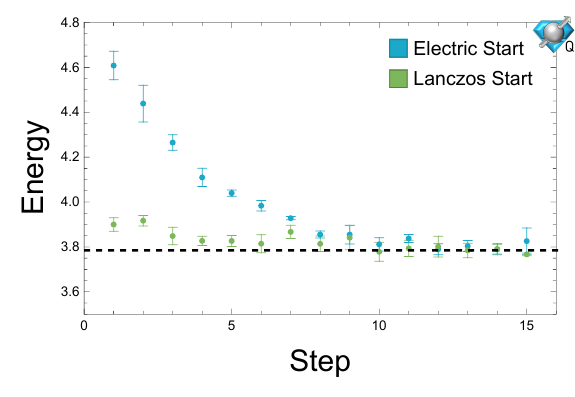}
	\caption{Variational state preparation of the vacuum state for a single plaquette truncated at $\mathbf{6^+}$ in the color parity basis with $g=0.8$ run on the {\tt Manila} quantum processor. The blue points show the result of gradient descent beginning at the electric vacuum and the green points begin at the state obtained using the Lanczos algorithm with a Krylov dimension of two. The data in this figure is available in Table \ref{tab:OnePlaq6Data}.}
	\label{fig:VQE_one_plaq_6}
\end{figure}

As discussed in Section \ref{section:VQEInitial}, the Lanczos algorithm can be used to obtain an initial ansatz for the VQE algorithm. At a coupling of $g=1$, the vacuum state obtained using a two dimensional Krylov subspace has an overlap with the true vacuum within experimental errors on the {\tt Manila} chip \cite{ibmManila}. To accurately reproduce physics at a lower coupling, more electric field representations must be included. This can be done without increasing the qubit count by performing a calculation in the color parity basis. Using two qubits, the color parity basis allows the $\mathbf{6}$ and $\mathbf{\bar{6}}$ representations to be included, which is sufficient to accurately describe a plaquette with a coupling of $g=0.8$. Fig. \ref{fig:VQE_one_plaq_6} shows the results of performing VQE for a single plaquette with $g=0.8$ in the color parity basis, beginning both at the electric vacuum and the vacuum obtained using a Krylov subspace of dimension two. As this figure shows, pre-conditioning with the vacuum obtained using the Lanczos algorithm allows one to begin closer to the actual vacuum and to converge to the true vacuum faster. Note that in both Fig. \ref{fig:VQE_one_plaq_8} and \ref{fig:VQE_one_plaq_6}, the energy computed fluctuates at late steps in the gradient descent instead of converging. This is because the gradient is computed on the {\tt Manila} chip with both statistical and systematic errors. As the optimizer approaches the vacuum state, the magnitude of the gradient vector decreases. Once the size of the gradient vector is comparable to the device errors, it can no longer be reliably computed and the updates to the circuit parameters are random noise which leads to the displayed fluctuations. This is a generic feature of having uncertainties in the computation of the gradient and will have to be considered when devising stopping criterion for VQE calculations of larger systems.

\FloatBarrier

\section{Multiple Plaquettes}
\FloatBarrier
The single plaquette calculations in Section \ref{section:OnePlaq} provide insight into the requirements of state preparation in a simple system. To perform calculations at scale, these insights need to be combined with features that only occur on larger lattices, such as Gauss's law constraints that can't be solved exactly without sacrificing locality.
\begin{figure}
	\includegraphics[scale=0.25]{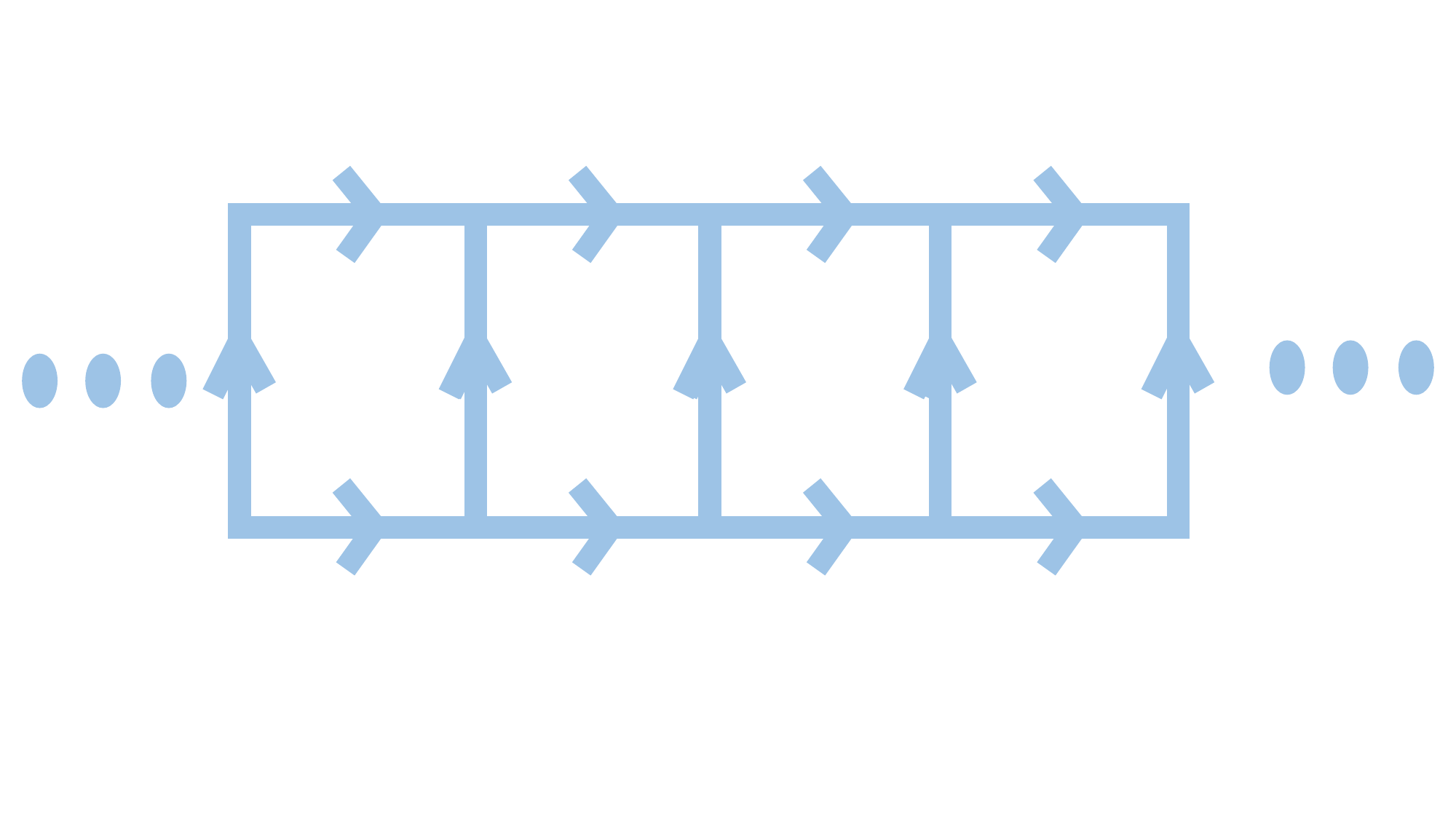}
	\caption{A lattice composed of a chain of plaquettes.}
	\label{fig:Plaq_Chain}
\end{figure}
The Lanczos algorithm provides a good starting ansatz for VQE on a single plaquette, but it is inefficient on larger lattices. This can be seen by using the electric vacuum as the initial state for a chain of $L$ plaquettes with periodic boundary conditions (PBC) as shown in Fig. \ref{fig:Plaq_Chain}. When a Krylov subspace with dimension $d$ is used, every basis state with $d$ plaquettes excited to have a loop of electric fields in the $\mathbf{3}$ representation will occur with equal amplitude. There are ${L \choose d}$ of these states and their superposition requires non-local circuits to capture the non-local correlations in the state. This leads to the circuit required to prepare the state given by the Lanczos algorithm growing exponentially in size with the Krylov dimension, and therefore no quantum advantage. An alternative approach is to use a form of domain decomposition.

In lattice QCD calculations on classical computers, a large amount of time is spent solving discretized versions of the Dirac equation. These calculations have been accelerated by making use of a domain decomposition \cite{LUSCHER2004209,frommer2014adaptive,Heybrock_2014}. Domain decomposition accelerates the calculation by solving the Dirac equation in separate sub-regions and then stitching the solutions together. Similar to solving the Dirac equation, directly preparing the vacuum state for a theory on a large lattice is difficult because the Hilbert space associated with the entire lattice is too large to efficiently work with. The ideas behind domain decomposition can be applied in a VQE calculation by splitting the lattice into separate disconnected sub-regions and preparing each sub-region in its vacuum state (note that there will be links between these regions that will remain unexcited). The vacuum state of each sub-region can be computed classically or in another VQE calculation. The VQE algorithm can then be used to excite links in-between the sub-regions and stitch together the sub-regions to form the vacuum state for the entire lattice. SU(3) Yang-Mills is a theory with spatial correlations that decay exponentially fast with distance, so it is anticipated that the domain decomposition ansatz should converge exponentially fast to the true vacuum as the domain size is increased. 

Conceptually, this approach to vacuum state preparation is similar to the density matrix renormalization group (DMRG) algorithm on classical computers \cite{PhysRevLett.69.2863}. In DMRG, the vacuum state of a lattice is prepared, and the density matrix of a sub-region is diagonalized. The eigenstates of the density matrix with largest weight are then used as the local basis for a calculation on a larger lattice. In this manner, DMRG constructs the vacuum state for a large lattice from the vacuum state for smaller regions. This is analogous to beginning the VQE optimization in a domain-decomposed vacuum, except the calculation on the quantum computer has no need to extract eigenstates of the density matrix for subregions. Once the desired lattice length is achieved, DMRG optimizes the approximation to the vacuum state by decomposing the system into left and right blocks and using the eigenstates of the density matrix of the subregions to generate a new basis for the regions. By growing and shrinking the size of the left and right blocks, DMRG is able to converge to the true vacuum state. The process of growing and shrinking the blocks used is analogous to the stitching procedure described in this work to improve the overlap with the true vacuum, except, once again, the quantum calculation does not require the diagonalization of density matrices.

While this stitching procedure will be explicitly demonstrated for a quasi one dimensional system, it can be performed in higher dimensions as well. For a system with three spatial dimensions, the sub-regions initialized in their vacuum state will be cubes of some size. Unlike in one dimension, the number of links left unexcited between the initial subregions will scale as the surface area of the subregions. A sequence of unitary transformations acting on the individual unexcited links, controlled by their neighboring links on the two cubes they connect, can be optimized using VQE to get closer to the vacuum state of the entire lattice. By limiting the number of links each unitary acts on in this manner, the number of free parameters in the VQE ansatz circuit can be restricted to grow linearly with the surface area instead of exponentially as it could if all links were allowed to be acted on simultaneously.

\subsection{Domain Decomposition on Plaquette Chains}
A lattice composed of a chain of plaquettes as shown in Fig. \ref{fig:Plaq_Chain} with PBC displays many of the complications inherent to larger lattices while still being tractable to simulate on classical computers. A domain decomposition of a plaquette chain can be performed by breaking up the lattice into separate sub-chains, preparing each subchain in its vacuum state and using VQE to optimize circuits that act on the boundaries and space between the domains to stitch them together.

\begin{table}
	\begin{tabular}{| c | c | c |} 
		\hline
		& State 1 & State 2  \\ [0.5ex] 
		\hline
		$R_1$ & $
		\Bigg | \chi\begin{pmatrix}\mathbf{1}, \mathbf{1},\mathbf{1} \\  \mathbf{1}, \mathbf{1} \\
			\mathbf{1}, \mathbf{1}, \mathbf{1}\end{pmatrix} \Bigg\rangle \nonumber$ & $
		\Bigg | \chi\begin{pmatrix}\mathbf{1}, \mathbf{3},\mathbf{1} \\  \mathbf{3}, \mathbf{\bar{3}} \\
			\mathbf{1}, \mathbf{\bar{3}}, \mathbf{1}\end{pmatrix} \Bigg\rangle \nonumber$  \\ 
		\hline
		$R_2$ & $
		\Bigg | \chi\begin{pmatrix}\mathbf{3}, \mathbf{1},\mathbf{1} \\  \mathbf{\bar{3}}, \mathbf{1} \\
			\mathbf{\bar{3}}, \mathbf{1}, \mathbf{1}\end{pmatrix} \Bigg\rangle \nonumber$ & $
		\Bigg | \chi\begin{pmatrix}\mathbf{3}, \mathbf{3},\mathbf{1} \\  \mathbf{1}, \mathbf{\bar{3}} \\
			\mathbf{\bar{3}}, \mathbf{\bar{3}}, \mathbf{1}\end{pmatrix} \Bigg\rangle \nonumber$  \\
		\hline
		$R_3$ & $
		\Bigg | \chi\begin{pmatrix}\mathbf{3}, \mathbf{1},\mathbf{1} \\  \mathbf{\bar{3}}, \mathbf{1} \\
			\mathbf{\bar{3}}, \mathbf{1}, \mathbf{1}\end{pmatrix} \Bigg\rangle \nonumber$ & $
		\Bigg | \chi\begin{pmatrix}\mathbf{3}, \mathbf{\bar{3}},\mathbf{1} \\  \mathbf{3}, \mathbf{\bar{3}} \\
			\mathbf{\bar{3}}, \mathbf{3}, \mathbf{1}\end{pmatrix} \Bigg\rangle \nonumber$  \\
		\hline
		$R_4$ & $
		\Bigg | \chi\begin{pmatrix}\mathbf{1}, \mathbf{1},\mathbf{3} \\  \mathbf{1}, \mathbf{\bar{3}} \\
			\mathbf{1}, \mathbf{1}, \mathbf{\bar{3}}\end{pmatrix} \Bigg\rangle \nonumber$ & $
		\Bigg | \chi\begin{pmatrix}\mathbf{1}, \mathbf{3},\mathbf{3} \\  \mathbf{\bar{3}}, \mathbf{1} \\
			\mathbf{1}, \mathbf{\bar{3}}, \mathbf{\bar{3}}\end{pmatrix} \Bigg\rangle \nonumber$  \\
		\hline
		$R_5$ & $
		\Bigg | \chi\begin{pmatrix}\mathbf{1}, \mathbf{1},\mathbf{3} \\  \mathbf{1}, \mathbf{\bar{3}} \\
			\mathbf{1}, \mathbf{1}, \mathbf{\bar{3}}\end{pmatrix} \Bigg\rangle \nonumber$ & $
		\Bigg | \chi\begin{pmatrix}\mathbf{1}, \mathbf{\bar{3}},\mathbf{3} \\  \mathbf{\bar{3}}, \mathbf{3} \\
			\mathbf{1}, \mathbf{3}, \mathbf{\bar{3}}\end{pmatrix} \Bigg\rangle \nonumber$  \\
		\hline
		$R_6$ & $
		\Bigg | \chi\begin{pmatrix}\mathbf{3}, \mathbf{3},\mathbf{3} \\  \mathbf{1}, \mathbf{1} \\
			\mathbf{\bar{3}}, \mathbf{\bar{3}}, \mathbf{\bar{3}}\end{pmatrix} \Bigg\rangle \nonumber$ & $
		\Bigg | \chi\begin{pmatrix}\mathbf{3}, \mathbf{1},\mathbf{3} \\  \mathbf{\bar{3}}, \mathbf{3} \\
			\mathbf{\bar{3}}, \mathbf{1}, \mathbf{\bar{3}}\end{pmatrix} \Bigg\rangle \nonumber$  \\ 
		\hline
		$R_7$ & $
		\Bigg | \chi\begin{pmatrix}\mathbf{3}, \mathbf{\bar{3}},\mathbf{\bar{3}} \\  \mathbf{3}, \mathbf{1} \\
			\mathbf{\bar{3}}, \mathbf{3}, \mathbf{3}\end{pmatrix} \Bigg\rangle \nonumber$ & $
		\Bigg | \chi\begin{pmatrix}\mathbf{3}, \mathbf{1},\mathbf{\bar{3}} \\  \mathbf{\bar{3}}, \mathbf{\bar{3}} \\
			\mathbf{\bar{3}}, \mathbf{1}, \mathbf{3}\end{pmatrix} \Bigg\rangle \nonumber$  \\ 
		\hline
	\end{tabular}
	\caption{This table enumerates the local Givens rotations required to initialize a domain vacuum on the plaquette chain truncated at an electric field representation of $\mathbf{3}$, (up to CP conjugates of the rotations listed here). The basis states are defined in the same way as the states in Fig. \ref{fig:plaq_controls}. The first column labels the rotation and the other two columns specify the basis states being rotated. $R_1$ excites a single plaquette loop of electric flux. $R_2$ through $R_5$ stretch the length of a loop of electric flux by one plaquette. $R_6$ and $R_7$ break a loop of electric flux into two loops. The basis labels used here were introduced in Ref. \cite{Ciavarella_2021}.}
	\label{tab:GivensRot3}
\end{table}
To be useful as an initial state for VQE, a quantum circuit for the preparation of these domain-decomposed vacuums must be designed. The circuit to prepare the vacuum state for a domain of length $l$ can be constructed recursively from the circuit to prepare the vacuum state for a domain of length $l-1$ as follows. A single plaquette state can be constructed by performing an $R_1$ rotation from Table \ref{tab:GivensRot3} and its CP conjugate on the qubits that make up the links in the plaquette. The two plaquette state can be prepared by applying $R_1$ rotations on two neighboring plaquettes and then applying $R_3$ and $R_4$ rotations on one of the plaquettes. The circuit that prepares the three plaquette vacuum state can then be constructed by exciting a third plaquette (i.e. apply an $R_1$ rotation), stretching over the previous two plaquettes (i.e. apply $R_3$ and $R_4$ rotations to the plaquettes that have been excited), and performing a rotation on the center plaquette to de-excite it (i.e. apply $R_6$ and $R_7$ rotations to the center plaquette). In general, the circuit to prepare a domain of size $l$ can be constructed from the circuit for a domain of size $l-1$ by exciting a neighboring plaquette, stretching it over the previous domain, and then de-exciting plaquettes in the center. In general, this approach to constructing circuits for a domain state scales exponentially with the size of the domain. 

The initial domain decomposition ansatz can be improved upon by stitching together the different domains. More specifically, in the circuit that prepares the vacuum ansatz, gates $R_1$ through $R_7$, along with their CP conjugates, can be applied to the plaquettes in-between the domains and VQE can be used to optimize the rotation angles. This stitching procedure can also be used to construct the vacuum for a larger domain instead of using the generic state preparation circuit. After performing the stitching, the overlap with the true vacuum can be increased further by layering another block of gates on the original domains and optimizing the angles with VQE again. Explicitly, if the state obtained from the VQE algorithm is $S(\vec{\theta}_2) D(\vec{\theta}_1)\ket{0}$, where $D(\vec{\theta}_1)$ prepares the states on the domain and $S(\vec{\theta}_2)$ stitches the domains together, then the ansatz state
\begin{equation}
	C(\vec{\theta}_1,\vec{\theta}_2,\vec{\theta}_3)\ket{0} = D(\vec{\theta}_3) S(\vec{\theta}_2) D(\vec{\theta}_1)\ket{0} 
\end{equation}
can be prepared on the quantum processor and the energy can be minimized as a function of $\vec{\theta}_1$, $\vec{\theta}_2$ and $\vec{\theta}_3$ using the VQE algorithm. Due to the exponentially decaying correlations in SU(3) Yang-Mills, the overlap with the true vacuum should increase exponentially with the number of additional gate layers stacked on the domains and their boundaries. 

\begin{figure}
	\includegraphics{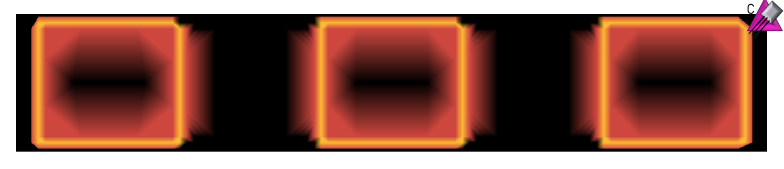}
	\includegraphics{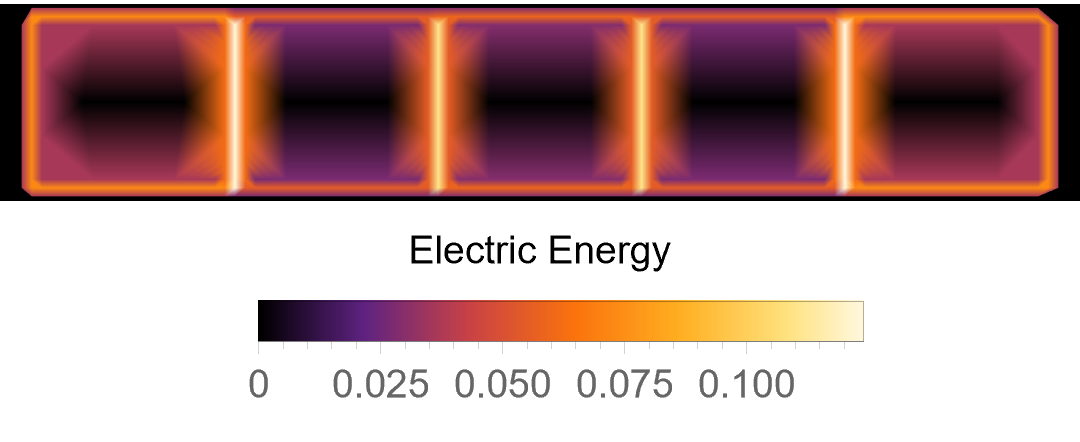}
	\caption{The top panel shows the expectation of the electric energy for a five plaquette chain with open boundary conditions where every other plaquette has been initialized to the single plaquette vacuum. The bottom panel shows the expectation of the electric energy after the boundaries of the initial domains have been stitched together with VQE.}
	\label{fig:Plaq5_HeatMap}
\end{figure}

A plaquette chain simulated in the multiplet basis with chromo-electric fields truncated at the $\mathbf{3}$ representation will be used to test the performance of the domain decomposition ansatz. Finite and infinite plaquette chains were studied using an MPS representation of states in the TEBD algorithm as described in Appendix \ref{appendix:PlaqTN}. Fig. \ref{fig:domain_decomp_finite} shows the results of optimizing different domain decomposition ansatzes for a chain of five plaquettes with $g=0.9$ and open boundary conditions. Fig. \ref{fig:Plaq5_HeatMap} shows the expectation of the electric energy for the initial single plaquette ansatz and the state obtained after stitching the boundaries together with VQE. As the size of the initial domains is increased, the overlap with the actual vacuum increases.  However, the improvement eventually saturates due to boundary effects. Due to the short correlation length at this coupling, even a single layer of stitching is able to achieve a high overlap with the actual vacuum.
\begin{figure}
	\includegraphics{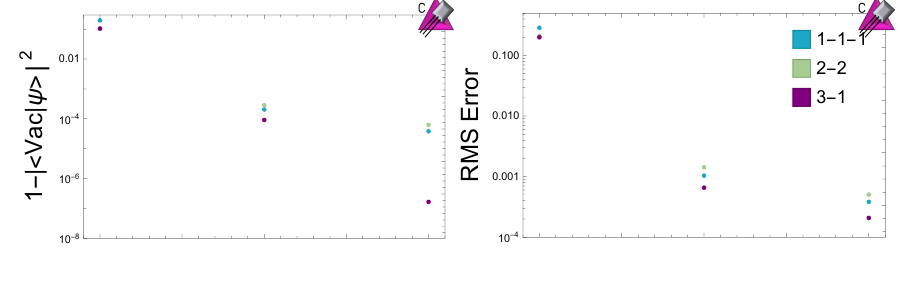}
	\caption{The left panel shows the overlap of different domain decompositions with the true vacuum. The right panel shows the RMS error in the expectation of the different single plaquette operators on the five plaquette lattice with open boundary conditions. The left-most points show the results for the initial domain decomposition, the middle points show the result after using VQE to stitch the boundaries of the domains together, and the right points show the results after using VQE to optimize another layer of circuits on the domains after stitching.}
	\label{fig:domain_decomp_finite}
\end{figure}

\begin{figure}
	\includegraphics{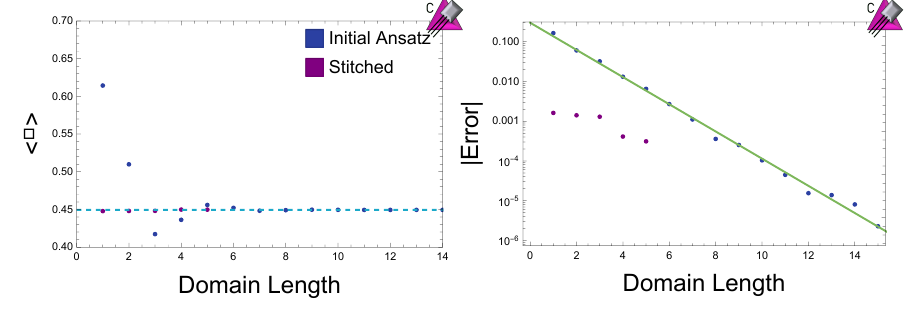}
	\caption{The left panel shows the expectation of a plaquette operator at the center of a domain as a function of domain length for both the initial ansatz and the state after using VQE to stitch domains together. The dashed blue line shows the vacuum expectation of a single plaquette operator on an infinite chain of plaquettes with $g=0.9$. The right panel shows the error in the vacuum plaquette expectation as a function of the domain size.}
	\label{fig:domain_decomp_infinite}
\end{figure}

To understand how the domain decomposition VQE ansatz performs for a large lattice, the time evolving block decimation algorithm was used to prepare the vacuum state and simulate VQE on an infinite plaquette chain as described in Appendix \ref{appendix:PlaqTN}. VQE was performed using gradient descent as the classical optimizer. The vacuum expectation of a single plaquette operator was chosen as a test observable to study the convergence to the true vacuum. As Fig. \ref{fig:domain_decomp_infinite} shows, the vacuum expectation of the plaquette operator converges exponentially fast with the domain size. A classically simulated version of VQE was used to simulate the stitching of small domains together. For domains of lengths 1-4 plaquettes, the initial domain vacuum was prepared using a generic state preparation circuit. For the initial domain of length five, the circuit to prepare the vacuum was constructed by stitching together a vacuum state preparation circuit for a domain of length three plaquettes and of length one plaquette. The circuit optimized in VQE consisted of the initial domain vacuum circuit, along with all rotations in Table \ref{tab:GivensRot3} with all rotation angles allowed to vary freely. For each domain size, the optimization of the stitching improved the estimation of the vacuum plaquette expectation by at least an order of magnitude.

\FloatBarrier
\subsection{Hardware Implementation}
As with the single plaquette case, it is instructive to study multiple plaquettes on existing quantum hardware. Unfortunately, simulating multiple plaquettes in a local basis as described in the previous section is beyond the reach of existing hardware. However, these techniques can be applied to state preparation in a global basis. IBM's {\tt Manila} quantum processor was used to simulate a two plaquette system truncated at an electric field representation of $\mathbf{3}$ in the global CP invariant basis \cite{ibmManila}. For this simple system, preparing the single plaquette vacuum is equivalent to using the vacuum state obtained using the Lanczos algorithm with a Krylov dimension of two. The results of performing VQE with the error mitigation procedures described in Appendix \ref{appendix:hardware} are shown in Fig. \ref{fig:VQE_two_plaq}. As this figure shows, the VQE algorithm is able to converge to the true vacuum energy whether it begins in the electric or single plaquette vacuum. However, by initializing the state in the single plaquette vacuum, the VQE algorithm is able to converge to the true vacuum state faster.
\begin{figure}
	\includegraphics{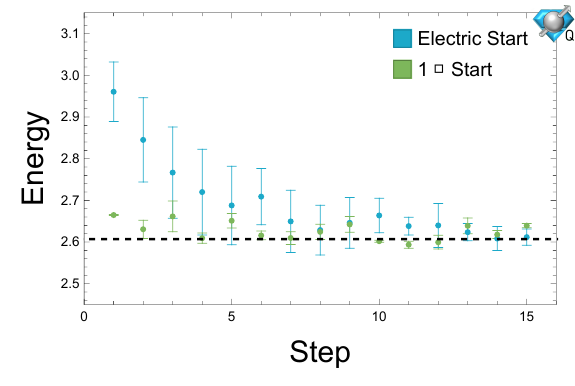}
	\caption{Variational state preparation of the vacuum state for a two plaquette system with $g=1$ and PBC run on the IBM {\tt Manila} quantum processor. The blue points show the results of performing gradient descent beginning at the electric vacuum and the green points show the results for beginning with the single plaquette vacuum.  The data in this figure is available in Table \ref{tab:TwoPlaqData}}
	\label{fig:VQE_two_plaq}
\end{figure}
While the two initial states converge to the same vacuum state, the uncertainties in the vacuum energy they converge to are quite different. This is due to the circuit ansatz used to initialize the state having redundancies in the angle parametrization of the state, leading to the two initial ansatzes converging to different sets of angles describing the same state. In the absence of noise on the quantum processor, these parametrizations would be equivalent. However, existing quantum processors are noisy and there are systematic errors with angle dependence leading to the different error bars shown in Fig. \ref{fig:VQE_two_plaq}.

\FloatBarrier
	
\section{Discussion}
Achieving a quantum advantage in the simulation of lattice gauge theories requires the preparation of physically interesting states, such as the vacuum. In the NISQ era, hybrid algorithms such as VQE will be essential. To make use of VQE, an appropriate classical optimizer and ansatz circuit must be chosen. In this work, state preparation on simple SU(3) lattice gauge theories has been performed with an eye towards scalability. In the variational state preparation of single plaquette systems, we showed that Bayesian optimization suffers from convergence issues as the coupling $g$ is decreased, while gradient descent methods suffer from no such issue. This suggests that VQE calculations at scale may need to make use of gradient descent methods in order to converge, despite the increase in computational overhead required to compute the gradient. Note that gradient based methods may converge to a local minimum instead of the true vacuum. This has not occurred for the simple systems studied in this work, but may need to be considered when performing calculations at scale. 

Calculations at scale will also require appropriate ansatz circuits to perform VQE. Due to the exponential growth of the Hilbert space with lattice size, circuits capable of preparing a generic state on the lattice will not be able to go to scale. In this work, it was demonstrated that in a quasi-1D SU(3) lattice gauge theory, VQE can be used to stitch together domains in their vacuum state to prepare the vaccum state of a larger lattice. The exponential convergence with domain size on an infinite lattice suggests that even shallow circuits may be able to achieve a large overlap with the true vacuum state at scale. The calculations on IBM's {\tt Manila} quantum processor showed that circuit ansatzes that respect a global symmetry will still respect the global symmetry on existing hardware despite the presence of noise and imperfect gates. This allows global symmetries to be used to construct circuit ansatzes that have fewer free degrees of freedom which makes them easier to optimize. 

While the computations in this work are encouraging, preparing a vacuum state for QCD with VQE will require significant developments in the application of quantum algorithms to lattice gauge theories. The calculations performed in this work were for a one dimensional string of plaquettes, but QCD is a three-dimensional theory. In a 3D theory, the domains being initialized in their vacuum state will be 3D blocks and the number of circuits required to stitch them together will scale with the surface area of the domain blocks. Additionally, a QCD calculation that can be taken to the continuum limit may require more electric field representations to be included, which will increase the number of possible local rotations in the VQE stitching circuit. It is conceivable that it is possible to reach the continuum limit without increasing the field truncation, but this remains to be investigated. Regardless, as the continuum limit is approached, the correlation length of the system will diverge and more layers of circuits will be required in the VQE stitching to accurately prepare the vacuum state. Matter will also need to be included at the sites, which will complicate the integrating out of the internal gauge space. In addition to these conceptual complications, achieving a quantum advantage in the simulation of lattice QCD will require quantum hardware with more qubits and a lower error rate, in order to enable the simulation of a large lattice in a local basis. While scaling up quantum hardware is challenging, the rapid improvement in quantum hardware and recent proposals for co-design \cite{Ciavarella_2021,zhang2021simulating,andrade2021engineering} of quantum processors suggest that it can be done in a manner that will allow the simulation of lattice QCD on quantum computers in the near future.

\begin{acknowledgments}
	The authors would like to thank Martin Savage, Alessandro Roggero, Natalie Klco, Stephen Caspar and Hersh Singh for useful discussions. The views expressed are those of the authors, and do not reflect the official policy or position of IBM or the IBM Quantum team. This work was facilitated through the use of advanced computational, storage, and networking infrastructure provided by the Hyak supercomputer system at the University of Washington. The tensor network simulations were performed using modified versions of code available at tensors.net written by Glen Evenbly. Data presented throughout this manuscript is available upon email request. This work is supported by the DOE QuantISED program through the theory consortium ``Intersections of QIS and Theoretical Particle Physics" at Fermilab. AC is supported in part by Fermi National Accelerator Laboratory PO No. 652197. IC is supported in part by the U.S. Department of Energy, Office of Science, Office of Nuclear Physics, InQubator for Quantum Simulation (IQuS) under Award Number DOE (NP) Award DE-SC0020970 through the Quantum Horizons: QIS Research and Innovation for Nuclear Science Initiative.
\end{acknowledgments}

\FloatBarrier

\appendix
\section{Hardware Calculations}
\label{appendix:hardware}
\begin{figure}
	\includegraphics{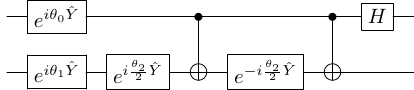}
	
	\includegraphics{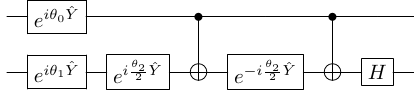}
	
	\includegraphics{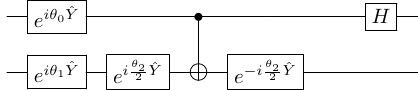}
	
	\caption{The top circuit is used to compute the expectation of $H_1$, the second circuit is used to compute the expectation of $H_2$, and the bottom circuit is used to compute the expectation of $H_3$.}
	\label{fig:VQECircuits}
\end{figure}

To perform VQE on a quantum computer, a circuit must be designed to prepare the ansatz state. For the calculations demonstrated here, only two qubits were used, so the circuit used to construct the state was capable of preparing an arbitrary 2 qubit state whose wavefunction has only real coefficients. Once the ansatz state has been prepared on the quantum computer, the energy of the state must also be computed. This can efficiently be done by breaking the Hamiltonian up into a sum over tractable terms, applying gates that diagonalize each term of the Hamiltonian, and performing measurements in the computational basis. This approach to computing the energy will require one circuit per term in the Hamiltonian. Each of the Hamiltonians studied in this work can be written in the form
\begin{align}
	\hat{H} &= \hat{H}_1 + \hat{H}_2 + \hat{H}_3 \nonumber \\
	\hat{H}_1 &= h_{11} \hat{1} \otimes \hat{Z} + h_{12} \hat{X} \otimes \hat{1}  + h_{13} \hat{X} \otimes \hat{Z}  \nonumber \\
	\hat{H}_2 &= h_{21} \hat{Z} \otimes \hat{1} +h_{22} \hat{1} \otimes \hat{X} \nonumber \\
	\hat{H}_3 &= h_{31} \hat{X} \otimes \hat{X} + h_{32} \hat{Y} \otimes \hat{Y} + h_{33} \hat{Z} \otimes \hat{Z}\ \ \ . 
\end{align}
These Hamiltonians can be diagonalized using the circuits shown in Fig. \ref{fig:VQECircuits}. To use gradient descent based methods in the classical optimization step of VQE, the gradient for the energy of the state as a function of the rotation angles in the ansatz circuit must be computed on the quantum computer. Due to the periodicity of $\sin$ and $\cos$, the gradient can be computed exactly using a symmetric finite difference formula with a shift of $\frac{\pi}{4}$. Explicitly, components of the gradient are computed using
\begin{equation}
	\partial_i E\left(\vec{\theta}\right) = E\left(\vec{\theta} + \frac{\pi}{4} \hat{i}\right) - E\left(\vec{\theta} - \frac{\pi}{4} \hat{i}\right) \ \ \ ,
\end{equation}
where $E\left(\vec{\theta}\right)$ is the energy as a function of the angles in the ansatz circuit and $\hat{i}$ is a unit vector pointing in the $i$-th direction. Therefore the gradient can be computed on the quantum computer using a number of circuits equal to two times the number of parameters in the ansatz circuit.
The calculation  of the energy on a real quantum computer suffers from systematic errors due to errors in the implementation of the gates on the computer and errors in the measurement process. The measurement errors can be mitigated by using Qiskit's {\tt measurement filter} subroutine, which removes the leading order measurement errors by optimizing an approximate inverse of the calculated all-to-all measurement matrix \cite{ibmmeaserror}. The dominant gate errors come from the implementation of CNOT gates. The errors associated with CNOT gates are mitigated using an extrapolation procedure \cite{PhysRevX.7.021050,PhysRevLett.119.180509}. Each CNOT in the circuit is replaced with an odd number $r$ of CNOT gates ($r=3,5,7$) and a linear extrapolation is performed to $r=0$.
\FloatBarrier

\section{Bayesian Optimization}
\label{app:bayes}

Bayesian optimization is a classical optimizer that can be used in the VQE algorithm.  Bayesian optimization uses the data already collected to create a Gaussian process-based surrogate function that approximates the function, $f$, being optimized.  This surrogate function is then used to create an acquisition function, which is then optimized to find a new trial point for the location of $f$'s minimum.  $f$ is then evaluated at that new point and the result is incorporated into the data for the next iteration \cite{bayesopt_presentation}. The Gaussian process used requires both a mean and covariance matrix for the function $f$. The covariance matrix used in this work is constructed from the Gaussian kernel \cite{gprocess_presentation}, which defines the covariance between $f({\bf x}_1)$ and $f({\bf x}_2)$ to be
\begin{equation}
	\label{eq:gaussker}
	K({\bf x}_1,  {\bf x}_2) = e^{-\Sigma_{i=1}^d \frac{({x_1}_i-{x_2}_i)^2}{l_i^2}} \ \ \ , 
\end{equation}
where d is the number of dimensions of the inputted point and $l_i$ are hyperparameters specifying the width of the Gaussian for each component of $\bf{x}$. The mean of $f$ is generically unknown, but given the covariance matrix the mean can be approximated by the best linear unbiased predictor,
\begin{equation}
	\label{eq:blupmean}
	\mu = ({\bf{1}^T} {\bf C}^{-1} {\bf 1})^{-1} {\bf{1}^T}  {\bf C}^{-1} {\bf Z} \ \ \,
\end{equation}
where $\bf{1}$ is a vector with all entries equal to 1,  $\bf{C}$ is the covariance matrix with matrix elements given by ${\bf C}_{ij} = K({\bf x}_i,  {\bf x}_j)$,  and $\bf{Z}$ is a vector with entries given by the value of the function at the evaluated points, ${\bf{Z}}_i = f({\bf x}_i)$ \cite{Cressie1990}.

Given the mean and variance of the Gaussian process, the value of $f$ at a point ${\bf x_{posterior}}$ that has not already been evaluated follows a Gaussian distribution with a mean and variance given by
\begin{align}
	\label{eq:ok}
	\mu_{posterior} &= {\bf{c}^T} {\bf C}^{-1} {\bf Z} - (1 - {\bf{c}^T} {\bf C}^{-1} {\bf 1}) ({\bf{1}^T} {\bf C}^{-1} {\bf 1})^{-1} {\bf{1}^T}  {\bf C}^{-1} {\bf Z} \nonumber \\	
	\sigma^{2}_{posterior} &= K({\bf{x}_{posterior}},  {\bf{x}_{posterior}}) - {\bf{c}^T} {\bf C}^{-1} {\bf c} + {(1 - {\bf{c}^T} {\bf C}^{-1} {\bf 1})^{2}} ({\bf {1}^T} {\bf C}^{-1} {\bf 1})^{-1} \ \ \ ,
\end{align}
where $\bf{c}$ is a vector with entries ${\bf{c}}_i = K({\bf x_{posterior}},  {\bf x}_i)$ \cite{Cressie1990}. Eq. (\ref{eq:ok}) expresses the posterior mean and variance under the assumption that $f$ can be evaluated without error. In order to incorporate errors,  the variance of the data must be added to the diagonal elements of the covariance matrix $\bf{C}$ and to $\sigma^{2}_{posterior}$ \cite{gprocess_presentation}.

To use a Gaussian process in practice, the hyperparameters of the kernel must be selected. In this work, this was done by maximizing the likelihood of the data under a multivariate Gaussian model with a mean equal to the best linear unbiased predictor's mean and with a covariance equal to $\bf{C}$ (with the variance of the data added to its diagonal elements) from Eq. (\ref{eq:blupmean}). Another issue with practical implementation that arises is that $\bf{C}$ often ends up singular as the Gaussian process is iterated. This issue is known as multicollinearity and it occurs when one of the points used to construct $\bf{C}$ can be exactly predicted from the other points leading to zero being an eigenvalue of $\bf{C}$.  This can be remedied by using Tikohonov regularization where a fake ``data variance'' distinct from the real data variance is added to $\bf{C}$ but not to $\sigma^{2}_{posterior}$ \cite{mittikhonovnotes}.

The probability distribution of $f$ at unevaluated points is used to construct an acquisition function, whose job it is to balance exploration and exploitation. The acquisition function is optimized to find the minimum of $f$. In this work, probability of improvement \cite{bayesopt_presentation} was used as the acquisition function, ie. the probability that the minimum of $f$ is smaller than the previously found minimum is maximized.  This is equivalent to minimizing
\begin{equation}
	\label{eq:acqpofi_used}
	acq({\bf x})_{PI} = \frac{\mu_{posterior}({\bf x}) - f_{min}}{\sigma_{posterior}({\bf x})}
\end{equation}
where $f_{min}$ is the previously found minimum of $f$.

\section{Plaquette Chain Tensor Network}
\label{appendix:PlaqTN}
\begin{figure}
	\includegraphics[scale=0.25]{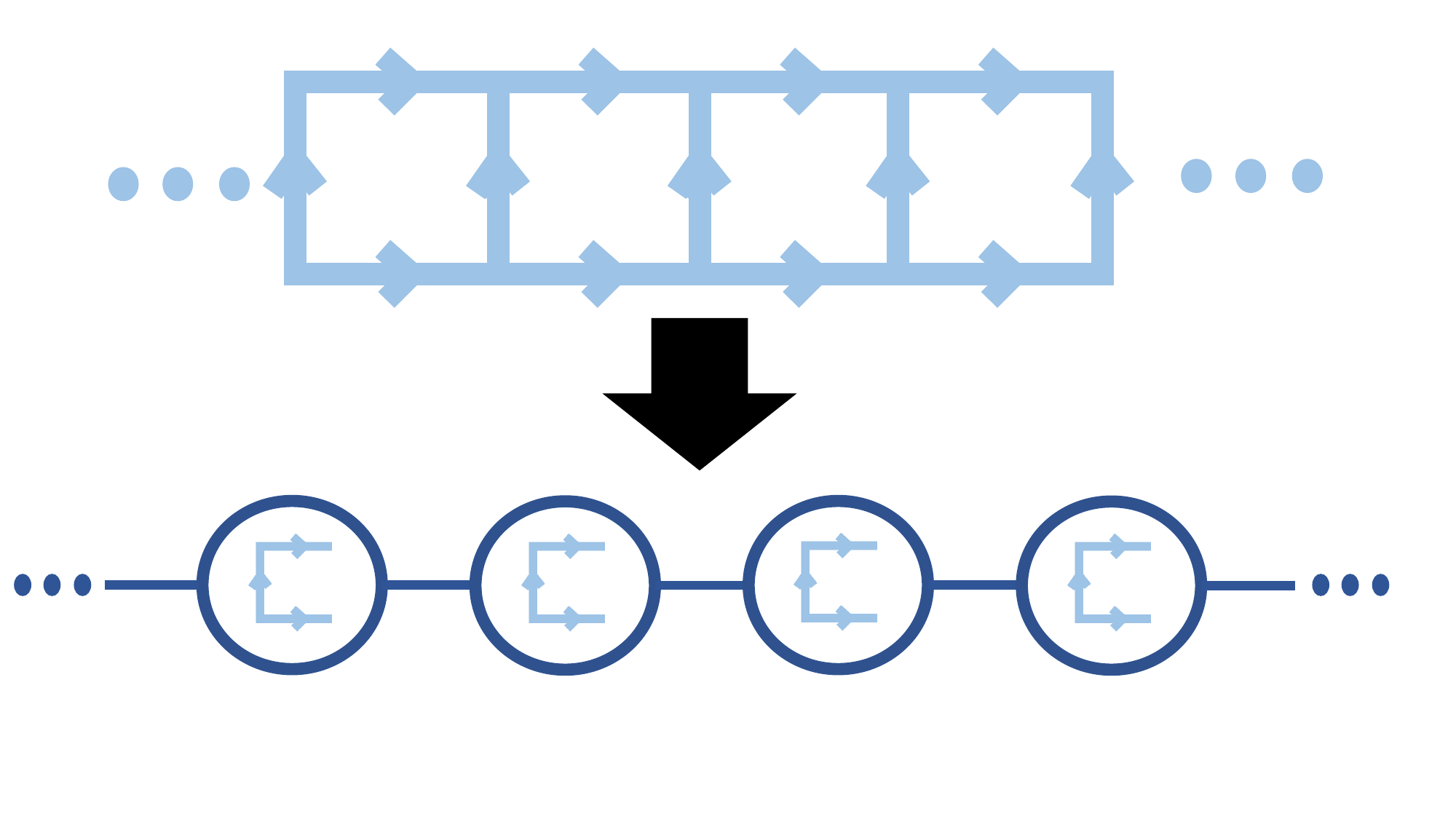}
	\caption{An infinite chain of SU(3) plaquettes can be mapped onto a 1D quantum system whose state can be represented with MPS by blocking sets of 3 links together as shown.}
	\label{fig:plaq_chain_tensor}
\end{figure}

The time evolving block decimation (TEBD) algorithm can be used to simulate the time evolution of an infinite translationally invariant quantum system by Trotterizing the time evolution operator \cite{2003,2004Eff,2004MPS}. The vacuum state of a system can be prepared by performing imaginary time evolution. This algorithm was developed for the simulation of systems whose Hamiltonian only consists of 2-site nearest-neighbor couplings, so its application to the simulation of a plaquette chain requires nonstandard modifications. Fig. \ref{fig:plaq_chain_tensor} shows how the links in the plaquette chain can be blocked together to form a 1D quantum system whose state can be described with MPS.
\FloatBarrier

In this blocking, the electric field operator on a single link becomes a single site operator, the plaquette operator becomes a three site operator, and the Gauss's law constraint become a constraint on neighboring sites. The Gauss's law constraint can be enforced by adding an energy penalty for violating Gauss's law.

\begin{figure}
	\includegraphics[scale=0.25]{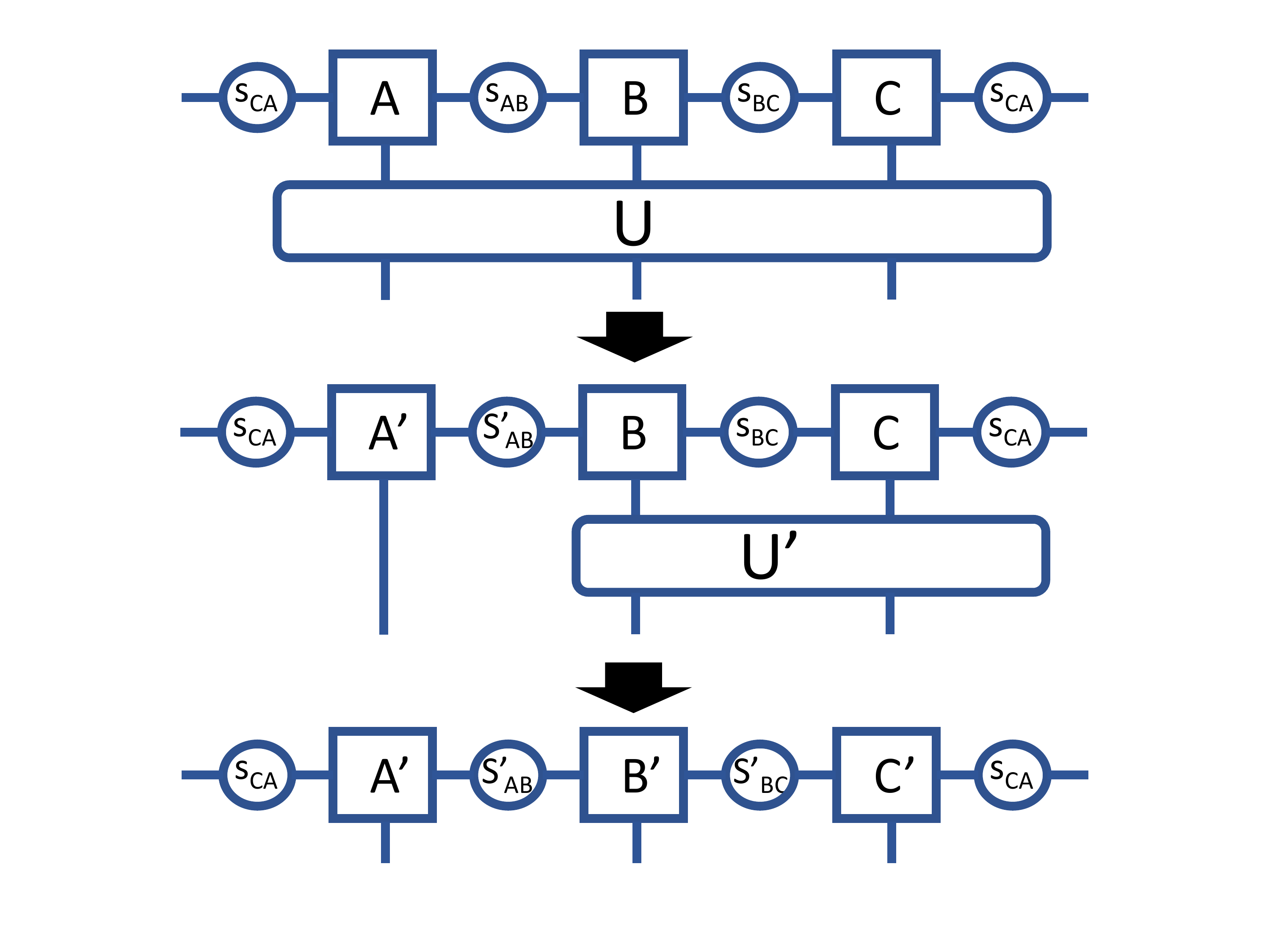}
	\caption{This figure shows the required sequence of SVDs that must be performed to return an MPS tensor network to MPS form after applying a 3 site gate.}
	\label{fig:plaq_chain_SVD}
\end{figure}

The TEBD algorithm finds the vacuum by applying a Trotterized version of the imaginary time evolution operator to a translationally invariant state. For a 2-site Hamiltonian, this is accomplished by storing a unit cell of 2 sites and performing an SVD after applying each gate to keep the most relevant states. For a 3-site Hamiltonian, such as the Hamiltonian obtained for the plaquette chain, a unit cell of 3 sites must be stored and two SVD's must be performed to obtain the most relevant local states as shown in Fig. \ref{fig:plaq_chain_SVD}. The approach used to perform time evolution in TEBD can also be used to apply arbitrary gates. To represent the ansatz states obtained using domains of $l$ plaquettes, a unit cell of length $l+1$ had to be stored and the state was prepared by applying gates and performing a SVD to return to MPS form as in the case of time evolution.
\FloatBarrier

\section{Data from IBM's Manila Processor}
\label{appendix:Data}
The following tables in this appendix contain the energies that were computed on IBM's {\tt Manila} quantum processor. All error bars were computed from the uncertainty in the linear CNOT extrapolation as described in Appendix \ref{appendix:hardware}.
\begin{table}
	\begin{tabular}{| c | c | c |} 
		\hline
		Step Number & CP Symmetry Enforced & CP Symmetry Unenforced  \\ [0.5ex] 
		\hline
		1 & $2.957 \pm 0.025$ & $2.941 \pm 0.017$ \\ 
		\hline
		2 & $2.973 \pm 0.028$ & $2.91 \pm 0.04$ \\ 
		\hline
		3 & $2.93 \pm 0.04$ & $2.931 \pm 0.032$ \\ 
		\hline
		4 & $2.891 \pm 0.013$ & $2.928 \pm 0.016$ \\ 
		\hline
		5 & $2.905 \pm 0.028$ & $2.877 \pm 0.017$ \\ 
		\hline
		6 & $2.88 \pm 0.05$ & $2.830 \pm 0.009$ \\ 
		\hline
		7 & $2.868 \pm 0.019$ & $2.825 \pm 0.023$ \\ 
		\hline
		8 & $2.806 \pm 0.025$ &  $2.84 \pm 0.04$\\ 
		\hline
		9 & $2.826 \pm 0.016$ & $2.847 \pm 0.027$ \\ 
		\hline
		10 & $2.83 \pm 0.017$ & $2.822 \pm 0.006$ \\ 
		\hline
		11 & $2.87 \pm 0.04$ & $2.882 \pm 0.025$ \\ 
		\hline
		12 & $2.824 \pm 0.02$ & $2.823 \pm 0.024$ \\ 
		\hline
		13 & $2.806 \pm 0.026$ & $2.826 \pm 0.015$ \\ 
		\hline
		14 & $2.834 \pm 0.007$ & $2.846 \pm 0.021$ \\ 
		\hline
		15 & $2.808 \pm 0.019$ & $2.833 \pm 0.015$ \\ 
		\hline
		16 & $2.783 \pm 0.004$ & $2.812 \pm 0.004$ \\ 
		\hline
		17 & $2.843 \pm 0.007$ & $2.819 \pm 0.006$ \\ 
		\hline
		18 & $2.808 \pm 0.016$ & $2.801 \pm 0.013$ \\ 
		\hline
	\end{tabular}
	\caption{This table lists the data shown in Fig. \ref{fig:VQE_one_plaq_8}. The left column states the number of times gradient descent was applied, the center column contains the energies computed for the circuit that had the CP symmetry explicitly enforced, and the right column contains the energies computed for the circuit without the CP symmetry enforced.}
	\label{tab:OnePlaq8Data}
\end{table}

\begin{table}
	\begin{tabular}{| c | c | c |} 
		\hline
		Step Number & Electric Start & Krylov Start  \\ [0.5ex] 
		\hline
		1 & $4.61 \pm 0.06$ & $3.900 \pm 0.030$ \\ 
		\hline
		2 & $4.44 \pm 0.08$ & $3.917 \pm 0.024$ \\ 
		\hline
		3 & $4.265 \pm 0.034$ & $3.85 \pm 0.04$ \\ 
		\hline
		4 & $4.11 \pm 0.04$ & $3.827 \pm 0.020$ \\ 
		\hline
		5 & $4.040 \pm 0.014$ & $3.826 \pm 0.025$ \\ 
		\hline
		6 & $3.984 \pm 0.024$ & $3.81 \pm 0.04$ \\ 
		\hline
		7 & $3.928 \pm 0.007$ & $3.867 \pm 0.030$ \\ 
		\hline
		8 & $3.855 \pm 0.016$ &  $3.814 \pm 0.034$\\ 
		\hline
		9 & $3.85 \pm 0.04$ & $3.84 \pm 0.05$ \\ 
		\hline
		10 & $3.811 \pm 0.030$ & $3.78 \pm 0.04$ \\ 
		\hline
		11 & $3.837 \pm 0.018$ & $3.79 \pm 0.04$ \\ 
		\hline
		12 & $3.790 \pm 0.024$ & $3.80 \pm 0.05$ \\ 
		\hline
		13 & $3.804 \pm 0.025$ & $3.785 \pm 0.032$ \\ 
		\hline
		14 & $3.789 \pm 0.024$ & $3.790 \pm 0.022$ \\ 
		\hline
		15 & $3.83 \pm 0.06$ & $3.767 \pm 0.007$ \\ 
		\hline
	\end{tabular}
	\caption{This table lists the data shown in Fig. \ref{fig:VQE_one_plaq_6}. The left column states the number of times gradient descent was applied, the center column contains the energies computed for the gradient descent that began at the electric vacuum, and the right column contains the energies computed for the gradient descent that began at the state obtained from the Lanczos algorithm with a Krylov dimension of 2.}
	\label{tab:OnePlaq6Data}
\end{table}

\begin{table}
	\begin{tabular}{| c | c | c |} 
		\hline
		Step Number & Electric Start & One Plaquette Start  \\ [0.5ex] 
		\hline
		1 & $2.96 \pm 0.07$ & $2.6648 \pm 0.0013$ \\ 
		\hline
		2 & $2.85 \pm 0.10$ & $2.631 \pm 0.022$ \\ 
		\hline
		3 & $2.77 \pm 0.11$ & $2.66 \pm 0.04$ \\ 
		\hline
		4 & $2.72 \pm 0.10$ & $2.609 \pm 0.012$ \\ 
		\hline
		5 & $2.69 \pm 0.09$ & $2.651 \pm 0.017$ \\ 
		\hline
		6 & $2.71 \pm 0.07$ & $2.616 \pm 0.010$ \\ 
		\hline
		7 & $2.65 \pm 0.07$ & $2.610 \pm 0.015$ \\ 
		\hline
		8 & $2.63 \pm 0.06$ &  $2.625 \pm 0.018$\\ 
		\hline
		9 & $2.65 \pm 0.06$ & $2.642 \pm 0.019$ \\ 
		\hline
		10 & $2.66 \pm 0.04$ & $2.6021 \pm 0.0023$ \\ 
		\hline
		11 & $2.638 \pm 0.021$ & $2.594 \pm 0.008$ \\ 
		\hline
		12 & $2.64 \pm 0.05$ & $2.600 \pm 0.017$ \\ 
		\hline
		13 & $2.624 \pm 0.021$ & $2.639 \pm 0.019$ \\ 
		\hline
		14 & $2.608 \pm 0.029$ & $2.618 \pm 0.010$ \\ 
		\hline
		15 & $2.612 \pm 0.019$ & $2.639 \pm 0.005$ \\ 
		\hline
	\end{tabular}
	\caption{This table lists the data shown in Fig. \ref{fig:VQE_two_plaq}. The left column states the number of times gradient descent was applied. The center column contains the energies computed for the gradient descent that began at the electric vacuum.  The right column contains the energies computed for the gradient descent that began at the single plaquette vacuum.}
	\label{tab:TwoPlaqData}
\end{table}

\FloatBarrier

\bibliography{bibVQE}

\begin{thebibliography}{149}%
\makeatletter
\providecommand \@ifxundefined [1]{%
 \@ifx{#1\undefined}
}%
\providecommand \@ifnum [1]{%
 \ifnum #1\expandafter \@firstoftwo
 \else \expandafter \@secondoftwo
 \fi
}%
\providecommand \@ifx [1]{%
 \ifx #1\expandafter \@firstoftwo
 \else \expandafter \@secondoftwo
 \fi
}%
\providecommand \natexlab [1]{#1}%
\providecommand \enquote  [1]{``#1''}%
\providecommand \bibnamefont  [1]{#1}%
\providecommand \bibfnamefont [1]{#1}%
\providecommand \citenamefont [1]{#1}%
\providecommand \href@noop [0]{\@secondoftwo}%
\providecommand \href [0]{\begingroup \@sanitize@url \@href}%
\providecommand \@href[1]{\@@startlink{#1}\@@href}%
\providecommand \@@href[1]{\endgroup#1\@@endlink}%
\providecommand \@sanitize@url [0]{\catcode `\\12\catcode `\$12\catcode
  `\&12\catcode `\#12\catcode `\^12\catcode `\_12\catcode `\%12\relax}%
\providecommand \@@startlink[1]{}%
\providecommand \@@endlink[0]{}%
\providecommand \url  [0]{\begingroup\@sanitize@url \@url }%
\providecommand \@url [1]{\endgroup\@href {#1}{\urlprefix }}%
\providecommand \urlprefix  [0]{URL }%
\providecommand \Eprint [0]{\href }%
\providecommand \doibase [0]{https://doi.org/}%
\providecommand \selectlanguage [0]{\@gobble}%
\providecommand \bibinfo  [0]{\@secondoftwo}%
\providecommand \bibfield  [0]{\@secondoftwo}%
\providecommand \translation [1]{[#1]}%
\providecommand \BibitemOpen [0]{}%
\providecommand \bibitemStop [0]{}%
\providecommand \bibitemNoStop [0]{.\EOS\space}%
\providecommand \EOS [0]{\spacefactor3000\relax}%
\providecommand \BibitemShut  [1]{\csname bibitem#1\endcsname}%
\let\auto@bib@innerbib\@empty
\bibitem [{\citenamefont {Wilson}(1974)}]{PhysRevD.10.2445}%
  \BibitemOpen
  \bibfield  {author} {\bibinfo {author} {\bibfnamefont {K.~G.}\ \bibnamefont
  {Wilson}},\ }\bibfield  {title} {\bibinfo {title} {Confinement of quarks},\
  }\href {https://doi.org/10.1103/PhysRevD.10.2445} {\bibfield  {journal}
  {\bibinfo  {journal} {Phys. Rev. D}\ }\textbf {\bibinfo {volume} {10}},\
  \bibinfo {pages} {2445} (\bibinfo {year} {1974})}\BibitemShut {NoStop}%
\bibitem [{\citenamefont {Creutz}(1980)}]{PhysRevD.21.2308}%
  \BibitemOpen
  \bibfield  {author} {\bibinfo {author} {\bibfnamefont {M.}~\bibnamefont
  {Creutz}},\ }\bibfield  {title} {\bibinfo {title} {Monte carlo study of
  quantized su(2) gauge theory},\ }\href
  {https://doi.org/10.1103/PhysRevD.21.2308} {\bibfield  {journal} {\bibinfo
  {journal} {Phys. Rev. D}\ }\textbf {\bibinfo {volume} {21}},\ \bibinfo
  {pages} {2308} (\bibinfo {year} {1980})}\BibitemShut {NoStop}%
\bibitem [{\citenamefont {Borsanyi}\ \emph {et~al.}(2015)\citenamefont
  {Borsanyi}, \citenamefont {Durr}, \citenamefont {Fodor}, \citenamefont
  {Hoelbling}, \citenamefont {Katz}, \citenamefont {Krieg}, \citenamefont
  {Lellouch}, \citenamefont {Lippert}, \citenamefont {Portelli}, \citenamefont
  {Szabo},\ and\ \citenamefont {Toth}}]{doi:10.1126/science.1257050}%
  \BibitemOpen
  \bibfield  {author} {\bibinfo {author} {\bibfnamefont {S.}~\bibnamefont
  {Borsanyi}}, \bibinfo {author} {\bibfnamefont {S.}~\bibnamefont {Durr}},
  \bibinfo {author} {\bibfnamefont {Z.}~\bibnamefont {Fodor}}, \bibinfo
  {author} {\bibfnamefont {C.}~\bibnamefont {Hoelbling}}, \bibinfo {author}
  {\bibfnamefont {S.~D.}\ \bibnamefont {Katz}}, \bibinfo {author}
  {\bibfnamefont {S.}~\bibnamefont {Krieg}}, \bibinfo {author} {\bibfnamefont
  {L.}~\bibnamefont {Lellouch}}, \bibinfo {author} {\bibfnamefont
  {T.}~\bibnamefont {Lippert}}, \bibinfo {author} {\bibfnamefont
  {A.}~\bibnamefont {Portelli}}, \bibinfo {author} {\bibfnamefont {K.~K.}\
  \bibnamefont {Szabo}},\ and\ \bibinfo {author} {\bibfnamefont {B.~C.}\
  \bibnamefont {Toth}},\ }\bibfield  {title} {\bibinfo {title} {Ab initio
  calculation of the neutron-proton mass difference},\ }\href
  {https://doi.org/10.1126/science.1257050} {\bibfield  {journal} {\bibinfo
  {journal} {Science}\ }\textbf {\bibinfo {volume} {347}},\ \bibinfo {pages}
  {1452} (\bibinfo {year} {2015})}\BibitemShut {NoStop}%
\bibitem [{\citenamefont {Majumder}\ and\ \citenamefont
  {M\"uller}(2010)}]{PhysRevLett.105.252002}%
  \BibitemOpen
  \bibfield  {author} {\bibinfo {author} {\bibfnamefont {A.}~\bibnamefont
  {Majumder}}\ and\ \bibinfo {author} {\bibfnamefont {B.}~\bibnamefont
  {M\"uller}},\ }\bibfield  {title} {\bibinfo {title} {Hadron mass spectrum
  from lattice qcd},\ }\href {https://doi.org/10.1103/PhysRevLett.105.252002}
  {\bibfield  {journal} {\bibinfo  {journal} {Phys. Rev. Lett.}\ }\textbf
  {\bibinfo {volume} {105}},\ \bibinfo {pages} {252002} (\bibinfo {year}
  {2010})}\BibitemShut {NoStop}%
\bibitem [{\citenamefont {Karsch}\ \emph {et~al.}(2003)\citenamefont {Karsch},
  \citenamefont {Redlich},\ and\ \citenamefont {Tawfik}}]{2003resonance}%
  \BibitemOpen
  \bibfield  {author} {\bibinfo {author} {\bibfnamefont {F.}~\bibnamefont
  {Karsch}}, \bibinfo {author} {\bibfnamefont {K.}~\bibnamefont {Redlich}},\
  and\ \bibinfo {author} {\bibfnamefont {A.}~\bibnamefont {Tawfik}},\
  }\bibfield  {title} {\bibinfo {title} {Hadron resonance mass spectrum and
  lattice qcd thermodynamics},\ }\href
  {https://doi.org/10.1140/epjc/s2003-01228-y} {\bibfield  {journal} {\bibinfo
  {journal} {The European Physical Journal C}\ }\textbf {\bibinfo {volume}
  {29}},\ \bibinfo {pages} {549–556} (\bibinfo {year} {2003})}\BibitemShut
  {NoStop}%
\bibitem [{\citenamefont {Tiburzi}\ \emph {et~al.}(2017)\citenamefont
  {Tiburzi}, \citenamefont {Wagman}, \citenamefont {Winter}, \citenamefont
  {Chang}, \citenamefont {Davoudi}, \citenamefont {Detmold}, \citenamefont
  {Orginos}, \citenamefont {Savage},\ and\ \citenamefont
  {Shanahan}}]{PhysRevD.96.054505}%
  \BibitemOpen
  \bibfield  {author} {\bibinfo {author} {\bibfnamefont {B.~C.}\ \bibnamefont
  {Tiburzi}}, \bibinfo {author} {\bibfnamefont {M.~L.}\ \bibnamefont {Wagman}},
  \bibinfo {author} {\bibfnamefont {F.}~\bibnamefont {Winter}}, \bibinfo
  {author} {\bibfnamefont {E.}~\bibnamefont {Chang}}, \bibinfo {author}
  {\bibfnamefont {Z.}~\bibnamefont {Davoudi}}, \bibinfo {author} {\bibfnamefont
  {W.}~\bibnamefont {Detmold}}, \bibinfo {author} {\bibfnamefont
  {K.}~\bibnamefont {Orginos}}, \bibinfo {author} {\bibfnamefont {M.~J.}\
  \bibnamefont {Savage}},\ and\ \bibinfo {author} {\bibfnamefont {P.~E.}\
  \bibnamefont {Shanahan}} (\bibinfo {collaboration} {NPLQCD Collaboration}),\
  }\bibfield  {title} {\bibinfo {title} {Double-$\ensuremath{\beta}$ decay
  matrix elements from lattice quantum chromodynamics},\ }\href
  {https://doi.org/10.1103/PhysRevD.96.054505} {\bibfield  {journal} {\bibinfo
  {journal} {Phys. Rev. D}\ }\textbf {\bibinfo {volume} {96}},\ \bibinfo
  {pages} {054505} (\bibinfo {year} {2017})}\BibitemShut {NoStop}%
\bibitem [{\citenamefont {Beane}\ \emph {et~al.}(2015)\citenamefont {Beane},
  \citenamefont {Chang}, \citenamefont {Detmold}, \citenamefont {Orginos},
  \citenamefont {Parre\~no}, \citenamefont {Savage},\ and\ \citenamefont
  {Tiburzi}}]{PhysRevLett.115.132001}%
  \BibitemOpen
  \bibfield  {author} {\bibinfo {author} {\bibfnamefont {S.~R.}\ \bibnamefont
  {Beane}}, \bibinfo {author} {\bibfnamefont {E.}~\bibnamefont {Chang}},
  \bibinfo {author} {\bibfnamefont {W.}~\bibnamefont {Detmold}}, \bibinfo
  {author} {\bibfnamefont {K.}~\bibnamefont {Orginos}}, \bibinfo {author}
  {\bibfnamefont {A.}~\bibnamefont {Parre\~no}}, \bibinfo {author}
  {\bibfnamefont {M.~J.}\ \bibnamefont {Savage}},\ and\ \bibinfo {author}
  {\bibfnamefont {B.~C.}\ \bibnamefont {Tiburzi}} (\bibinfo {collaboration}
  {NPLQCD Collaboration}),\ }\bibfield  {title} {\bibinfo {title} {Ab initio
  calculation of the $np\ensuremath{\rightarrow}d\ensuremath{\gamma}$ radiative
  capture process},\ }\href {https://doi.org/10.1103/PhysRevLett.115.132001}
  {\bibfield  {journal} {\bibinfo  {journal} {Phys. Rev. Lett.}\ }\textbf
  {\bibinfo {volume} {115}},\ \bibinfo {pages} {132001} (\bibinfo {year}
  {2015})}\BibitemShut {NoStop}%
\bibitem [{\citenamefont {Bazavov}\ \emph
  {et~al.}(2014{\natexlab{a}})\citenamefont {Bazavov}, \citenamefont {Bernard},
  \citenamefont {Bouchard}, \citenamefont {DeTar}, \citenamefont {Du},
  \citenamefont {El-Khadra}, \citenamefont {Foley}, \citenamefont {Freeland},
  \citenamefont {G\'amiz}, \citenamefont {Gottlieb}, \citenamefont {Heller},
  \citenamefont {Kim}, \citenamefont {Kronfeld}, \citenamefont {Laiho},
  \citenamefont {Levkova}, \citenamefont {Mackenzie}, \citenamefont {Neil},
  \citenamefont {Oktay}, \citenamefont {Qiu}, \citenamefont {Simone},
  \citenamefont {Sugar}, \citenamefont {Toussaint}, \citenamefont {Van~de
  Water},\ and\ \citenamefont {Zhou}}]{PhysRevLett.112.112001}%
  \BibitemOpen
  \bibfield  {author} {\bibinfo {author} {\bibfnamefont {A.}~\bibnamefont
  {Bazavov}}, \bibinfo {author} {\bibfnamefont {C.}~\bibnamefont {Bernard}},
  \bibinfo {author} {\bibfnamefont {C.~M.}\ \bibnamefont {Bouchard}}, \bibinfo
  {author} {\bibfnamefont {C.}~\bibnamefont {DeTar}}, \bibinfo {author}
  {\bibfnamefont {D.}~\bibnamefont {Du}}, \bibinfo {author} {\bibfnamefont
  {A.~X.}\ \bibnamefont {El-Khadra}}, \bibinfo {author} {\bibfnamefont
  {J.}~\bibnamefont {Foley}}, \bibinfo {author} {\bibfnamefont {E.~D.}\
  \bibnamefont {Freeland}}, \bibinfo {author} {\bibfnamefont {E.}~\bibnamefont
  {G\'amiz}}, \bibinfo {author} {\bibfnamefont {S.}~\bibnamefont {Gottlieb}},
  \bibinfo {author} {\bibfnamefont {U.~M.}\ \bibnamefont {Heller}}, \bibinfo
  {author} {\bibfnamefont {J.}~\bibnamefont {Kim}}, \bibinfo {author}
  {\bibfnamefont {A.~S.}\ \bibnamefont {Kronfeld}}, \bibinfo {author}
  {\bibfnamefont {J.}~\bibnamefont {Laiho}}, \bibinfo {author} {\bibfnamefont
  {L.}~\bibnamefont {Levkova}}, \bibinfo {author} {\bibfnamefont {P.~B.}\
  \bibnamefont {Mackenzie}}, \bibinfo {author} {\bibfnamefont {E.~T.}\
  \bibnamefont {Neil}}, \bibinfo {author} {\bibfnamefont {M.~B.}\ \bibnamefont
  {Oktay}}, \bibinfo {author} {\bibfnamefont {S.-W.}\ \bibnamefont {Qiu}},
  \bibinfo {author} {\bibfnamefont {J.~N.}\ \bibnamefont {Simone}}, \bibinfo
  {author} {\bibfnamefont {R.}~\bibnamefont {Sugar}}, \bibinfo {author}
  {\bibfnamefont {D.}~\bibnamefont {Toussaint}}, \bibinfo {author}
  {\bibfnamefont {R.~S.}\ \bibnamefont {Van~de Water}},\ and\ \bibinfo {author}
  {\bibfnamefont {R.}~\bibnamefont {Zhou}} (\bibinfo {collaboration} {Fermilab
  Lattice and MILC Collaborations}),\ }\bibfield  {title} {\bibinfo {title}
  {Determination of $|{V}_{us}|$ from a lattice qcd calculation of the
  $k\ensuremath{\rightarrow}\ensuremath{\pi}\ensuremath{\ell}\ensuremath{\nu}$
  semileptonic form factor with physical quark masses},\ }\href
  {https://doi.org/10.1103/PhysRevLett.112.112001} {\bibfield  {journal}
  {\bibinfo  {journal} {Phys. Rev. Lett.}\ }\textbf {\bibinfo {volume} {112}},\
  \bibinfo {pages} {112001} (\bibinfo {year} {2014}{\natexlab{a}})}\BibitemShut
  {NoStop}%
\bibitem [{\citenamefont {Bazavov}\ \emph {et~al.}(2019)\citenamefont
  {Bazavov}, \citenamefont {Bernard}, \citenamefont {DeTar}, \citenamefont
  {Du}, \citenamefont {El-Khadra}, \citenamefont {Freeland}, \citenamefont
  {G\'amiz}, \citenamefont {Gottlieb}, \citenamefont {Heller}, \citenamefont
  {Komijani}, \citenamefont {Kronfeld}, \citenamefont {Laiho}, \citenamefont
  {Mackenzie}, \citenamefont {Neil}, \citenamefont {Primer}, \citenamefont
  {Simone}, \citenamefont {Sugar}, \citenamefont {Toussaint},\ and\
  \citenamefont {Van~de Water}}]{PhysRevD.99.114509}%
  \BibitemOpen
  \bibfield  {author} {\bibinfo {author} {\bibfnamefont {A.}~\bibnamefont
  {Bazavov}}, \bibinfo {author} {\bibfnamefont {C.}~\bibnamefont {Bernard}},
  \bibinfo {author} {\bibfnamefont {C.}~\bibnamefont {DeTar}}, \bibinfo
  {author} {\bibfnamefont {D.}~\bibnamefont {Du}}, \bibinfo {author}
  {\bibfnamefont {A.~X.}\ \bibnamefont {El-Khadra}}, \bibinfo {author}
  {\bibfnamefont {E.~D.}\ \bibnamefont {Freeland}}, \bibinfo {author}
  {\bibfnamefont {E.}~\bibnamefont {G\'amiz}}, \bibinfo {author} {\bibfnamefont
  {S.}~\bibnamefont {Gottlieb}}, \bibinfo {author} {\bibfnamefont {U.~M.}\
  \bibnamefont {Heller}}, \bibinfo {author} {\bibfnamefont {J.}~\bibnamefont
  {Komijani}}, \bibinfo {author} {\bibfnamefont {A.~S.}\ \bibnamefont
  {Kronfeld}}, \bibinfo {author} {\bibfnamefont {J.}~\bibnamefont {Laiho}},
  \bibinfo {author} {\bibfnamefont {P.~B.}\ \bibnamefont {Mackenzie}}, \bibinfo
  {author} {\bibfnamefont {E.~T.}\ \bibnamefont {Neil}}, \bibinfo {author}
  {\bibfnamefont {T.}~\bibnamefont {Primer}}, \bibinfo {author} {\bibfnamefont
  {J.~N.}\ \bibnamefont {Simone}}, \bibinfo {author} {\bibfnamefont
  {R.}~\bibnamefont {Sugar}}, \bibinfo {author} {\bibfnamefont
  {D.}~\bibnamefont {Toussaint}},\ and\ \bibinfo {author} {\bibfnamefont
  {R.~S.}\ \bibnamefont {Van~de Water}} (\bibinfo {collaboration} {Fermilab
  Lattice and MILC Collaborations}),\ }\bibfield  {title} {\bibinfo {title}
  {$|{\mathrm{v}}_{us}|$ from ${K}_{\ensuremath{\ell}3}$ decay and four-flavor
  lattice qcd},\ }\href {https://doi.org/10.1103/PhysRevD.99.114509} {\bibfield
   {journal} {\bibinfo  {journal} {Phys. Rev. D}\ }\textbf {\bibinfo {volume}
  {99}},\ \bibinfo {pages} {114509} (\bibinfo {year} {2019})}\BibitemShut
  {NoStop}%
\bibitem [{\citenamefont {Gámiz}\ \emph {et~al.}(2014)\citenamefont {Gámiz},
  \citenamefont {Bazavov}, \citenamefont {Bernard}, \citenamefont {Bouchard},
  \citenamefont {DeTar}, \citenamefont {Du}, \citenamefont {El-Khadra},
  \citenamefont {Foley}, \citenamefont {Freeland}, \citenamefont {Gottlieb},
  \citenamefont {Heller}, \citenamefont {Kim}, \citenamefont {Kronfeld},
  \citenamefont {Laiho}, \citenamefont {Levkova}, \citenamefont {Mackenzie},
  \citenamefont {Neil}, \citenamefont {Oktay}, \citenamefont {Qiu},
  \citenamefont {Simone}, \citenamefont {Sugar}, \citenamefont {Toussaint},
  \citenamefont {de~Water},\ and\ \citenamefont {Zhou}}]{gamiz2014}%
  \BibitemOpen
  \bibfield  {author} {\bibinfo {author} {\bibfnamefont {E.}~\bibnamefont
  {Gámiz}}, \bibinfo {author} {\bibfnamefont {A.}~\bibnamefont {Bazavov}},
  \bibinfo {author} {\bibfnamefont {C.}~\bibnamefont {Bernard}}, \bibinfo
  {author} {\bibfnamefont {C.}~\bibnamefont {Bouchard}}, \bibinfo {author}
  {\bibfnamefont {C.}~\bibnamefont {DeTar}}, \bibinfo {author} {\bibfnamefont
  {D.}~\bibnamefont {Du}}, \bibinfo {author} {\bibfnamefont {A.~X.}\
  \bibnamefont {El-Khadra}}, \bibinfo {author} {\bibfnamefont {J.}~\bibnamefont
  {Foley}}, \bibinfo {author} {\bibfnamefont {E.~D.}\ \bibnamefont {Freeland}},
  \bibinfo {author} {\bibfnamefont {S.}~\bibnamefont {Gottlieb}}, \bibinfo
  {author} {\bibfnamefont {U.~M.}\ \bibnamefont {Heller}}, \bibinfo {author}
  {\bibfnamefont {J.}~\bibnamefont {Kim}}, \bibinfo {author} {\bibfnamefont
  {A.~S.}\ \bibnamefont {Kronfeld}}, \bibinfo {author} {\bibfnamefont
  {J.}~\bibnamefont {Laiho}}, \bibinfo {author} {\bibfnamefont
  {L.}~\bibnamefont {Levkova}}, \bibinfo {author} {\bibfnamefont {P.~B.}\
  \bibnamefont {Mackenzie}}, \bibinfo {author} {\bibfnamefont {E.~T.}\
  \bibnamefont {Neil}}, \bibinfo {author} {\bibfnamefont {M.~B.}\ \bibnamefont
  {Oktay}}, \bibinfo {author} {\bibfnamefont {S.-W.}\ \bibnamefont {Qiu}},
  \bibinfo {author} {\bibfnamefont {J.~N.}\ \bibnamefont {Simone}}, \bibinfo
  {author} {\bibfnamefont {R.}~\bibnamefont {Sugar}}, \bibinfo {author}
  {\bibfnamefont {D.}~\bibnamefont {Toussaint}}, \bibinfo {author}
  {\bibfnamefont {R.~S.~V.}\ \bibnamefont {de~Water}},\ and\ \bibinfo {author}
  {\bibfnamefont {R.}~\bibnamefont {Zhou}},\ }\href@noop {} {\bibinfo {title}
  {K semileptonic form factor with hisq fermions at the physical point}}
  (\bibinfo {year} {2014}),\ \Eprint {https://arxiv.org/abs/1311.7264}
  {arXiv:1311.7264 [hep-lat]} \BibitemShut {NoStop}%
\bibitem [{\citenamefont {Bazavov}\ \emph {et~al.}(2018)\citenamefont
  {Bazavov}, \citenamefont {Bernard}, \citenamefont {Brown}, \citenamefont
  {DeTar}, \citenamefont {El-Khadra}, \citenamefont {G\'amiz}, \citenamefont
  {Gottlieb}, \citenamefont {Heller}, \citenamefont {Komijani}, \citenamefont
  {Kronfeld}, \citenamefont {Laiho}, \citenamefont {Mackenzie}, \citenamefont
  {Neil}, \citenamefont {Simone}, \citenamefont {Sugar}, \citenamefont
  {Toussaint},\ and\ \citenamefont {Van~de Water}}]{PhysRevD.98.074512}%
  \BibitemOpen
  \bibfield  {author} {\bibinfo {author} {\bibfnamefont {A.}~\bibnamefont
  {Bazavov}}, \bibinfo {author} {\bibfnamefont {C.}~\bibnamefont {Bernard}},
  \bibinfo {author} {\bibfnamefont {N.}~\bibnamefont {Brown}}, \bibinfo
  {author} {\bibfnamefont {C.}~\bibnamefont {DeTar}}, \bibinfo {author}
  {\bibfnamefont {A.~X.}\ \bibnamefont {El-Khadra}}, \bibinfo {author}
  {\bibfnamefont {E.}~\bibnamefont {G\'amiz}}, \bibinfo {author} {\bibfnamefont
  {S.}~\bibnamefont {Gottlieb}}, \bibinfo {author} {\bibfnamefont {U.~M.}\
  \bibnamefont {Heller}}, \bibinfo {author} {\bibfnamefont {J.}~\bibnamefont
  {Komijani}}, \bibinfo {author} {\bibfnamefont {A.~S.}\ \bibnamefont
  {Kronfeld}}, \bibinfo {author} {\bibfnamefont {J.}~\bibnamefont {Laiho}},
  \bibinfo {author} {\bibfnamefont {P.~B.}\ \bibnamefont {Mackenzie}}, \bibinfo
  {author} {\bibfnamefont {E.~T.}\ \bibnamefont {Neil}}, \bibinfo {author}
  {\bibfnamefont {J.~N.}\ \bibnamefont {Simone}}, \bibinfo {author}
  {\bibfnamefont {R.~L.}\ \bibnamefont {Sugar}}, \bibinfo {author}
  {\bibfnamefont {D.}~\bibnamefont {Toussaint}},\ and\ \bibinfo {author}
  {\bibfnamefont {R.~S.}\ \bibnamefont {Van~de Water}} (\bibinfo
  {collaboration} {Fermilab Lattice and MILC Collaborations}),\ }\bibfield
  {title} {\bibinfo {title} {$b$- and $d$-meson leptonic decay constants from
  four-flavor lattice qcd},\ }\href
  {https://doi.org/10.1103/PhysRevD.98.074512} {\bibfield  {journal} {\bibinfo
  {journal} {Phys. Rev. D}\ }\textbf {\bibinfo {volume} {98}},\ \bibinfo
  {pages} {074512} (\bibinfo {year} {2018})}\BibitemShut {NoStop}%
\bibitem [{\citenamefont {Bazavov}\ \emph
  {et~al.}(2014{\natexlab{b}})\citenamefont {Bazavov}, \citenamefont {Bernard},
  \citenamefont {Komijani}, \citenamefont {Bouchard}, \citenamefont {DeTar},
  \citenamefont {Foley}, \citenamefont {Levkova}, \citenamefont {Du},
  \citenamefont {Laiho}, \citenamefont {El-Khadra}, \citenamefont {Freeland},
  \citenamefont {G\'amiz}, \citenamefont {Gottlieb}, \citenamefont {Heller},
  \citenamefont {Kim}, \citenamefont {Toussaint}, \citenamefont {Kronfeld},
  \citenamefont {Mackenzie}, \citenamefont {Simone}, \citenamefont {Van~de
  Water}, \citenamefont {Zhou}, \citenamefont {Neil},\ and\ \citenamefont
  {Sugar}}]{PhysRevD.90.074509}%
  \BibitemOpen
  \bibfield  {author} {\bibinfo {author} {\bibfnamefont {A.}~\bibnamefont
  {Bazavov}}, \bibinfo {author} {\bibfnamefont {C.}~\bibnamefont {Bernard}},
  \bibinfo {author} {\bibfnamefont {J.}~\bibnamefont {Komijani}}, \bibinfo
  {author} {\bibfnamefont {C.~M.}\ \bibnamefont {Bouchard}}, \bibinfo {author}
  {\bibfnamefont {C.}~\bibnamefont {DeTar}}, \bibinfo {author} {\bibfnamefont
  {J.}~\bibnamefont {Foley}}, \bibinfo {author} {\bibfnamefont
  {L.}~\bibnamefont {Levkova}}, \bibinfo {author} {\bibfnamefont
  {D.}~\bibnamefont {Du}}, \bibinfo {author} {\bibfnamefont {J.}~\bibnamefont
  {Laiho}}, \bibinfo {author} {\bibfnamefont {A.~X.}\ \bibnamefont
  {El-Khadra}}, \bibinfo {author} {\bibfnamefont {E.~D.}\ \bibnamefont
  {Freeland}}, \bibinfo {author} {\bibfnamefont {E.}~\bibnamefont {G\'amiz}},
  \bibinfo {author} {\bibfnamefont {S.}~\bibnamefont {Gottlieb}}, \bibinfo
  {author} {\bibfnamefont {U.~M.}\ \bibnamefont {Heller}}, \bibinfo {author}
  {\bibfnamefont {J.}~\bibnamefont {Kim}}, \bibinfo {author} {\bibfnamefont
  {D.}~\bibnamefont {Toussaint}}, \bibinfo {author} {\bibfnamefont {A.~S.}\
  \bibnamefont {Kronfeld}}, \bibinfo {author} {\bibfnamefont {P.~B.}\
  \bibnamefont {Mackenzie}}, \bibinfo {author} {\bibfnamefont {J.~N.}\
  \bibnamefont {Simone}}, \bibinfo {author} {\bibfnamefont {R.~S.}\
  \bibnamefont {Van~de Water}}, \bibinfo {author} {\bibfnamefont
  {R.}~\bibnamefont {Zhou}}, \bibinfo {author} {\bibfnamefont {E.~T.}\
  \bibnamefont {Neil}},\ and\ \bibinfo {author} {\bibfnamefont
  {R.}~\bibnamefont {Sugar}} (\bibinfo {collaboration} {Fermilab Lattice and
  MILC Collaborations}),\ }\bibfield  {title} {\bibinfo {title} {Charmed and
  light pseudoscalar meson decay constants from four-flavor lattice qcd with
  physical light quarks},\ }\href {https://doi.org/10.1103/PhysRevD.90.074509}
  {\bibfield  {journal} {\bibinfo  {journal} {Phys. Rev. D}\ }\textbf {\bibinfo
  {volume} {90}},\ \bibinfo {pages} {074509} (\bibinfo {year}
  {2014}{\natexlab{b}})}\BibitemShut {NoStop}%
\bibitem [{\citenamefont {Chang}\ \emph {et~al.}(2018)\citenamefont {Chang},
  \citenamefont {Nicholson}, \citenamefont {Rinaldi}, \citenamefont
  {Berkowitz}, \citenamefont {Garron}, \citenamefont {Brantley}, \citenamefont
  {Monge-Camacho}, \citenamefont {Monahan}, \citenamefont {Bouchard},
  \citenamefont {Clark},\ and\ \citenamefont {et~al.}}]{2018axial}%
  \BibitemOpen
  \bibfield  {author} {\bibinfo {author} {\bibfnamefont {C.~C.}\ \bibnamefont
  {Chang}}, \bibinfo {author} {\bibfnamefont {A.~N.}\ \bibnamefont
  {Nicholson}}, \bibinfo {author} {\bibfnamefont {E.}~\bibnamefont {Rinaldi}},
  \bibinfo {author} {\bibfnamefont {E.}~\bibnamefont {Berkowitz}}, \bibinfo
  {author} {\bibfnamefont {N.}~\bibnamefont {Garron}}, \bibinfo {author}
  {\bibfnamefont {D.~A.}\ \bibnamefont {Brantley}}, \bibinfo {author}
  {\bibfnamefont {H.}~\bibnamefont {Monge-Camacho}}, \bibinfo {author}
  {\bibfnamefont {C.~J.}\ \bibnamefont {Monahan}}, \bibinfo {author}
  {\bibfnamefont {C.}~\bibnamefont {Bouchard}}, \bibinfo {author}
  {\bibfnamefont {M.~A.}\ \bibnamefont {Clark}},\ and\ \bibinfo {author}
  {\bibnamefont {et~al.}},\ }\bibfield  {title} {\bibinfo {title} {A
  per-cent-level determination of the nucleon axial coupling from quantum
  chromodynamics},\ }\href {https://doi.org/10.1038/s41586-018-0161-8}
  {\bibfield  {journal} {\bibinfo  {journal} {Nature}\ }\textbf {\bibinfo
  {volume} {558}},\ \bibinfo {pages} {91–94} (\bibinfo {year}
  {2018})}\BibitemShut {NoStop}%
\bibitem [{\citenamefont {Bhattacharya}\ \emph {et~al.}(2016)\citenamefont
  {Bhattacharya}, \citenamefont {Cirigliano}, \citenamefont {Cohen},
  \citenamefont {Gupta}, \citenamefont {Lin},\ and\ \citenamefont
  {Yoon}}]{PhysRevD.94.054508}%
  \BibitemOpen
  \bibfield  {author} {\bibinfo {author} {\bibfnamefont {T.}~\bibnamefont
  {Bhattacharya}}, \bibinfo {author} {\bibfnamefont {V.}~\bibnamefont
  {Cirigliano}}, \bibinfo {author} {\bibfnamefont {S.~D.}\ \bibnamefont
  {Cohen}}, \bibinfo {author} {\bibfnamefont {R.}~\bibnamefont {Gupta}},
  \bibinfo {author} {\bibfnamefont {H.-W.}\ \bibnamefont {Lin}},\ and\ \bibinfo
  {author} {\bibfnamefont {B.}~\bibnamefont {Yoon}} (\bibinfo {collaboration}
  {Precision Neutron Decay Matrix Elements (PNDME) Collaboration}),\ }\bibfield
   {title} {\bibinfo {title} {Axial, scalar, and tensor charges of the nucleon
  from $2+1+1$-flavor lattice qcd},\ }\href
  {https://doi.org/10.1103/PhysRevD.94.054508} {\bibfield  {journal} {\bibinfo
  {journal} {Phys. Rev. D}\ }\textbf {\bibinfo {volume} {94}},\ \bibinfo
  {pages} {054508} (\bibinfo {year} {2016})}\BibitemShut {NoStop}%
\bibitem [{\citenamefont {Savage}\ \emph {et~al.}(2017)\citenamefont {Savage},
  \citenamefont {Shanahan}, \citenamefont {Tiburzi}, \citenamefont {Wagman},
  \citenamefont {Winter}, \citenamefont {Beane}, \citenamefont {Chang},
  \citenamefont {Davoudi}, \citenamefont {Detmold},\ and\ \citenamefont
  {Orginos}}]{PhysRevLett.119.062002}%
  \BibitemOpen
  \bibfield  {author} {\bibinfo {author} {\bibfnamefont {M.~J.}\ \bibnamefont
  {Savage}}, \bibinfo {author} {\bibfnamefont {P.~E.}\ \bibnamefont
  {Shanahan}}, \bibinfo {author} {\bibfnamefont {B.~C.}\ \bibnamefont
  {Tiburzi}}, \bibinfo {author} {\bibfnamefont {M.~L.}\ \bibnamefont {Wagman}},
  \bibinfo {author} {\bibfnamefont {F.}~\bibnamefont {Winter}}, \bibinfo
  {author} {\bibfnamefont {S.~R.}\ \bibnamefont {Beane}}, \bibinfo {author}
  {\bibfnamefont {E.}~\bibnamefont {Chang}}, \bibinfo {author} {\bibfnamefont
  {Z.}~\bibnamefont {Davoudi}}, \bibinfo {author} {\bibfnamefont
  {W.}~\bibnamefont {Detmold}},\ and\ \bibinfo {author} {\bibfnamefont
  {K.}~\bibnamefont {Orginos}} (\bibinfo {collaboration} {NPLQCD
  Collaboration}),\ }\bibfield  {title} {\bibinfo {title} {Proton-proton fusion
  and tritium $\ensuremath{\beta}$ decay from lattice quantum chromodynamics},\
  }\href {https://doi.org/10.1103/PhysRevLett.119.062002} {\bibfield  {journal}
  {\bibinfo  {journal} {Phys. Rev. Lett.}\ }\textbf {\bibinfo {volume} {119}},\
  \bibinfo {pages} {062002} (\bibinfo {year} {2017})}\BibitemShut {NoStop}%
\bibitem [{\citenamefont {Beane}\ \emph {et~al.}(2008)\citenamefont {Beane},
  \citenamefont {Luu}, \citenamefont {Orginos}, \citenamefont {Parre\~no},
  \citenamefont {Savage}, \citenamefont {Torok},\ and\ \citenamefont
  {Walker-Loud}}]{PhysRevD.77.094507}%
  \BibitemOpen
  \bibfield  {author} {\bibinfo {author} {\bibfnamefont {S.~R.}\ \bibnamefont
  {Beane}}, \bibinfo {author} {\bibfnamefont {T.~C.}\ \bibnamefont {Luu}},
  \bibinfo {author} {\bibfnamefont {K.}~\bibnamefont {Orginos}}, \bibinfo
  {author} {\bibfnamefont {A.}~\bibnamefont {Parre\~no}}, \bibinfo {author}
  {\bibfnamefont {M.~J.}\ \bibnamefont {Savage}}, \bibinfo {author}
  {\bibfnamefont {A.}~\bibnamefont {Torok}},\ and\ \bibinfo {author}
  {\bibfnamefont {A.}~\bibnamefont {Walker-Loud}} (\bibinfo {collaboration}
  {NPLQCD Collaboration}),\ }\bibfield  {title} {\bibinfo {title}
  {${K}^{+}{K}^{+}$ scattering length from lattice qcd},\ }\href
  {https://doi.org/10.1103/PhysRevD.77.094507} {\bibfield  {journal} {\bibinfo
  {journal} {Phys. Rev. D}\ }\textbf {\bibinfo {volume} {77}},\ \bibinfo
  {pages} {094507} (\bibinfo {year} {2008})}\BibitemShut {NoStop}%
\bibitem [{\citenamefont {Beane}\ \emph {et~al.}(2007)\citenamefont {Beane},
  \citenamefont {Bedaque}, \citenamefont {Luu}, \citenamefont {Orginos},
  \citenamefont {Pallante}, \citenamefont {Parreño},\ and\ \citenamefont
  {Savage}}]{BEANE200762}%
  \BibitemOpen
  \bibfield  {author} {\bibinfo {author} {\bibfnamefont {S.~R.}\ \bibnamefont
  {Beane}}, \bibinfo {author} {\bibfnamefont {P.~F.}\ \bibnamefont {Bedaque}},
  \bibinfo {author} {\bibfnamefont {T.~C.}\ \bibnamefont {Luu}}, \bibinfo
  {author} {\bibfnamefont {K.}~\bibnamefont {Orginos}}, \bibinfo {author}
  {\bibfnamefont {E.}~\bibnamefont {Pallante}}, \bibinfo {author}
  {\bibfnamefont {A.}~\bibnamefont {Parreño}},\ and\ \bibinfo {author}
  {\bibfnamefont {M.~J.}\ \bibnamefont {Savage}},\ }\bibfield  {title}
  {\bibinfo {title} {Hyperon–nucleon scattering from fully-dynamical lattice
  qcd},\ }\href
  {https://doi.org/https://doi.org/10.1016/j.nuclphysa.2007.07.006} {\bibfield
  {journal} {\bibinfo  {journal} {Nuclear Physics A}\ }\textbf {\bibinfo
  {volume} {794}},\ \bibinfo {pages} {62} (\bibinfo {year} {2007})}\BibitemShut
  {NoStop}%
\bibitem [{\citenamefont {Blanton}\ \emph {et~al.}(2020)\citenamefont
  {Blanton}, \citenamefont {Romero-L\'opez},\ and\ \citenamefont
  {Sharpe}}]{PhysRevLett.124.032001}%
  \BibitemOpen
  \bibfield  {author} {\bibinfo {author} {\bibfnamefont {T.~D.}\ \bibnamefont
  {Blanton}}, \bibinfo {author} {\bibfnamefont {F.}~\bibnamefont
  {Romero-L\'opez}},\ and\ \bibinfo {author} {\bibfnamefont {S.~R.}\
  \bibnamefont {Sharpe}},\ }\bibfield  {title} {\bibinfo {title} {$i=3$
  three-pion scattering amplitude from lattice qcd},\ }\href
  {https://doi.org/10.1103/PhysRevLett.124.032001} {\bibfield  {journal}
  {\bibinfo  {journal} {Phys. Rev. Lett.}\ }\textbf {\bibinfo {volume} {124}},\
  \bibinfo {pages} {032001} (\bibinfo {year} {2020})}\BibitemShut {NoStop}%
\bibitem [{\citenamefont {Davoudi}\ \emph
  {et~al.}(2021{\natexlab{a}})\citenamefont {Davoudi}, \citenamefont {Detmold},
  \citenamefont {Shanahan}, \citenamefont {Orginos}, \citenamefont {Parreño},
  \citenamefont {Savage},\ and\ \citenamefont {Wagman}}]{DAVOUDI20211}%
  \BibitemOpen
  \bibfield  {author} {\bibinfo {author} {\bibfnamefont {Z.}~\bibnamefont
  {Davoudi}}, \bibinfo {author} {\bibfnamefont {W.}~\bibnamefont {Detmold}},
  \bibinfo {author} {\bibfnamefont {P.}~\bibnamefont {Shanahan}}, \bibinfo
  {author} {\bibfnamefont {K.}~\bibnamefont {Orginos}}, \bibinfo {author}
  {\bibfnamefont {A.}~\bibnamefont {Parreño}}, \bibinfo {author}
  {\bibfnamefont {M.~J.}\ \bibnamefont {Savage}},\ and\ \bibinfo {author}
  {\bibfnamefont {M.~L.}\ \bibnamefont {Wagman}},\ }\bibfield  {title}
  {\bibinfo {title} {Nuclear matrix elements from lattice qcd for electroweak
  and beyond-standard-model processes},\ }\href
  {https://doi.org/https://doi.org/10.1016/j.physrep.2020.10.004} {\bibfield
  {journal} {\bibinfo  {journal} {Physics Reports}\ }\textbf {\bibinfo {volume}
  {900}},\ \bibinfo {pages} {1} (\bibinfo {year} {2021}{\natexlab{a}})},\
  \bibinfo {note} {nuclear matrix elements from lattice QCD for electroweak and
  beyond–Standard-Model processes}\BibitemShut {NoStop}%
\bibitem [{\citenamefont {Aoki}\ \emph {et~al.}(2020)\citenamefont {Aoki},
  \citenamefont {Aoki}, \citenamefont {Bečirević}, \citenamefont {Blum},
  \citenamefont {Colangelo}, \citenamefont {Collins}, \citenamefont
  {Della~Morte}, \citenamefont {Dimopoulos}, \citenamefont {Dürr},
  \citenamefont {Fukaya},\ and\ \citenamefont {et~al.}}]{Aoki_2020}%
  \BibitemOpen
  \bibfield  {author} {\bibinfo {author} {\bibfnamefont {S.}~\bibnamefont
  {Aoki}}, \bibinfo {author} {\bibfnamefont {Y.}~\bibnamefont {Aoki}}, \bibinfo
  {author} {\bibfnamefont {D.}~\bibnamefont {Bečirević}}, \bibinfo {author}
  {\bibfnamefont {T.}~\bibnamefont {Blum}}, \bibinfo {author} {\bibfnamefont
  {G.}~\bibnamefont {Colangelo}}, \bibinfo {author} {\bibfnamefont
  {S.}~\bibnamefont {Collins}}, \bibinfo {author} {\bibfnamefont
  {M.}~\bibnamefont {Della~Morte}}, \bibinfo {author} {\bibfnamefont
  {P.}~\bibnamefont {Dimopoulos}}, \bibinfo {author} {\bibfnamefont
  {S.}~\bibnamefont {Dürr}}, \bibinfo {author} {\bibfnamefont
  {H.}~\bibnamefont {Fukaya}},\ and\ \bibinfo {author} {\bibnamefont
  {et~al.}},\ }\bibfield  {title} {\bibinfo {title} {Flag review 2019},\
  }\bibfield  {journal} {\bibinfo  {journal} {The European Physical Journal C}\
  }\textbf {\bibinfo {volume} {80}},\ \href
  {https://doi.org/10.1140/epjc/s10052-019-7354-7}
  {10.1140/epjc/s10052-019-7354-7} (\bibinfo {year} {2020})\BibitemShut
  {NoStop}%
\bibitem [{\citenamefont {Allton}\ \emph {et~al.}(2002)\citenamefont {Allton},
  \citenamefont {Ejiri}, \citenamefont {Hands}, \citenamefont {Kaczmarek},
  \citenamefont {Karsch}, \citenamefont {Laermann}, \citenamefont {Schmidt},\
  and\ \citenamefont {Scorzato}}]{PhysRevD.66.074507}%
  \BibitemOpen
  \bibfield  {author} {\bibinfo {author} {\bibfnamefont {C.~R.}\ \bibnamefont
  {Allton}}, \bibinfo {author} {\bibfnamefont {S.}~\bibnamefont {Ejiri}},
  \bibinfo {author} {\bibfnamefont {S.~J.}\ \bibnamefont {Hands}}, \bibinfo
  {author} {\bibfnamefont {O.}~\bibnamefont {Kaczmarek}}, \bibinfo {author}
  {\bibfnamefont {F.}~\bibnamefont {Karsch}}, \bibinfo {author} {\bibfnamefont
  {E.}~\bibnamefont {Laermann}}, \bibinfo {author} {\bibfnamefont
  {C.}~\bibnamefont {Schmidt}},\ and\ \bibinfo {author} {\bibfnamefont
  {L.}~\bibnamefont {Scorzato}},\ }\bibfield  {title} {\bibinfo {title} {Qcd
  thermal phase transition in the presence of a small chemical potential},\
  }\href {https://doi.org/10.1103/PhysRevD.66.074507} {\bibfield  {journal}
  {\bibinfo  {journal} {Phys. Rev. D}\ }\textbf {\bibinfo {volume} {66}},\
  \bibinfo {pages} {074507} (\bibinfo {year} {2002})}\BibitemShut {NoStop}%
\bibitem [{\citenamefont {de~Forcrand}\ and\ \citenamefont
  {Philipsen}(2002)}]{DEFORCRAND2002290}%
  \BibitemOpen
  \bibfield  {author} {\bibinfo {author} {\bibfnamefont {P.}~\bibnamefont
  {de~Forcrand}}\ and\ \bibinfo {author} {\bibfnamefont {O.}~\bibnamefont
  {Philipsen}},\ }\bibfield  {title} {\bibinfo {title} {The qcd phase diagram
  for small densities from imaginary chemical potential},\ }\href
  {https://doi.org/https://doi.org/10.1016/S0550-3213(02)00626-0} {\bibfield
  {journal} {\bibinfo  {journal} {Nuclear Physics B}\ }\textbf {\bibinfo
  {volume} {642}},\ \bibinfo {pages} {290} (\bibinfo {year}
  {2002})}\BibitemShut {NoStop}%
\bibitem [{\citenamefont {Seiler}\ \emph {et~al.}(2013)\citenamefont {Seiler},
  \citenamefont {Sexty},\ and\ \citenamefont {Stamatescu}}]{SEILER2013213}%
  \BibitemOpen
  \bibfield  {author} {\bibinfo {author} {\bibfnamefont {E.}~\bibnamefont
  {Seiler}}, \bibinfo {author} {\bibfnamefont {D.}~\bibnamefont {Sexty}},\ and\
  \bibinfo {author} {\bibfnamefont {I.-O.}\ \bibnamefont {Stamatescu}},\
  }\bibfield  {title} {\bibinfo {title} {Gauge cooling in complex langevin for
  lattice qcd with heavy quarks},\ }\href
  {https://doi.org/https://doi.org/10.1016/j.physletb.2013.04.062} {\bibfield
  {journal} {\bibinfo  {journal} {Physics Letters B}\ }\textbf {\bibinfo
  {volume} {723}},\ \bibinfo {pages} {213} (\bibinfo {year}
  {2013})}\BibitemShut {NoStop}%
\bibitem [{\citenamefont {Hasenfratz}\ and\ \citenamefont
  {Toussaint}(1992)}]{HASENFRATZ1992539}%
  \BibitemOpen
  \bibfield  {author} {\bibinfo {author} {\bibfnamefont {A.}~\bibnamefont
  {Hasenfratz}}\ and\ \bibinfo {author} {\bibfnamefont {D.}~\bibnamefont
  {Toussaint}},\ }\bibfield  {title} {\bibinfo {title} {Canonical ensembles and
  nonzero density quantum chromodynamics},\ }\href
  {https://doi.org/https://doi.org/10.1016/0550-3213(92)90247-9} {\bibfield
  {journal} {\bibinfo  {journal} {Nuclear Physics B}\ }\textbf {\bibinfo
  {volume} {371}},\ \bibinfo {pages} {539} (\bibinfo {year}
  {1992})}\BibitemShut {NoStop}%
\bibitem [{\citenamefont {Giordano}\ \emph {et~al.}(2020)\citenamefont
  {Giordano}, \citenamefont {Kapas}, \citenamefont {Katz}, \citenamefont
  {Nogradi},\ and\ \citenamefont {Pasztor}}]{Density2020}%
  \BibitemOpen
  \bibfield  {author} {\bibinfo {author} {\bibfnamefont {M.}~\bibnamefont
  {Giordano}}, \bibinfo {author} {\bibfnamefont {K.}~\bibnamefont {Kapas}},
  \bibinfo {author} {\bibfnamefont {S.~D.}\ \bibnamefont {Katz}}, \bibinfo
  {author} {\bibfnamefont {D.}~\bibnamefont {Nogradi}},\ and\ \bibinfo {author}
  {\bibfnamefont {A.}~\bibnamefont {Pasztor}},\ }\bibfield  {title} {\bibinfo
  {title} {New approach to lattice qcd at finite density; results for the
  critical end point on coarse lattices},\ }\bibfield  {journal} {\bibinfo
  {journal} {Journal of High Energy Physics}\ }\textbf {\bibinfo {volume}
  {2020}},\ \href {https://doi.org/10.1007/jhep05(2020)088}
  {10.1007/jhep05(2020)088} (\bibinfo {year} {2020})\BibitemShut {NoStop}%
\bibitem [{\citenamefont {\"Unsal}(2012)}]{PhysRevD.86.105012}%
  \BibitemOpen
  \bibfield  {author} {\bibinfo {author} {\bibfnamefont {M.}~\bibnamefont
  {\"Unsal}},\ }\bibfield  {title} {\bibinfo {title} {Theta dependence, sign
  problems, and topological interference},\ }\href
  {https://doi.org/10.1103/PhysRevD.86.105012} {\bibfield  {journal} {\bibinfo
  {journal} {Phys. Rev. D}\ }\textbf {\bibinfo {volume} {86}},\ \bibinfo
  {pages} {105012} (\bibinfo {year} {2012})}\BibitemShut {NoStop}%
\bibitem [{\citenamefont {Feynman}(1982)}]{Feynman1982}%
  \BibitemOpen
  \bibfield  {author} {\bibinfo {author} {\bibfnamefont {R.~P.}\ \bibnamefont
  {Feynman}},\ }\bibfield  {title} {\bibinfo {title} {Simulating physics with
  computers},\ }\href {https://doi.org/10.1007/BF02650179} {\bibfield
  {journal} {\bibinfo  {journal} {International Journal of Theoretical
  Physics}\ }\textbf {\bibinfo {volume} {21}},\ \bibinfo {pages} {467}
  (\bibinfo {year} {1982})}\BibitemShut {NoStop}%
\bibitem [{\citenamefont {Benioff}(1980)}]{Benioff:1980}%
  \BibitemOpen
  \bibfield  {author} {\bibinfo {author} {\bibfnamefont {P.}~\bibnamefont
  {Benioff}},\ }\bibfield  {title} {\bibinfo {title} {The computer as a
  physical system: A microscopic quantum mechanical hamiltonian model of
  computers as represented by turing machines},\ }\href@noop {} {\bibfield
  {journal} {\bibinfo  {journal} {Journal of statistical physics}\ }\textbf
  {\bibinfo {volume} {22}},\ \bibinfo {pages} {563} (\bibinfo {year}
  {1980})}\BibitemShut {NoStop}%
\bibitem [{\citenamefont {Huang}\ \emph {et~al.}(2020)\citenamefont {Huang},
  \citenamefont {Wu}, \citenamefont {Fan},\ and\ \citenamefont
  {Zhu}}]{huang2020superconducting}%
  \BibitemOpen
  \bibfield  {author} {\bibinfo {author} {\bibfnamefont {H.-L.}\ \bibnamefont
  {Huang}}, \bibinfo {author} {\bibfnamefont {D.}~\bibnamefont {Wu}}, \bibinfo
  {author} {\bibfnamefont {D.}~\bibnamefont {Fan}},\ and\ \bibinfo {author}
  {\bibfnamefont {X.}~\bibnamefont {Zhu}},\ }\href@noop {} {\bibinfo {title}
  {Superconducting quantum computing: A review}} (\bibinfo {year} {2020}),\
  \Eprint {https://arxiv.org/abs/2006.10433} {arXiv:2006.10433 [quant-ph]}
  \BibitemShut {NoStop}%
\bibitem [{\citenamefont {Bruzewicz}\ \emph {et~al.}(2019)\citenamefont
  {Bruzewicz}, \citenamefont {Chiaverini}, \citenamefont {McConnell},\ and\
  \citenamefont {Sage}}]{2019Trap}%
  \BibitemOpen
  \bibfield  {author} {\bibinfo {author} {\bibfnamefont {C.~D.}\ \bibnamefont
  {Bruzewicz}}, \bibinfo {author} {\bibfnamefont {J.}~\bibnamefont
  {Chiaverini}}, \bibinfo {author} {\bibfnamefont {R.}~\bibnamefont
  {McConnell}},\ and\ \bibinfo {author} {\bibfnamefont {J.~M.}\ \bibnamefont
  {Sage}},\ }\bibfield  {title} {\bibinfo {title} {Trapped-ion quantum
  computing: Progress and challenges},\ }\href
  {https://doi.org/10.1063/1.5088164} {\bibfield  {journal} {\bibinfo
  {journal} {Applied Physics Reviews}\ }\textbf {\bibinfo {volume} {6}},\
  \bibinfo {pages} {021314} (\bibinfo {year} {2019})}\BibitemShut {NoStop}%
\bibitem [{\citenamefont {Slussarenko}\ and\ \citenamefont
  {Pryde}(2019)}]{2019Photon}%
  \BibitemOpen
  \bibfield  {author} {\bibinfo {author} {\bibfnamefont {S.}~\bibnamefont
  {Slussarenko}}\ and\ \bibinfo {author} {\bibfnamefont {G.~J.}\ \bibnamefont
  {Pryde}},\ }\bibfield  {title} {\bibinfo {title} {Photonic quantum
  information processing: A concise review},\ }\href
  {https://doi.org/10.1063/1.5115814} {\bibfield  {journal} {\bibinfo
  {journal} {Applied Physics Reviews}\ }\textbf {\bibinfo {volume} {6}},\
  \bibinfo {pages} {041303} (\bibinfo {year} {2019})}\BibitemShut {NoStop}%
\bibitem [{\citenamefont {Preskill}(2018)}]{preskill2018simulating}%
  \BibitemOpen
  \bibfield  {author} {\bibinfo {author} {\bibfnamefont {J.}~\bibnamefont
  {Preskill}},\ }\href@noop {} {\bibinfo {title} {Simulating quantum field
  theory with a quantum computer}} (\bibinfo {year} {2018}),\ \Eprint
  {https://arxiv.org/abs/1811.10085} {arXiv:1811.10085 [hep-lat]} \BibitemShut
  {NoStop}%
\bibitem [{\citenamefont {Jordan}\ \emph {et~al.}(2019)\citenamefont {Jordan},
  \citenamefont {Lee},\ and\ \citenamefont {Preskill}}]{jordan2019quantum}%
  \BibitemOpen
  \bibfield  {author} {\bibinfo {author} {\bibfnamefont {S.~P.}\ \bibnamefont
  {Jordan}}, \bibinfo {author} {\bibfnamefont {K.~S.~M.}\ \bibnamefont {Lee}},\
  and\ \bibinfo {author} {\bibfnamefont {J.}~\bibnamefont {Preskill}},\
  }\href@noop {} {\bibinfo {title} {Quantum computation of scattering in scalar
  quantum field theories}} (\bibinfo {year} {2019}),\ \Eprint
  {https://arxiv.org/abs/1112.4833} {arXiv:1112.4833 [hep-th]} \BibitemShut
  {NoStop}%
\bibitem [{\citenamefont {Somma}(2016)}]{10.5555/3179430.3179434}%
  \BibitemOpen
  \bibfield  {author} {\bibinfo {author} {\bibfnamefont {R.~D.}\ \bibnamefont
  {Somma}},\ }\bibfield  {title} {\bibinfo {title} {Quantum simulations of one
  dimensional quantum systems},\ }\href@noop {} {\bibfield  {journal} {\bibinfo
   {journal} {Quantum Info. Comput.}\ }\textbf {\bibinfo {volume} {16}},\
  \bibinfo {pages} {1125–1168} (\bibinfo {year} {2016})}\BibitemShut
  {NoStop}%
\bibitem [{\citenamefont {Macridin}\ \emph {et~al.}(2018)\citenamefont
  {Macridin}, \citenamefont {Spentzouris}, \citenamefont {Amundson},\ and\
  \citenamefont {Harnik}}]{PhysRevA.98.042312}%
  \BibitemOpen
  \bibfield  {author} {\bibinfo {author} {\bibfnamefont {A.}~\bibnamefont
  {Macridin}}, \bibinfo {author} {\bibfnamefont {P.}~\bibnamefont
  {Spentzouris}}, \bibinfo {author} {\bibfnamefont {J.}~\bibnamefont
  {Amundson}},\ and\ \bibinfo {author} {\bibfnamefont {R.}~\bibnamefont
  {Harnik}},\ }\bibfield  {title} {\bibinfo {title} {Digital quantum
  computation of fermion-boson interacting systems},\ }\href
  {https://doi.org/10.1103/PhysRevA.98.042312} {\bibfield  {journal} {\bibinfo
  {journal} {Phys. Rev. A}\ }\textbf {\bibinfo {volume} {98}},\ \bibinfo
  {pages} {042312} (\bibinfo {year} {2018})}\BibitemShut {NoStop}%
\bibitem [{\citenamefont {Klco}\ and\ \citenamefont
  {Savage}(2019)}]{Klco:2018zqz}%
  \BibitemOpen
  \bibfield  {author} {\bibinfo {author} {\bibfnamefont {N.}~\bibnamefont
  {Klco}}\ and\ \bibinfo {author} {\bibfnamefont {M.~J.}\ \bibnamefont
  {Savage}},\ }\bibfield  {title} {\bibinfo {title} {Digitization of scalar
  fields for quantum computing},\ }\href
  {https://doi.org/10.1103/PhysRevA.99.052335} {\bibfield  {journal} {\bibinfo
  {journal} {Phys. Rev.}\ }\textbf {\bibinfo {volume} {A99}},\ \bibinfo {pages}
  {052335} (\bibinfo {year} {2019})},\ \Eprint
  {https://arxiv.org/abs/1808.10378} {arXiv:1808.10378 [quant-ph]} \BibitemShut
  {NoStop}%
\bibitem [{\citenamefont {Yeter-Aydeniz}\ \emph {et~al.}(2019)\citenamefont
  {Yeter-Aydeniz}, \citenamefont {Dumitrescu}, \citenamefont {McCaskey},
  \citenamefont {Bennink}, \citenamefont {Pooser},\ and\ \citenamefont
  {Siopsis}}]{Yeter_Aydeniz_2019}%
  \BibitemOpen
  \bibfield  {author} {\bibinfo {author} {\bibfnamefont {K.}~\bibnamefont
  {Yeter-Aydeniz}}, \bibinfo {author} {\bibfnamefont {E.~F.}\ \bibnamefont
  {Dumitrescu}}, \bibinfo {author} {\bibfnamefont {A.~J.}\ \bibnamefont
  {McCaskey}}, \bibinfo {author} {\bibfnamefont {R.~S.}\ \bibnamefont
  {Bennink}}, \bibinfo {author} {\bibfnamefont {R.~C.}\ \bibnamefont
  {Pooser}},\ and\ \bibinfo {author} {\bibfnamefont {G.}~\bibnamefont
  {Siopsis}},\ }\bibfield  {title} {\bibinfo {title} {Scalar quantum field
  theories as a benchmark for near-term quantum computers},\ }\bibfield
  {journal} {\bibinfo  {journal} {Physical Review A}\ }\textbf {\bibinfo
  {volume} {99}},\ \href {https://doi.org/10.1103/physreva.99.032306}
  {10.1103/physreva.99.032306} (\bibinfo {year} {2019})\BibitemShut {NoStop}%
\bibitem [{\citenamefont {Barata}\ \emph {et~al.}(2021)\citenamefont {Barata},
  \citenamefont {Mueller}, \citenamefont {Tarasov},\ and\ \citenamefont
  {Venugopalan}}]{PhysRevA.103.042410}%
  \BibitemOpen
  \bibfield  {author} {\bibinfo {author} {\bibfnamefont {J.~a.}\ \bibnamefont
  {Barata}}, \bibinfo {author} {\bibfnamefont {N.}~\bibnamefont {Mueller}},
  \bibinfo {author} {\bibfnamefont {A.}~\bibnamefont {Tarasov}},\ and\ \bibinfo
  {author} {\bibfnamefont {R.}~\bibnamefont {Venugopalan}},\ }\bibfield
  {title} {\bibinfo {title} {Single-particle digitization strategy for quantum
  computation of a ${\ensuremath{\phi}}^{4}$ scalar field theory},\ }\href
  {https://doi.org/10.1103/PhysRevA.103.042410} {\bibfield  {journal} {\bibinfo
   {journal} {Phys. Rev. A}\ }\textbf {\bibinfo {volume} {103}},\ \bibinfo
  {pages} {042410} (\bibinfo {year} {2021})}\BibitemShut {NoStop}%
\bibitem [{\citenamefont {Jordan}\ \emph {et~al.}(2014)\citenamefont {Jordan},
  \citenamefont {Lee},\ and\ \citenamefont {Preskill}}]{jordan2014quantum}%
  \BibitemOpen
  \bibfield  {author} {\bibinfo {author} {\bibfnamefont {S.~P.}\ \bibnamefont
  {Jordan}}, \bibinfo {author} {\bibfnamefont {K.~S.~M.}\ \bibnamefont {Lee}},\
  and\ \bibinfo {author} {\bibfnamefont {J.}~\bibnamefont {Preskill}},\
  }\href@noop {} {\bibinfo {title} {Quantum algorithms for fermionic quantum
  field theories}} (\bibinfo {year} {2014}),\ \Eprint
  {https://arxiv.org/abs/1404.7115} {arXiv:1404.7115 [hep-th]} \BibitemShut
  {NoStop}%
\bibitem [{\citenamefont {Lamm}\ \emph {et~al.}(2020)\citenamefont {Lamm},
  \citenamefont {Lawrence},\ and\ \citenamefont {Yamauchi}}]{Lamm_2020}%
  \BibitemOpen
  \bibfield  {author} {\bibinfo {author} {\bibfnamefont {H.}~\bibnamefont
  {Lamm}}, \bibinfo {author} {\bibfnamefont {S.}~\bibnamefont {Lawrence}},\
  and\ \bibinfo {author} {\bibfnamefont {Y.}~\bibnamefont {Yamauchi}},\
  }\bibfield  {title} {\bibinfo {title} {Parton physics on a quantum
  computer},\ }\bibfield  {journal} {\bibinfo  {journal} {Physical Review
  Research}\ }\textbf {\bibinfo {volume} {2}},\ \href
  {https://doi.org/10.1103/physrevresearch.2.013272}
  {10.1103/physrevresearch.2.013272} (\bibinfo {year} {2020})\BibitemShut
  {NoStop}%
\bibitem [{\citenamefont {Mazza}\ \emph {et~al.}(2012)\citenamefont {Mazza},
  \citenamefont {Bermudez}, \citenamefont {Goldman}, \citenamefont {Rizzi},
  \citenamefont {Martin-Delgado},\ and\ \citenamefont
  {Lewenstein}}]{Mazza_2012}%
  \BibitemOpen
  \bibfield  {author} {\bibinfo {author} {\bibfnamefont {L.}~\bibnamefont
  {Mazza}}, \bibinfo {author} {\bibfnamefont {A.}~\bibnamefont {Bermudez}},
  \bibinfo {author} {\bibfnamefont {N.}~\bibnamefont {Goldman}}, \bibinfo
  {author} {\bibfnamefont {M.}~\bibnamefont {Rizzi}}, \bibinfo {author}
  {\bibfnamefont {M.~A.}\ \bibnamefont {Martin-Delgado}},\ and\ \bibinfo
  {author} {\bibfnamefont {M.}~\bibnamefont {Lewenstein}},\ }\bibfield  {title}
  {\bibinfo {title} {An optical-lattice-based quantum simulator for
  relativistic field theories and topological insulators},\ }\href
  {https://doi.org/10.1088/1367-2630/14/1/015007} {\bibfield  {journal}
  {\bibinfo  {journal} {New Journal of Physics}\ }\textbf {\bibinfo {volume}
  {14}},\ \bibinfo {pages} {015007} (\bibinfo {year} {2012})}\BibitemShut
  {NoStop}%
\bibitem [{\citenamefont {Alexandru}\ \emph
  {et~al.}(2019{\natexlab{a}})\citenamefont {Alexandru}, \citenamefont
  {Bedaque}, \citenamefont {Lamm},\ and\ \citenamefont
  {Lawrence}}]{Alexandru_2019_2}%
  \BibitemOpen
  \bibfield  {author} {\bibinfo {author} {\bibfnamefont {A.}~\bibnamefont
  {Alexandru}}, \bibinfo {author} {\bibfnamefont {P.~F.}\ \bibnamefont
  {Bedaque}}, \bibinfo {author} {\bibfnamefont {H.}~\bibnamefont {Lamm}},\ and\
  \bibinfo {author} {\bibfnamefont {S.}~\bibnamefont {Lawrence}},\ }\bibfield
  {title} {\bibinfo {title} {Sigma models on quantum computers},\ }\bibfield
  {journal} {\bibinfo  {journal} {Physical Review Letters}\ }\textbf {\bibinfo
  {volume} {123}},\ \href {https://doi.org/10.1103/physrevlett.123.090501}
  {10.1103/physrevlett.123.090501} (\bibinfo {year}
  {2019}{\natexlab{a}})\BibitemShut {NoStop}%
\bibitem [{\citenamefont {Singh}\ and\ \citenamefont
  {Chandrasekharan}(2019)}]{Singh:2019uwd}%
  \BibitemOpen
  \bibfield  {author} {\bibinfo {author} {\bibfnamefont {H.}~\bibnamefont
  {Singh}}\ and\ \bibinfo {author} {\bibfnamefont {S.}~\bibnamefont
  {Chandrasekharan}},\ }\bibfield  {title} {\bibinfo {title} {Qubit
  regularization of the $o(3)$ sigma model},\ }\href
  {https://doi.org/10.1103/PhysRevD.100.054505} {\bibfield  {journal} {\bibinfo
   {journal} {Phys. Rev. D}\ }\textbf {\bibinfo {volume} {100}},\ \bibinfo
  {pages} {054505} (\bibinfo {year} {2019})},\ \Eprint
  {https://arxiv.org/abs/1905.13204} {arXiv:1905.13204 [hep-lat]} \BibitemShut
  {NoStop}%
\bibitem [{\citenamefont {Bhattacharya}\ \emph {et~al.}(2021)\citenamefont
  {Bhattacharya}, \citenamefont {Buser}, \citenamefont {Chandrasekharan},
  \citenamefont {Gupta},\ and\ \citenamefont {Singh}}]{Bhattacharya:2020gpm}%
  \BibitemOpen
  \bibfield  {author} {\bibinfo {author} {\bibfnamefont {T.}~\bibnamefont
  {Bhattacharya}}, \bibinfo {author} {\bibfnamefont {A.~J.}\ \bibnamefont
  {Buser}}, \bibinfo {author} {\bibfnamefont {S.}~\bibnamefont
  {Chandrasekharan}}, \bibinfo {author} {\bibfnamefont {R.}~\bibnamefont
  {Gupta}},\ and\ \bibinfo {author} {\bibfnamefont {H.}~\bibnamefont {Singh}},\
  }\bibfield  {title} {\bibinfo {title} {Qubit regularization of asymptotic
  freedom},\ }\href@noop {} {\bibfield  {journal} {\bibinfo  {journal} {Phys.
  Rev. Lett.}\ }\textbf {\bibinfo {volume} {126}} (\bibinfo {year} {2021})},\
  \Eprint {https://arxiv.org/abs/2012.02153} {arXiv:2012.02153 [hep-lat]}
  \BibitemShut {NoStop}%
\bibitem [{\citenamefont {Hostetler}\ \emph {et~al.}(2021)\citenamefont
  {Hostetler}, \citenamefont {Zhang}, \citenamefont {Sakai}, \citenamefont
  {Unmuth-Yockey}, \citenamefont {Bazavov},\ and\ \citenamefont
  {Meurice}}]{Hostetler_2021}%
  \BibitemOpen
  \bibfield  {author} {\bibinfo {author} {\bibfnamefont {L.}~\bibnamefont
  {Hostetler}}, \bibinfo {author} {\bibfnamefont {J.}~\bibnamefont {Zhang}},
  \bibinfo {author} {\bibfnamefont {R.}~\bibnamefont {Sakai}}, \bibinfo
  {author} {\bibfnamefont {J.}~\bibnamefont {Unmuth-Yockey}}, \bibinfo {author}
  {\bibfnamefont {A.}~\bibnamefont {Bazavov}},\ and\ \bibinfo {author}
  {\bibfnamefont {Y.}~\bibnamefont {Meurice}},\ }\bibfield  {title} {\bibinfo
  {title} {Clock model interpolation and symmetry breaking in o(2) models},\
  }\bibfield  {journal} {\bibinfo  {journal} {Physical Review D}\ }\textbf
  {\bibinfo {volume} {104}},\ \href
  {https://doi.org/10.1103/physrevd.104.054505} {10.1103/physrevd.104.054505}
  (\bibinfo {year} {2021})\BibitemShut {NoStop}%
\bibitem [{\citenamefont {Gharibyan}\ \emph {et~al.}(2021)\citenamefont
  {Gharibyan}, \citenamefont {Hanada}, \citenamefont {Honda},\ and\
  \citenamefont {Liu}}]{Gharibyan_2021}%
  \BibitemOpen
  \bibfield  {author} {\bibinfo {author} {\bibfnamefont {H.}~\bibnamefont
  {Gharibyan}}, \bibinfo {author} {\bibfnamefont {M.}~\bibnamefont {Hanada}},
  \bibinfo {author} {\bibfnamefont {M.}~\bibnamefont {Honda}},\ and\ \bibinfo
  {author} {\bibfnamefont {J.}~\bibnamefont {Liu}},\ }\bibfield  {title}
  {\bibinfo {title} {Toward simulating superstring/m-theory on a quantum
  computer},\ }\bibfield  {journal} {\bibinfo  {journal} {Journal of High
  Energy Physics}\ }\textbf {\bibinfo {volume} {2021}},\ \href
  {https://doi.org/10.1007/jhep07(2021)140} {10.1007/jhep07(2021)140} (\bibinfo
  {year} {2021})\BibitemShut {NoStop}%
\bibitem [{\citenamefont {Klco}\ \emph {et~al.}(2018)\citenamefont {Klco},
  \citenamefont {Dumitrescu}, \citenamefont {McCaskey}, \citenamefont {Morris},
  \citenamefont {Pooser}, \citenamefont {Sanz}, \citenamefont {Solano},
  \citenamefont {Lougovski},\ and\ \citenamefont {Savage}}]{Klco:2018kyo}%
  \BibitemOpen
  \bibfield  {author} {\bibinfo {author} {\bibfnamefont {N.}~\bibnamefont
  {Klco}}, \bibinfo {author} {\bibfnamefont {E.~F.}\ \bibnamefont
  {Dumitrescu}}, \bibinfo {author} {\bibfnamefont {A.~J.}\ \bibnamefont
  {McCaskey}}, \bibinfo {author} {\bibfnamefont {T.~D.}\ \bibnamefont
  {Morris}}, \bibinfo {author} {\bibfnamefont {R.~C.}\ \bibnamefont {Pooser}},
  \bibinfo {author} {\bibfnamefont {M.}~\bibnamefont {Sanz}}, \bibinfo {author}
  {\bibfnamefont {E.}~\bibnamefont {Solano}}, \bibinfo {author} {\bibfnamefont
  {P.}~\bibnamefont {Lougovski}},\ and\ \bibinfo {author} {\bibfnamefont
  {M.~J.}\ \bibnamefont {Savage}},\ }\bibfield  {title} {\bibinfo {title}
  {Quantum-classical computation of schwinger model dynamics using quantum
  computers},\ }\href {https://doi.org/10.1103/PhysRevA.98.032331} {\bibfield
  {journal} {\bibinfo  {journal} {Phys. Rev.}\ }\textbf {\bibinfo {volume}
  {A98}},\ \bibinfo {pages} {032331} (\bibinfo {year} {2018})},\ \Eprint
  {https://arxiv.org/abs/1803.03326} {arXiv:1803.03326 [quant-ph]} \BibitemShut
  {NoStop}%
\bibitem [{\citenamefont {Kokail}\ \emph {et~al.}(2019)\citenamefont {Kokail},
  \citenamefont {Maier}, \citenamefont {van Bijnen}, \citenamefont {Brydges},
  \citenamefont {Joshi}, \citenamefont {Jurcevic}, \citenamefont {Muschik},
  \citenamefont {Silvi}, \citenamefont {Blatt}, \citenamefont {Roos},\ and\
  \citenamefont {et~al.}}]{Kokail_2019}%
  \BibitemOpen
  \bibfield  {author} {\bibinfo {author} {\bibfnamefont {C.}~\bibnamefont
  {Kokail}}, \bibinfo {author} {\bibfnamefont {C.}~\bibnamefont {Maier}},
  \bibinfo {author} {\bibfnamefont {R.}~\bibnamefont {van Bijnen}}, \bibinfo
  {author} {\bibfnamefont {T.}~\bibnamefont {Brydges}}, \bibinfo {author}
  {\bibfnamefont {M.~K.}\ \bibnamefont {Joshi}}, \bibinfo {author}
  {\bibfnamefont {P.}~\bibnamefont {Jurcevic}}, \bibinfo {author}
  {\bibfnamefont {C.~A.}\ \bibnamefont {Muschik}}, \bibinfo {author}
  {\bibfnamefont {P.}~\bibnamefont {Silvi}}, \bibinfo {author} {\bibfnamefont
  {R.}~\bibnamefont {Blatt}}, \bibinfo {author} {\bibfnamefont {C.~F.}\
  \bibnamefont {Roos}},\ and\ \bibinfo {author} {\bibnamefont {et~al.}},\
  }\bibfield  {title} {\bibinfo {title} {Self-verifying variational quantum
  simulation of lattice models},\ }\href
  {https://doi.org/10.1038/s41586-019-1177-4} {\bibfield  {journal} {\bibinfo
  {journal} {Nature}\ }\textbf {\bibinfo {volume} {569}},\ \bibinfo {pages}
  {355–360} (\bibinfo {year} {2019})}\BibitemShut {NoStop}%
\bibitem [{\citenamefont {Kharzeev}\ and\ \citenamefont
  {Kikuchi}(2020)}]{Kharzeev_2020}%
  \BibitemOpen
  \bibfield  {author} {\bibinfo {author} {\bibfnamefont {D.~E.}\ \bibnamefont
  {Kharzeev}}\ and\ \bibinfo {author} {\bibfnamefont {Y.}~\bibnamefont
  {Kikuchi}},\ }\bibfield  {title} {\bibinfo {title} {Real-time chiral dynamics
  from a digital quantum simulation},\ }\bibfield  {journal} {\bibinfo
  {journal} {Physical Review Research}\ }\textbf {\bibinfo {volume} {2}},\
  \href {https://doi.org/10.1103/physrevresearch.2.023342}
  {10.1103/physrevresearch.2.023342} (\bibinfo {year} {2020})\BibitemShut
  {NoStop}%
\bibitem [{\citenamefont {Lu}\ \emph {et~al.}(2019)\citenamefont {Lu},
  \citenamefont {Klco}, \citenamefont {Lukens}, \citenamefont {Morris},
  \citenamefont {Bansal}, \citenamefont {Ekström}, \citenamefont {Hagen},
  \citenamefont {Papenbrock}, \citenamefont {Weiner}, \citenamefont {Savage},\
  and\ \citenamefont {et~al.}}]{Lu_2019}%
  \BibitemOpen
  \bibfield  {author} {\bibinfo {author} {\bibfnamefont {H.-H.}\ \bibnamefont
  {Lu}}, \bibinfo {author} {\bibfnamefont {N.}~\bibnamefont {Klco}}, \bibinfo
  {author} {\bibfnamefont {J.~M.}\ \bibnamefont {Lukens}}, \bibinfo {author}
  {\bibfnamefont {T.~D.}\ \bibnamefont {Morris}}, \bibinfo {author}
  {\bibfnamefont {A.}~\bibnamefont {Bansal}}, \bibinfo {author} {\bibfnamefont
  {A.}~\bibnamefont {Ekström}}, \bibinfo {author} {\bibfnamefont
  {G.}~\bibnamefont {Hagen}}, \bibinfo {author} {\bibfnamefont
  {T.}~\bibnamefont {Papenbrock}}, \bibinfo {author} {\bibfnamefont {A.~M.}\
  \bibnamefont {Weiner}}, \bibinfo {author} {\bibfnamefont {M.~J.}\
  \bibnamefont {Savage}},\ and\ \bibinfo {author} {\bibnamefont {et~al.}},\
  }\bibfield  {title} {\bibinfo {title} {Simulations of subatomic many-body
  physics on a quantum frequency processor},\ }\bibfield  {journal} {\bibinfo
  {journal} {Physical Review A}\ }\textbf {\bibinfo {volume} {100}},\ \href
  {https://doi.org/10.1103/physreva.100.012320} {10.1103/physreva.100.012320}
  (\bibinfo {year} {2019})\BibitemShut {NoStop}%
\bibitem [{\citenamefont {Chakraborty}\ \emph {et~al.}(2020)\citenamefont
  {Chakraborty}, \citenamefont {Honda}, \citenamefont {Izubuchi}, \citenamefont
  {Kikuchi},\ and\ \citenamefont {Tomiya}}]{chakraborty2020digital}%
  \BibitemOpen
  \bibfield  {author} {\bibinfo {author} {\bibfnamefont {B.}~\bibnamefont
  {Chakraborty}}, \bibinfo {author} {\bibfnamefont {M.}~\bibnamefont {Honda}},
  \bibinfo {author} {\bibfnamefont {T.}~\bibnamefont {Izubuchi}}, \bibinfo
  {author} {\bibfnamefont {Y.}~\bibnamefont {Kikuchi}},\ and\ \bibinfo {author}
  {\bibfnamefont {A.}~\bibnamefont {Tomiya}},\ }\href@noop {} {\bibinfo {title}
  {Digital quantum simulation of the schwinger model with topological term via
  adiabatic state preparation}} (\bibinfo {year} {2020}),\ \Eprint
  {https://arxiv.org/abs/2001.00485} {arXiv:2001.00485 [hep-lat]} \BibitemShut
  {NoStop}%
\bibitem [{\citenamefont {Shaw}\ \emph {et~al.}(2020)\citenamefont {Shaw},
  \citenamefont {Lougovski}, \citenamefont {Stryker},\ and\ \citenamefont
  {Wiebe}}]{Shaw_2020}%
  \BibitemOpen
  \bibfield  {author} {\bibinfo {author} {\bibfnamefont {A.~F.}\ \bibnamefont
  {Shaw}}, \bibinfo {author} {\bibfnamefont {P.}~\bibnamefont {Lougovski}},
  \bibinfo {author} {\bibfnamefont {J.~R.}\ \bibnamefont {Stryker}},\ and\
  \bibinfo {author} {\bibfnamefont {N.}~\bibnamefont {Wiebe}},\ }\bibfield
  {title} {\bibinfo {title} {Quantum algorithms for simulating the lattice
  schwinger model},\ }\href {https://doi.org/10.22331/q-2020-08-10-306}
  {\bibfield  {journal} {\bibinfo  {journal} {Quantum}\ }\textbf {\bibinfo
  {volume} {4}},\ \bibinfo {pages} {306} (\bibinfo {year} {2020})}\BibitemShut
  {NoStop}%
\bibitem [{\citenamefont {Bender}\ \emph {et~al.}(2018)\citenamefont {Bender},
  \citenamefont {Zohar}, \citenamefont {Farace},\ and\ \citenamefont
  {Cirac}}]{Bender_2018}%
  \BibitemOpen
  \bibfield  {author} {\bibinfo {author} {\bibfnamefont {J.}~\bibnamefont
  {Bender}}, \bibinfo {author} {\bibfnamefont {E.}~\bibnamefont {Zohar}},
  \bibinfo {author} {\bibfnamefont {A.}~\bibnamefont {Farace}},\ and\ \bibinfo
  {author} {\bibfnamefont {J.~I.}\ \bibnamefont {Cirac}},\ }\bibfield  {title}
  {\bibinfo {title} {Digital quantum simulation of lattice gauge theories in
  three spatial dimensions},\ }\href {https://doi.org/10.1088/1367-2630/aadb71}
  {\bibfield  {journal} {\bibinfo  {journal} {New Journal of Physics}\ }\textbf
  {\bibinfo {volume} {20}},\ \bibinfo {pages} {093001} (\bibinfo {year}
  {2018})}\BibitemShut {NoStop}%
\bibitem [{\citenamefont {Klco}\ \emph {et~al.}(2020)\citenamefont {Klco},
  \citenamefont {Savage},\ and\ \citenamefont {Stryker}}]{Klco_2020}%
  \BibitemOpen
  \bibfield  {author} {\bibinfo {author} {\bibfnamefont {N.}~\bibnamefont
  {Klco}}, \bibinfo {author} {\bibfnamefont {M.~J.}\ \bibnamefont {Savage}},\
  and\ \bibinfo {author} {\bibfnamefont {J.~R.}\ \bibnamefont {Stryker}},\
  }\bibfield  {title} {\bibinfo {title} {Su(2) non-abelian gauge field theory
  in one dimension on digital quantum computers},\ }\bibfield  {journal}
  {\bibinfo  {journal} {Physical Review D}\ }\textbf {\bibinfo {volume}
  {101}},\ \href {https://doi.org/10.1103/physrevd.101.074512}
  {10.1103/physrevd.101.074512} (\bibinfo {year} {2020})\BibitemShut {NoStop}%
\bibitem [{\citenamefont {Zohar}\ \emph
  {et~al.}(2013{\natexlab{a}})\citenamefont {Zohar}, \citenamefont {Cirac},\
  and\ \citenamefont {Reznik}}]{Zohar_2013}%
  \BibitemOpen
  \bibfield  {author} {\bibinfo {author} {\bibfnamefont {E.}~\bibnamefont
  {Zohar}}, \bibinfo {author} {\bibfnamefont {J.~I.}\ \bibnamefont {Cirac}},\
  and\ \bibinfo {author} {\bibfnamefont {B.}~\bibnamefont {Reznik}},\
  }\bibfield  {title} {\bibinfo {title} {Quantum simulations of gauge theories
  with ultracold atoms: Local gauge invariance from angular-momentum
  conservation},\ }\bibfield  {journal} {\bibinfo  {journal} {Physical Review
  A}\ }\textbf {\bibinfo {volume} {88}},\ \href
  {https://doi.org/10.1103/physreva.88.023617} {10.1103/physreva.88.023617}
  (\bibinfo {year} {2013}{\natexlab{a}})\BibitemShut {NoStop}%
\bibitem [{\citenamefont {Zohar}\ \emph {et~al.}(2015)\citenamefont {Zohar},
  \citenamefont {Cirac},\ and\ \citenamefont {Reznik}}]{Zohar_2015}%
  \BibitemOpen
  \bibfield  {author} {\bibinfo {author} {\bibfnamefont {E.}~\bibnamefont
  {Zohar}}, \bibinfo {author} {\bibfnamefont {J.~I.}\ \bibnamefont {Cirac}},\
  and\ \bibinfo {author} {\bibfnamefont {B.}~\bibnamefont {Reznik}},\
  }\bibfield  {title} {\bibinfo {title} {Quantum simulations of lattice gauge
  theories using ultracold atoms in optical lattices},\ }\href
  {https://doi.org/10.1088/0034-4885/79/1/014401} {\bibfield  {journal}
  {\bibinfo  {journal} {Reports on Progress in Physics}\ }\textbf {\bibinfo
  {volume} {79}},\ \bibinfo {pages} {014401} (\bibinfo {year}
  {2015})}\BibitemShut {NoStop}%
\bibitem [{\citenamefont {Zohar}\ and\ \citenamefont
  {Burrello}(2015)}]{Zohar_2015_2}%
  \BibitemOpen
  \bibfield  {author} {\bibinfo {author} {\bibfnamefont {E.}~\bibnamefont
  {Zohar}}\ and\ \bibinfo {author} {\bibfnamefont {M.}~\bibnamefont
  {Burrello}},\ }\bibfield  {title} {\bibinfo {title} {Formulation of lattice
  gauge theories for quantum simulations},\ }\bibfield  {journal} {\bibinfo
  {journal} {Physical Review D}\ }\textbf {\bibinfo {volume} {91}},\ \href
  {https://doi.org/10.1103/physrevd.91.054506} {10.1103/physrevd.91.054506}
  (\bibinfo {year} {2015})\BibitemShut {NoStop}%
\bibitem [{\citenamefont {Lamm}\ \emph {et~al.}(2019)\citenamefont {Lamm},
  \citenamefont {Lawrence},\ and\ \citenamefont {Yamauchi}}]{Lamm_2019}%
  \BibitemOpen
  \bibfield  {author} {\bibinfo {author} {\bibfnamefont {H.}~\bibnamefont
  {Lamm}}, \bibinfo {author} {\bibfnamefont {S.}~\bibnamefont {Lawrence}},\
  and\ \bibinfo {author} {\bibfnamefont {Y.}~\bibnamefont {Yamauchi}},\
  }\bibfield  {title} {\bibinfo {title} {General methods for digital quantum
  simulation of gauge theories},\ }\bibfield  {journal} {\bibinfo  {journal}
  {Physical Review D}\ }\textbf {\bibinfo {volume} {100}},\ \href
  {https://doi.org/10.1103/physrevd.100.034518} {10.1103/physrevd.100.034518}
  (\bibinfo {year} {2019})\BibitemShut {NoStop}%
\bibitem [{\citenamefont {Alexandru}\ \emph
  {et~al.}(2019{\natexlab{b}})\citenamefont {Alexandru}, \citenamefont
  {Bedaque}, \citenamefont {Harmalkar}, \citenamefont {Lamm}, \citenamefont
  {Lawrence},\ and\ \citenamefont {Warrington}}]{Alexandru_2019}%
  \BibitemOpen
  \bibfield  {author} {\bibinfo {author} {\bibfnamefont {A.}~\bibnamefont
  {Alexandru}}, \bibinfo {author} {\bibfnamefont {P.~F.}\ \bibnamefont
  {Bedaque}}, \bibinfo {author} {\bibfnamefont {S.}~\bibnamefont {Harmalkar}},
  \bibinfo {author} {\bibfnamefont {H.}~\bibnamefont {Lamm}}, \bibinfo {author}
  {\bibfnamefont {S.}~\bibnamefont {Lawrence}},\ and\ \bibinfo {author}
  {\bibfnamefont {N.~C.}\ \bibnamefont {Warrington}},\ }\bibfield  {title}
  {\bibinfo {title} {Gluon field digitization for quantum computers},\
  }\bibfield  {journal} {\bibinfo  {journal} {Physical Review D}\ }\textbf
  {\bibinfo {volume} {100}},\ \href
  {https://doi.org/10.1103/physrevd.100.114501} {10.1103/physrevd.100.114501}
  (\bibinfo {year} {2019}{\natexlab{b}})\BibitemShut {NoStop}%
\bibitem [{\citenamefont {Bañuls}\ \emph {et~al.}(2020)\citenamefont
  {Bañuls}, \citenamefont {Blatt}, \citenamefont {Catani}, \citenamefont
  {Celi}, \citenamefont {Cirac}, \citenamefont {Dalmonte}, \citenamefont
  {Fallani}, \citenamefont {Jansen}, \citenamefont {Lewenstein}, \citenamefont
  {Montangero},\ and\ \citenamefont {et~al.}}]{Ba_uls_2020}%
  \BibitemOpen
  \bibfield  {author} {\bibinfo {author} {\bibfnamefont {M.~C.}\ \bibnamefont
  {Bañuls}}, \bibinfo {author} {\bibfnamefont {R.}~\bibnamefont {Blatt}},
  \bibinfo {author} {\bibfnamefont {J.}~\bibnamefont {Catani}}, \bibinfo
  {author} {\bibfnamefont {A.}~\bibnamefont {Celi}}, \bibinfo {author}
  {\bibfnamefont {J.~I.}\ \bibnamefont {Cirac}}, \bibinfo {author}
  {\bibfnamefont {M.}~\bibnamefont {Dalmonte}}, \bibinfo {author}
  {\bibfnamefont {L.}~\bibnamefont {Fallani}}, \bibinfo {author} {\bibfnamefont
  {K.}~\bibnamefont {Jansen}}, \bibinfo {author} {\bibfnamefont
  {M.}~\bibnamefont {Lewenstein}}, \bibinfo {author} {\bibfnamefont
  {S.}~\bibnamefont {Montangero}},\ and\ \bibinfo {author} {\bibnamefont
  {et~al.}},\ }\bibfield  {title} {\bibinfo {title} {Simulating lattice gauge
  theories within quantum technologies},\ }\bibfield  {journal} {\bibinfo
  {journal} {The European Physical Journal D}\ }\textbf {\bibinfo {volume}
  {74}},\ \href {https://doi.org/10.1140/epjd/e2020-100571-8}
  {10.1140/epjd/e2020-100571-8} (\bibinfo {year} {2020})\BibitemShut {NoStop}%
\bibitem [{\citenamefont {Tagliacozzo}\ \emph
  {et~al.}(2013{\natexlab{a}})\citenamefont {Tagliacozzo}, \citenamefont
  {Celi}, \citenamefont {Orland}, \citenamefont {Mitchell},\ and\ \citenamefont
  {Lewenstein}}]{Tagliacozzo_2013}%
  \BibitemOpen
  \bibfield  {author} {\bibinfo {author} {\bibfnamefont {L.}~\bibnamefont
  {Tagliacozzo}}, \bibinfo {author} {\bibfnamefont {A.}~\bibnamefont {Celi}},
  \bibinfo {author} {\bibfnamefont {P.}~\bibnamefont {Orland}}, \bibinfo
  {author} {\bibfnamefont {M.~W.}\ \bibnamefont {Mitchell}},\ and\ \bibinfo
  {author} {\bibfnamefont {M.}~\bibnamefont {Lewenstein}},\ }\bibfield  {title}
  {\bibinfo {title} {Simulation of non-abelian gauge theories with optical
  lattices},\ }\bibfield  {journal} {\bibinfo  {journal} {Nature
  Communications}\ }\textbf {\bibinfo {volume} {4}},\ \href
  {https://doi.org/10.1038/ncomms3615} {10.1038/ncomms3615} (\bibinfo {year}
  {2013}{\natexlab{a}})\BibitemShut {NoStop}%
\bibitem [{\citenamefont {Tagliacozzo}\ \emph
  {et~al.}(2013{\natexlab{b}})\citenamefont {Tagliacozzo}, \citenamefont
  {Celi}, \citenamefont {Zamora},\ and\ \citenamefont
  {Lewenstein}}]{Tagliacozzo_2013_2}%
  \BibitemOpen
  \bibfield  {author} {\bibinfo {author} {\bibfnamefont {L.}~\bibnamefont
  {Tagliacozzo}}, \bibinfo {author} {\bibfnamefont {A.}~\bibnamefont {Celi}},
  \bibinfo {author} {\bibfnamefont {A.}~\bibnamefont {Zamora}},\ and\ \bibinfo
  {author} {\bibfnamefont {M.}~\bibnamefont {Lewenstein}},\ }\bibfield  {title}
  {\bibinfo {title} {Optical abelian lattice gauge theories},\ }\href
  {https://doi.org/10.1016/j.aop.2012.11.009} {\bibfield  {journal} {\bibinfo
  {journal} {Annals of Physics}\ }\textbf {\bibinfo {volume} {330}},\ \bibinfo
  {pages} {160–191} (\bibinfo {year} {2013}{\natexlab{b}})}\BibitemShut
  {NoStop}%
\bibitem [{\citenamefont {Bermudez}\ \emph {et~al.}(2010)\citenamefont
  {Bermudez}, \citenamefont {Mazza}, \citenamefont {Rizzi}, \citenamefont
  {Goldman}, \citenamefont {Lewenstein},\ and\ \citenamefont
  {Martin-Delgado}}]{PhysRevLett.105.190404}%
  \BibitemOpen
  \bibfield  {author} {\bibinfo {author} {\bibfnamefont {A.}~\bibnamefont
  {Bermudez}}, \bibinfo {author} {\bibfnamefont {L.}~\bibnamefont {Mazza}},
  \bibinfo {author} {\bibfnamefont {M.}~\bibnamefont {Rizzi}}, \bibinfo
  {author} {\bibfnamefont {N.}~\bibnamefont {Goldman}}, \bibinfo {author}
  {\bibfnamefont {M.}~\bibnamefont {Lewenstein}},\ and\ \bibinfo {author}
  {\bibfnamefont {M.~A.}\ \bibnamefont {Martin-Delgado}},\ }\bibfield  {title}
  {\bibinfo {title} {Wilson fermions and axion electrodynamics in optical
  lattices},\ }\href {https://doi.org/10.1103/PhysRevLett.105.190404}
  {\bibfield  {journal} {\bibinfo  {journal} {Phys. Rev. Lett.}\ }\textbf
  {\bibinfo {volume} {105}},\ \bibinfo {pages} {190404} (\bibinfo {year}
  {2010})}\BibitemShut {NoStop}%
\bibitem [{\citenamefont {Mathis}\ \emph {et~al.}(2020)\citenamefont {Mathis},
  \citenamefont {Mazzola},\ and\ \citenamefont
  {Tavernelli}}]{Tavernelli20_094501}%
  \BibitemOpen
  \bibfield  {author} {\bibinfo {author} {\bibfnamefont {S.~V.}\ \bibnamefont
  {Mathis}}, \bibinfo {author} {\bibfnamefont {G.}~\bibnamefont {Mazzola}},\
  and\ \bibinfo {author} {\bibfnamefont {I.}~\bibnamefont {Tavernelli}},\
  }\bibfield  {title} {\bibinfo {title} {Toward scalable simulations of lattice
  gauge theories on quantum computers},\ }\href
  {https://doi.org/10.1103/PhysRevD.102.094501} {\bibfield  {journal} {\bibinfo
   {journal} {Physical Review D}\ }\textbf {\bibinfo {volume} {102}},\ \bibinfo
  {pages} {094501} (\bibinfo {year} {2020})}\BibitemShut {NoStop}%
\bibitem [{\citenamefont {Byrnes}\ and\ \citenamefont
  {Yamamoto}(2006)}]{Byrnes_2006}%
  \BibitemOpen
  \bibfield  {author} {\bibinfo {author} {\bibfnamefont {T.}~\bibnamefont
  {Byrnes}}\ and\ \bibinfo {author} {\bibfnamefont {Y.}~\bibnamefont
  {Yamamoto}},\ }\bibfield  {title} {\bibinfo {title} {Simulating lattice gauge
  theories on a quantum computer},\ }\bibfield  {journal} {\bibinfo  {journal}
  {Physical Review A}\ }\textbf {\bibinfo {volume} {73}},\ \href
  {https://doi.org/10.1103/physreva.73.022328} {10.1103/physreva.73.022328}
  (\bibinfo {year} {2006})\BibitemShut {NoStop}%
\bibitem [{\citenamefont {Ciavarella}\ \emph {et~al.}(2021)\citenamefont
  {Ciavarella}, \citenamefont {Klco},\ and\ \citenamefont
  {Savage}}]{Ciavarella_2021}%
  \BibitemOpen
  \bibfield  {author} {\bibinfo {author} {\bibfnamefont {A.}~\bibnamefont
  {Ciavarella}}, \bibinfo {author} {\bibfnamefont {N.}~\bibnamefont {Klco}},\
  and\ \bibinfo {author} {\bibfnamefont {M.~J.}\ \bibnamefont {Savage}},\
  }\bibfield  {title} {\bibinfo {title} {Trailhead for quantum simulation of
  su(3) yang-mills lattice gauge theory in the local multiplet basis},\
  }\bibfield  {journal} {\bibinfo  {journal} {Physical Review D}\ }\textbf
  {\bibinfo {volume} {103}},\ \href
  {https://doi.org/10.1103/physrevd.103.094501} {10.1103/physrevd.103.094501}
  (\bibinfo {year} {2021})\BibitemShut {NoStop}%
\bibitem [{\citenamefont {Kan}\ and\ \citenamefont
  {Nam}(2021)}]{kan2021lattice}%
  \BibitemOpen
  \bibfield  {author} {\bibinfo {author} {\bibfnamefont {A.}~\bibnamefont
  {Kan}}\ and\ \bibinfo {author} {\bibfnamefont {Y.}~\bibnamefont {Nam}},\
  }\href@noop {} {\bibinfo {title} {Lattice quantum chromodynamics and
  electrodynamics on a universal quantum computer}} (\bibinfo {year} {2021}),\
  \Eprint {https://arxiv.org/abs/2107.12769} {arXiv:2107.12769 [quant-ph]}
  \BibitemShut {NoStop}%
\bibitem [{\citenamefont {Stryker}(2021)}]{stryker2021shearing}%
  \BibitemOpen
  \bibfield  {author} {\bibinfo {author} {\bibfnamefont {J.~R.}\ \bibnamefont
  {Stryker}},\ }\href@noop {} {\bibinfo {title} {Shearing approach to gauge
  invariant trotterization}} (\bibinfo {year} {2021}),\ \Eprint
  {https://arxiv.org/abs/2105.11548} {arXiv:2105.11548 [hep-lat]} \BibitemShut
  {NoStop}%
\bibitem [{\citenamefont {Paulson}\ \emph {et~al.}(2021)\citenamefont
  {Paulson}, \citenamefont {Dellantonio}, \citenamefont {Haase}, \citenamefont
  {Celi}, \citenamefont {Kan}, \citenamefont {Jena}, \citenamefont {Kokail},
  \citenamefont {van Bijnen}, \citenamefont {Jansen}, \citenamefont {Zoller},\
  and\ \citenamefont {et~al.}}]{Paulson_2021}%
  \BibitemOpen
  \bibfield  {author} {\bibinfo {author} {\bibfnamefont {D.}~\bibnamefont
  {Paulson}}, \bibinfo {author} {\bibfnamefont {L.}~\bibnamefont
  {Dellantonio}}, \bibinfo {author} {\bibfnamefont {J.~F.}\ \bibnamefont
  {Haase}}, \bibinfo {author} {\bibfnamefont {A.}~\bibnamefont {Celi}},
  \bibinfo {author} {\bibfnamefont {A.}~\bibnamefont {Kan}}, \bibinfo {author}
  {\bibfnamefont {A.}~\bibnamefont {Jena}}, \bibinfo {author} {\bibfnamefont
  {C.}~\bibnamefont {Kokail}}, \bibinfo {author} {\bibfnamefont
  {R.}~\bibnamefont {van Bijnen}}, \bibinfo {author} {\bibfnamefont
  {K.}~\bibnamefont {Jansen}}, \bibinfo {author} {\bibfnamefont
  {P.}~\bibnamefont {Zoller}},\ and\ \bibinfo {author} {\bibnamefont
  {et~al.}},\ }\bibfield  {title} {\bibinfo {title} {Simulating 2d effects in
  lattice gauge theories on a quantum computer},\ }\bibfield  {journal}
  {\bibinfo  {journal} {PRX Quantum}\ }\textbf {\bibinfo {volume} {2}},\ \href
  {https://doi.org/10.1103/prxquantum.2.030334} {10.1103/prxquantum.2.030334}
  (\bibinfo {year} {2021})\BibitemShut {NoStop}%
\bibitem [{\citenamefont {Davoudi}\ \emph
  {et~al.}(2021{\natexlab{b}})\citenamefont {Davoudi}, \citenamefont {Linke},\
  and\ \citenamefont {Pagano}}]{davoudi2021simulating}%
  \BibitemOpen
  \bibfield  {author} {\bibinfo {author} {\bibfnamefont {Z.}~\bibnamefont
  {Davoudi}}, \bibinfo {author} {\bibfnamefont {N.~M.}\ \bibnamefont {Linke}},\
  and\ \bibinfo {author} {\bibfnamefont {G.}~\bibnamefont {Pagano}},\
  }\href@noop {} {\bibinfo {title} {Toward simulating quantum field theories
  with controlled phonon-ion dynamics: A hybrid analog-digital approach}}
  (\bibinfo {year} {2021}{\natexlab{b}}),\ \Eprint
  {https://arxiv.org/abs/2104.09346} {arXiv:2104.09346 [quant-ph]} \BibitemShut
  {NoStop}%
\bibitem [{\citenamefont {Zohar}\ \emph {et~al.}(2012)\citenamefont {Zohar},
  \citenamefont {Cirac},\ and\ \citenamefont {Reznik}}]{Zohar:2012ay}%
  \BibitemOpen
  \bibfield  {author} {\bibinfo {author} {\bibfnamefont {E.}~\bibnamefont
  {Zohar}}, \bibinfo {author} {\bibfnamefont {J.~I.}\ \bibnamefont {Cirac}},\
  and\ \bibinfo {author} {\bibfnamefont {B.}~\bibnamefont {Reznik}},\
  }\bibfield  {title} {\bibinfo {title} {Simulating compact quantum
  electrodynamics with ultracold atoms: Probing confinement and nonperturbative
  effects},\ }\href {https://doi.org/10.1103/PhysRevLett.109.125302} {\bibfield
   {journal} {\bibinfo  {journal} {Phys. Rev. Lett.}\ }\textbf {\bibinfo
  {volume} {109}},\ \bibinfo {pages} {125302} (\bibinfo {year} {2012})},\
  \Eprint {https://arxiv.org/abs/1204.6574} {arXiv:1204.6574 [quant-ph]}
  \BibitemShut {NoStop}%
\bibitem [{\citenamefont {Zohar}\ \emph
  {et~al.}(2013{\natexlab{b}})\citenamefont {Zohar}, \citenamefont {Cirac},\
  and\ \citenamefont {Reznik}}]{Zohar:2012xf}%
  \BibitemOpen
  \bibfield  {author} {\bibinfo {author} {\bibfnamefont {E.}~\bibnamefont
  {Zohar}}, \bibinfo {author} {\bibfnamefont {J.~I.}\ \bibnamefont {Cirac}},\
  and\ \bibinfo {author} {\bibfnamefont {B.}~\bibnamefont {Reznik}},\
  }\bibfield  {title} {\bibinfo {title} {Cold-atom quantum simulator for su(2)
  yang-mills lattice gauge theory},\ }\href
  {https://doi.org/10.1103/PhysRevLett.110.125304} {\bibfield  {journal}
  {\bibinfo  {journal} {Phys. Rev. Lett.}\ }\textbf {\bibinfo {volume} {110}},\
  \bibinfo {pages} {125304} (\bibinfo {year} {2013}{\natexlab{b}})},\ \Eprint
  {https://arxiv.org/abs/1211.2241} {arXiv:1211.2241 [quant-ph]} \BibitemShut
  {NoStop}%
\bibitem [{\citenamefont {Banerjee}\ \emph {et~al.}(2013)\citenamefont
  {Banerjee}, \citenamefont {B\"{o}gli}, \citenamefont {Dalmonte},
  \citenamefont {Rico}, \citenamefont {Stebler}, \citenamefont {Wiese},\ and\
  \citenamefont {Zoller}}]{Banerjee:2012xg}%
  \BibitemOpen
  \bibfield  {author} {\bibinfo {author} {\bibfnamefont {D.}~\bibnamefont
  {Banerjee}}, \bibinfo {author} {\bibfnamefont {M.}~\bibnamefont {B\"{o}gli}},
  \bibinfo {author} {\bibfnamefont {M.}~\bibnamefont {Dalmonte}}, \bibinfo
  {author} {\bibfnamefont {E.}~\bibnamefont {Rico}}, \bibinfo {author}
  {\bibfnamefont {P.}~\bibnamefont {Stebler}}, \bibinfo {author} {\bibfnamefont
  {U.~J.}\ \bibnamefont {Wiese}},\ and\ \bibinfo {author} {\bibfnamefont
  {P.}~\bibnamefont {Zoller}},\ }\bibfield  {title} {\bibinfo {title} {Atomic
  quantum simulation of u(n) and su(n) non-abelian lattice gauge theories},\
  }\href {https://doi.org/10.1103/PhysRevLett.110.125303} {\bibfield  {journal}
  {\bibinfo  {journal} {Phys. Rev. Lett.}\ }\textbf {\bibinfo {volume} {110}},\
  \bibinfo {pages} {125303} (\bibinfo {year} {2013})},\ \Eprint
  {https://arxiv.org/abs/1211.2242} {arXiv:1211.2242 [cond-mat.quant-gas]}
  \BibitemShut {NoStop}%
\bibitem [{\citenamefont {Banerjee}\ \emph {et~al.}(2012)\citenamefont
  {Banerjee}, \citenamefont {Dalmonte}, \citenamefont {Muller}, \citenamefont
  {Rico}, \citenamefont {Stebler}, \citenamefont {Wiese},\ and\ \citenamefont
  {Zoller}}]{Banerjee:2012pg}%
  \BibitemOpen
  \bibfield  {author} {\bibinfo {author} {\bibfnamefont {D.}~\bibnamefont
  {Banerjee}}, \bibinfo {author} {\bibfnamefont {M.}~\bibnamefont {Dalmonte}},
  \bibinfo {author} {\bibfnamefont {M.}~\bibnamefont {Muller}}, \bibinfo
  {author} {\bibfnamefont {E.}~\bibnamefont {Rico}}, \bibinfo {author}
  {\bibfnamefont {P.}~\bibnamefont {Stebler}}, \bibinfo {author} {\bibfnamefont
  {U.~J.}\ \bibnamefont {Wiese}},\ and\ \bibinfo {author} {\bibfnamefont
  {P.}~\bibnamefont {Zoller}},\ }\bibfield  {title} {\bibinfo {title} {Atomic
  quantum simulation of dynamical gauge fields coupled to fermionic matter:
  From string breaking to evolution after a quench},\ }\href
  {https://doi.org/10.1103/PhysRevLett.109.175302} {\bibfield  {journal}
  {\bibinfo  {journal} {Phys. Rev. Lett.}\ }\textbf {\bibinfo {volume} {109}},\
  \bibinfo {pages} {175302} (\bibinfo {year} {2012})},\ \Eprint
  {https://arxiv.org/abs/1205.6366} {arXiv:1205.6366 [cond-mat.quant-gas]}
  \BibitemShut {NoStop}%
\bibitem [{\citenamefont {Martinez}\ \emph {et~al.}(2016)\citenamefont
  {Martinez}, \citenamefont {Muschik}, \citenamefont {Schindler}, \citenamefont
  {Nigg}, \citenamefont {Erhard}, \citenamefont {Heyl}, \citenamefont {Hauke},
  \citenamefont {Dalmonte}, \citenamefont {Monz}, \citenamefont {Zoller},\ and\
  \citenamefont {Blatt}}]{Martinez2016}%
  \BibitemOpen
  \bibfield  {author} {\bibinfo {author} {\bibfnamefont {E.~A.}\ \bibnamefont
  {Martinez}}, \bibinfo {author} {\bibfnamefont {C.~A.}\ \bibnamefont
  {Muschik}}, \bibinfo {author} {\bibfnamefont {P.}~\bibnamefont {Schindler}},
  \bibinfo {author} {\bibfnamefont {D.}~\bibnamefont {Nigg}}, \bibinfo {author}
  {\bibfnamefont {A.}~\bibnamefont {Erhard}}, \bibinfo {author} {\bibfnamefont
  {M.}~\bibnamefont {Heyl}}, \bibinfo {author} {\bibfnamefont {P.}~\bibnamefont
  {Hauke}}, \bibinfo {author} {\bibfnamefont {M.}~\bibnamefont {Dalmonte}},
  \bibinfo {author} {\bibfnamefont {T.}~\bibnamefont {Monz}}, \bibinfo {author}
  {\bibfnamefont {P.}~\bibnamefont {Zoller}},\ and\ \bibinfo {author}
  {\bibfnamefont {R.}~\bibnamefont {Blatt}},\ }\bibfield  {title} {\bibinfo
  {title} {Real-time dynamics of lattice gauge theories with a few-qubit
  quantum computer},\ }\href {http://dx.doi.org/10.1038/nature18318} {\bibfield
   {journal} {\bibinfo  {journal} {Nature}\ }\textbf {\bibinfo {volume}
  {534}},\ \bibinfo {pages} {516 EP } (\bibinfo {year} {2016})}\BibitemShut
  {NoStop}%
\bibitem [{\citenamefont {Muschik}\ \emph {et~al.}(2017)\citenamefont
  {Muschik}, \citenamefont {Heyl}, \citenamefont {Martinez}, \citenamefont
  {Monz}, \citenamefont {Schindler}, \citenamefont {Vogell}, \citenamefont
  {Dalmonte}, \citenamefont {Hauke}, \citenamefont {Blatt},\ and\ \citenamefont
  {Zoller}}]{Muschik:2016tws}%
  \BibitemOpen
  \bibfield  {author} {\bibinfo {author} {\bibfnamefont {C.}~\bibnamefont
  {Muschik}}, \bibinfo {author} {\bibfnamefont {M.}~\bibnamefont {Heyl}},
  \bibinfo {author} {\bibfnamefont {E.}~\bibnamefont {Martinez}}, \bibinfo
  {author} {\bibfnamefont {T.}~\bibnamefont {Monz}}, \bibinfo {author}
  {\bibfnamefont {P.}~\bibnamefont {Schindler}}, \bibinfo {author}
  {\bibfnamefont {B.}~\bibnamefont {Vogell}}, \bibinfo {author} {\bibfnamefont
  {M.}~\bibnamefont {Dalmonte}}, \bibinfo {author} {\bibfnamefont
  {P.}~\bibnamefont {Hauke}}, \bibinfo {author} {\bibfnamefont
  {R.}~\bibnamefont {Blatt}},\ and\ \bibinfo {author} {\bibfnamefont
  {P.}~\bibnamefont {Zoller}},\ }\bibfield  {title} {\bibinfo {title} {U(1)
  wilson lattice gauge theories in digital quantum simulators},\ }\href
  {https://doi.org/10.1088/1367-2630/aa89ab} {\bibfield  {journal} {\bibinfo
  {journal} {New J. Phys.}\ }\textbf {\bibinfo {volume} {19}},\ \bibinfo
  {pages} {103020} (\bibinfo {year} {2017})},\ \Eprint
  {https://arxiv.org/abs/1612.08653} {arXiv:1612.08653 [quant-ph]} \BibitemShut
  {NoStop}%
\bibitem [{\citenamefont {Zohar}\ \emph {et~al.}(2017)\citenamefont {Zohar},
  \citenamefont {Farace}, \citenamefont {Reznik},\ and\ \citenamefont
  {Cirac}}]{Zohar:2016iic}%
  \BibitemOpen
  \bibfield  {author} {\bibinfo {author} {\bibfnamefont {E.}~\bibnamefont
  {Zohar}}, \bibinfo {author} {\bibfnamefont {A.}~\bibnamefont {Farace}},
  \bibinfo {author} {\bibfnamefont {B.}~\bibnamefont {Reznik}},\ and\ \bibinfo
  {author} {\bibfnamefont {J.~I.}\ \bibnamefont {Cirac}},\ }\bibfield  {title}
  {\bibinfo {title} {Digital lattice gauge theories},\ }\href
  {https://doi.org/10.1103/PhysRevA.95.023604} {\bibfield  {journal} {\bibinfo
  {journal} {Phys. Rev.}\ }\textbf {\bibinfo {volume} {A95}},\ \bibinfo {pages}
  {023604} (\bibinfo {year} {2017})},\ \Eprint
  {https://arxiv.org/abs/1607.08121} {arXiv:1607.08121 [quant-ph]} \BibitemShut
  {NoStop}%
\bibitem [{\citenamefont {Banuls}\ \emph {et~al.}(2017)\citenamefont {Banuls},
  \citenamefont {Cichy}, \citenamefont {Cirac}, \citenamefont {Jansen},\ and\
  \citenamefont {Kuhn}}]{Banuls:2017ena}%
  \BibitemOpen
  \bibfield  {author} {\bibinfo {author} {\bibfnamefont {M.~C.}\ \bibnamefont
  {Banuls}}, \bibinfo {author} {\bibfnamefont {K.}~\bibnamefont {Cichy}},
  \bibinfo {author} {\bibfnamefont {J.~I.}\ \bibnamefont {Cirac}}, \bibinfo
  {author} {\bibfnamefont {K.}~\bibnamefont {Jansen}},\ and\ \bibinfo {author}
  {\bibfnamefont {S.}~\bibnamefont {Kuhn}},\ }\bibfield  {title} {\bibinfo
  {title} {Efficient basis formulation for 1+1 dimensional su(2) lattice gauge
  theory: Spectral calculations with matrix product states},\ }\href
  {https://doi.org/10.1103/PhysRevX.7.041046} {\bibfield  {journal} {\bibinfo
  {journal} {Phys. Rev.}\ }\textbf {\bibinfo {volume} {X7}},\ \bibinfo {pages}
  {041046} (\bibinfo {year} {2017})},\ \Eprint
  {https://arxiv.org/abs/1707.06434} {arXiv:1707.06434 [hep-lat]} \BibitemShut
  {NoStop}%
\bibitem [{\citenamefont {Kaplan}\ and\ \citenamefont
  {Stryker}(2020)}]{Kaplan:2018vnj}%
  \BibitemOpen
  \bibfield  {author} {\bibinfo {author} {\bibfnamefont {D.~B.}\ \bibnamefont
  {Kaplan}}\ and\ \bibinfo {author} {\bibfnamefont {J.~R.}\ \bibnamefont
  {Stryker}},\ }\bibfield  {title} {\bibinfo {title} {Gauss's law, duality, and
  the hamiltonian formulation of u(1) lattice gauge theory},\ }\href@noop {}
  {\bibfield  {journal} {\bibinfo  {journal} {Phys. Rev. D}\ }\textbf {\bibinfo
  {volume} {102}},\ \bibinfo {pages} {094515} (\bibinfo {year} {2020})},\
  \Eprint {https://arxiv.org/abs/1806.08797} {arXiv:1806.08797 [hep-lat]}
  \BibitemShut {NoStop}%
\bibitem [{\citenamefont {Zache}\ \emph {et~al.}(2018)\citenamefont {Zache},
  \citenamefont {Hebenstreit}, \citenamefont {Jendrzejewski}, \citenamefont
  {Oberthaler}, \citenamefont {Berges},\ and\ \citenamefont
  {Hauke}}]{Zache:2018jbt}%
  \BibitemOpen
  \bibfield  {author} {\bibinfo {author} {\bibfnamefont {T.~V.}\ \bibnamefont
  {Zache}}, \bibinfo {author} {\bibfnamefont {F.}~\bibnamefont {Hebenstreit}},
  \bibinfo {author} {\bibfnamefont {F.}~\bibnamefont {Jendrzejewski}}, \bibinfo
  {author} {\bibfnamefont {M.~K.}\ \bibnamefont {Oberthaler}}, \bibinfo
  {author} {\bibfnamefont {J.}~\bibnamefont {Berges}},\ and\ \bibinfo {author}
  {\bibfnamefont {P.}~\bibnamefont {Hauke}},\ }\bibfield  {title} {\bibinfo
  {title} {Quantum simulation of lattice gauge theories using wilson
  fermions},\ }\href {https://doi.org/10.1088/2058-9565/aac33b} {\bibfield
  {journal} {\bibinfo  {journal} {Sci. Technol.}\ }\textbf {\bibinfo {volume}
  {3}},\ \bibinfo {pages} {034010} (\bibinfo {year} {2018})},\ \Eprint
  {https://arxiv.org/abs/1802.06704} {arXiv:1802.06704 [cond-mat.quant-gas]}
  \BibitemShut {NoStop}%
\bibitem [{\citenamefont {Stryker}(2019)}]{Stryker:2018efp}%
  \BibitemOpen
  \bibfield  {author} {\bibinfo {author} {\bibfnamefont {J.~R.}\ \bibnamefont
  {Stryker}},\ }\bibfield  {title} {\bibinfo {title} {Oracles for gauss's law
  on digital quantum computers},\ }\href
  {https://doi.org/10.1103/PhysRevA.99.042301} {\bibfield  {journal} {\bibinfo
  {journal} {Phys. Rev.}\ }\textbf {\bibinfo {volume} {A99}},\ \bibinfo {pages}
  {042301} (\bibinfo {year} {2019})},\ \Eprint
  {https://arxiv.org/abs/1812.01617} {arXiv:1812.01617 [quant-ph]} \BibitemShut
  {NoStop}%
\bibitem [{\citenamefont {Raychowdhury}(2019)}]{Raychowdhury:2018tfj}%
  \BibitemOpen
  \bibfield  {author} {\bibinfo {author} {\bibfnamefont {I.}~\bibnamefont
  {Raychowdhury}},\ }\bibfield  {title} {\bibinfo {title} {Low energy spectrum
  of su(2) lattice gauge theory},\ }\href
  {https://doi.org/10.1140/epjc/s10052-019-6753-0} {\bibfield  {journal}
  {\bibinfo  {journal} {Eur. Phys. J.}\ }\textbf {\bibinfo {volume} {C79}},\
  \bibinfo {pages} {235} (\bibinfo {year} {2019})},\ \Eprint
  {https://arxiv.org/abs/1804.01304} {arXiv:1804.01304 [hep-lat]} \BibitemShut
  {NoStop}%
\bibitem [{\citenamefont {Davoudi}\ \emph {et~al.}(2020)\citenamefont
  {Davoudi}, \citenamefont {Hafezi}, \citenamefont {Monroe}, \citenamefont
  {Pagano}, \citenamefont {Seif},\ and\ \citenamefont
  {Shaw}}]{Davoudi:2019bhy}%
  \BibitemOpen
  \bibfield  {author} {\bibinfo {author} {\bibfnamefont {Z.}~\bibnamefont
  {Davoudi}}, \bibinfo {author} {\bibfnamefont {M.}~\bibnamefont {Hafezi}},
  \bibinfo {author} {\bibfnamefont {C.}~\bibnamefont {Monroe}}, \bibinfo
  {author} {\bibfnamefont {G.}~\bibnamefont {Pagano}}, \bibinfo {author}
  {\bibfnamefont {A.}~\bibnamefont {Seif}},\ and\ \bibinfo {author}
  {\bibfnamefont {A.}~\bibnamefont {Shaw}},\ }\bibfield  {title} {\bibinfo
  {title} {Towards analog quantum simulations of lattice gauge theories with
  trapped ions},\ }\href@noop {} {\bibfield  {journal} {\bibinfo  {journal}
  {Phys. Rev. Research}\ }\textbf {\bibinfo {volume} {2}} (\bibinfo {year}
  {2020})},\ \Eprint {https://arxiv.org/abs/1908.03210} {arXiv:1908.03210
  [quant-ph]} \BibitemShut {NoStop}%
\bibitem [{\citenamefont {Haase}\ \emph {et~al.}(2021)\citenamefont {Haase},
  \citenamefont {Dellantonio}, \citenamefont {Celi}, \citenamefont {Paulson},
  \citenamefont {Kan}, \citenamefont {Jansen},\ and\ \citenamefont
  {Muschik}}]{Haase:2020kaj}%
  \BibitemOpen
  \bibfield  {author} {\bibinfo {author} {\bibfnamefont {J.~F.}\ \bibnamefont
  {Haase}}, \bibinfo {author} {\bibfnamefont {L.}~\bibnamefont {Dellantonio}},
  \bibinfo {author} {\bibfnamefont {A.}~\bibnamefont {Celi}}, \bibinfo {author}
  {\bibfnamefont {D.}~\bibnamefont {Paulson}}, \bibinfo {author} {\bibfnamefont
  {A.}~\bibnamefont {Kan}}, \bibinfo {author} {\bibfnamefont {K.}~\bibnamefont
  {Jansen}},\ and\ \bibinfo {author} {\bibfnamefont {C.~A.}\ \bibnamefont
  {Muschik}},\ }\bibfield  {title} {\bibinfo {title} {A resource efficient
  approach for quantum and classical simulations of gauge theories in particle
  physics},\ }\href@noop {} {\bibfield  {journal} {\bibinfo  {journal}
  {Quantum}\ }\textbf {\bibinfo {volume} {5}} (\bibinfo {year} {2021})},\
  \Eprint {https://arxiv.org/abs/2006.14160} {arXiv:2006.14160 [quant-ph]}
  \BibitemShut {NoStop}%
\bibitem [{\citenamefont {Paulson}\ \emph {et~al.}(2020)\citenamefont {Paulson}
  \emph {et~al.}}]{Paulson:2020zjd}%
  \BibitemOpen
  \bibfield  {author} {\bibinfo {author} {\bibfnamefont {D.}~\bibnamefont
  {Paulson}} \emph {et~al.},\ }\href@noop {} {\bibinfo {title} {Towards
  simulating 2d effects in lattice gauge theories on a quantum computer}}
  (\bibinfo {year} {2020}),\ \Eprint {https://arxiv.org/abs/2008.09252}
  {arXiv:2008.09252 [quant-ph]} \BibitemShut {NoStop}%
\bibitem [{\citenamefont {Davoudi}\ \emph
  {et~al.}(2021{\natexlab{c}})\citenamefont {Davoudi}, \citenamefont
  {Raychowdhury},\ and\ \citenamefont {Shaw}}]{Davoudi:2020yln}%
  \BibitemOpen
  \bibfield  {author} {\bibinfo {author} {\bibfnamefont {Z.}~\bibnamefont
  {Davoudi}}, \bibinfo {author} {\bibfnamefont {I.}~\bibnamefont
  {Raychowdhury}},\ and\ \bibinfo {author} {\bibfnamefont {A.}~\bibnamefont
  {Shaw}},\ }\bibfield  {title} {\bibinfo {title} {Search for efficient
  formulations for hamiltonian simulation of non-abelian lattice gauge
  theories},\ }\href@noop {} {\bibfield  {journal} {\bibinfo  {journal} {Phys.
  Rev. D}\ }\textbf {\bibinfo {volume} {104}} (\bibinfo {year}
  {2021}{\natexlab{c}})},\ \Eprint {https://arxiv.org/abs/2009.11802}
  {arXiv:2009.11802 [hep-lat]} \BibitemShut {NoStop}%
\bibitem [{\citenamefont {Buser}\ \emph {et~al.}(2020)\citenamefont {Buser},
  \citenamefont {Gharibyan}, \citenamefont {Hanada}, \citenamefont {Honda},\
  and\ \citenamefont {Liu}}]{Buser:2020cvn}%
  \BibitemOpen
  \bibfield  {author} {\bibinfo {author} {\bibfnamefont {A.}~\bibnamefont
  {Buser}}, \bibinfo {author} {\bibfnamefont {H.}~\bibnamefont {Gharibyan}},
  \bibinfo {author} {\bibfnamefont {M.}~\bibnamefont {Hanada}}, \bibinfo
  {author} {\bibfnamefont {M.}~\bibnamefont {Honda}},\ and\ \bibinfo {author}
  {\bibfnamefont {J.}~\bibnamefont {Liu}},\ }\href@noop {} {\bibinfo {title}
  {Quantum simulation of gauge theory via orbifold lattice}} (\bibinfo {year}
  {2020}),\ \Eprint {https://arxiv.org/abs/2011.06576} {arXiv:2011.06576
  [hep-th]} \BibitemShut {NoStop}%
\bibitem [{\citenamefont {Raychowdhury}\ and\ \citenamefont
  {Stryker}(2018)}]{Raychowdhury:2018osk}%
  \BibitemOpen
  \bibfield  {author} {\bibinfo {author} {\bibfnamefont {I.}~\bibnamefont
  {Raychowdhury}}\ and\ \bibinfo {author} {\bibfnamefont {J.~R.}\ \bibnamefont
  {Stryker}},\ }\href@noop {} {\bibinfo {title} {Tailoring non-abelian lattice
  gauge theory for quantum simulation}} (\bibinfo {year} {2018}),\ \Eprint
  {https://arxiv.org/abs/1812.07554} {arXiv:1812.07554 [hep-lat]} \BibitemShut
  {NoStop}%
\bibitem [{\citenamefont {Raychowdhury}\ and\ \citenamefont
  {Stryker}(2020)}]{Raychowdhury:2019iki}%
  \BibitemOpen
  \bibfield  {author} {\bibinfo {author} {\bibfnamefont {I.}~\bibnamefont
  {Raychowdhury}}\ and\ \bibinfo {author} {\bibfnamefont {J.~R.}\ \bibnamefont
  {Stryker}},\ }\bibfield  {title} {\bibinfo {title} {Loop, string, and hadron
  dynamics in su(2) hamiltonian lattice gauge theories},\ }\href
  {https://doi.org/10.1103/PhysRevD.101.114502} {\bibfield  {journal} {\bibinfo
   {journal} {Phys. Rev. D}\ }\textbf {\bibinfo {volume} {101}},\ \bibinfo
  {pages} {114502} (\bibinfo {year} {2020})},\ \Eprint
  {https://arxiv.org/abs/1912.06133} {arXiv:1912.06133 [hep-lat]} \BibitemShut
  {NoStop}%
\bibitem [{\citenamefont {Tagliacozzo}\ \emph
  {et~al.}(2013{\natexlab{c}})\citenamefont {Tagliacozzo}, \citenamefont
  {Celi}, \citenamefont {Orland},\ and\ \citenamefont
  {Lewenstein}}]{Tagliacozzo:2012df}%
  \BibitemOpen
  \bibfield  {author} {\bibinfo {author} {\bibfnamefont {L.}~\bibnamefont
  {Tagliacozzo}}, \bibinfo {author} {\bibfnamefont {A.}~\bibnamefont {Celi}},
  \bibinfo {author} {\bibfnamefont {P.}~\bibnamefont {Orland}},\ and\ \bibinfo
  {author} {\bibfnamefont {M.}~\bibnamefont {Lewenstein}},\ }\bibfield  {title}
  {\bibinfo {title} {Simulations of non-abelian gauge theories with optical
  lattices},\ }\href {https://doi.org/10.1038/ncomms3615} {\bibfield  {journal}
  {\bibinfo  {journal} {Nature Commun.}\ }\textbf {\bibinfo {volume} {4}},\
  \bibinfo {pages} {2615} (\bibinfo {year} {2013}{\natexlab{c}})},\ \Eprint
  {https://arxiv.org/abs/1211.2704} {arXiv:1211.2704 [cond-mat.quant-gas]}
  \BibitemShut {NoStop}%
\bibitem [{\citenamefont {Ji}\ \emph {et~al.}(2020)\citenamefont {Ji},
  \citenamefont {Lamm},\ and\ \citenamefont {Zhu}}]{Ji:2020kjk}%
  \BibitemOpen
  \bibfield  {author} {\bibinfo {author} {\bibfnamefont {Y.}~\bibnamefont
  {Ji}}, \bibinfo {author} {\bibfnamefont {H.}~\bibnamefont {Lamm}},\ and\
  \bibinfo {author} {\bibfnamefont {S.}~\bibnamefont {Zhu}} (\bibinfo
  {collaboration} {NuQS}),\ }\bibfield  {title} {\bibinfo {title} {Gluon field
  digitization via group space decimation for quantum computers},\ }\href
  {https://doi.org/10.1103/PhysRevD.102.114513} {\bibfield  {journal} {\bibinfo
   {journal} {Phys. Rev. D}\ }\textbf {\bibinfo {volume} {102}},\ \bibinfo
  {pages} {114513} (\bibinfo {year} {2020})},\ \Eprint
  {https://arxiv.org/abs/2005.14221} {arXiv:2005.14221 [hep-lat]} \BibitemShut
  {NoStop}%
\bibitem [{\citenamefont {Chandrasekharan}\ and\ \citenamefont
  {Wiese}(1997)}]{Chandrasekharan:1996ih}%
  \BibitemOpen
  \bibfield  {author} {\bibinfo {author} {\bibfnamefont {S.}~\bibnamefont
  {Chandrasekharan}}\ and\ \bibinfo {author} {\bibfnamefont {U.}~\bibnamefont
  {Wiese}},\ }\bibfield  {title} {\bibinfo {title} {Quantum link models: A
  discrete approach to gauge theories},\ }\href
  {https://doi.org/10.1016/S0550-3213(97)00006-0} {\bibfield  {journal}
  {\bibinfo  {journal} {Nucl. Phys. B}\ }\textbf {\bibinfo {volume} {492}},\
  \bibinfo {pages} {455} (\bibinfo {year} {1997})},\ \Eprint
  {https://arxiv.org/abs/hep-lat/9609042} {arXiv:hep-lat/9609042} \BibitemShut
  {NoStop}%
\bibitem [{\citenamefont {Brower}\ \emph {et~al.}(2004)\citenamefont {Brower},
  \citenamefont {Chandrasekharan}, \citenamefont {Riederer},\ and\
  \citenamefont {Wiese}}]{Brower:2003vy}%
  \BibitemOpen
  \bibfield  {author} {\bibinfo {author} {\bibfnamefont {R.}~\bibnamefont
  {Brower}}, \bibinfo {author} {\bibfnamefont {S.}~\bibnamefont
  {Chandrasekharan}}, \bibinfo {author} {\bibfnamefont {S.}~\bibnamefont
  {Riederer}},\ and\ \bibinfo {author} {\bibfnamefont {U.}~\bibnamefont
  {Wiese}},\ }\bibfield  {title} {\bibinfo {title} {D theory: Field
  quantization by dimensional reduction of discrete variables},\ }\href
  {https://doi.org/10.1016/j.nuclphysb.2004.06.007} {\bibfield  {journal}
  {\bibinfo  {journal} {Nucl. Phys. B}\ }\textbf {\bibinfo {volume} {693}},\
  \bibinfo {pages} {149} (\bibinfo {year} {2004})},\ \Eprint
  {https://arxiv.org/abs/hep-lat/0309182} {arXiv:hep-lat/0309182} \BibitemShut
  {NoStop}%
\bibitem [{\citenamefont {Wiese}(2006)}]{Wiese:2006kp}%
  \BibitemOpen
  \bibfield  {author} {\bibinfo {author} {\bibfnamefont {U.}~\bibnamefont
  {Wiese}},\ }\bibfield  {title} {\bibinfo {title} {D-theory: A quest for
  nature's regularization},\ }\href
  {https://doi.org/10.1016/j.nuclphysbps.2006.01.027} {\bibfield  {journal}
  {\bibinfo  {journal} {Nucl. Phys. B Proc. Suppl.}\ }\textbf {\bibinfo
  {volume} {153}},\ \bibinfo {pages} {336} (\bibinfo {year}
  {2006})}\BibitemShut {NoStop}%
\bibitem [{\citenamefont {Atas}\ \emph {et~al.}(2021)\citenamefont {Atas},
  \citenamefont {Zhang}, \citenamefont {Lewis}, \citenamefont {Jahanpour},
  \citenamefont {Haase},\ and\ \citenamefont {Muschik}}]{atas2021su2}%
  \BibitemOpen
  \bibfield  {author} {\bibinfo {author} {\bibfnamefont {Y.}~\bibnamefont
  {Atas}}, \bibinfo {author} {\bibfnamefont {J.}~\bibnamefont {Zhang}},
  \bibinfo {author} {\bibfnamefont {R.}~\bibnamefont {Lewis}}, \bibinfo
  {author} {\bibfnamefont {A.}~\bibnamefont {Jahanpour}}, \bibinfo {author}
  {\bibfnamefont {J.~F.}\ \bibnamefont {Haase}},\ and\ \bibinfo {author}
  {\bibfnamefont {C.~A.}\ \bibnamefont {Muschik}},\ }\href@noop {} {\bibinfo
  {title} {Su(2) hadrons on a quantum computer}} (\bibinfo {year} {2021}),\
  \Eprint {https://arxiv.org/abs/2102.08920} {arXiv:2102.08920 [quant-ph]}
  \BibitemShut {NoStop}%
\bibitem [{\citenamefont {Klco}\ \emph {et~al.}(2021)\citenamefont {Klco},
  \citenamefont {Roggero},\ and\ \citenamefont {Savage}}]{klco2021standard}%
  \BibitemOpen
  \bibfield  {author} {\bibinfo {author} {\bibfnamefont {N.}~\bibnamefont
  {Klco}}, \bibinfo {author} {\bibfnamefont {A.}~\bibnamefont {Roggero}},\ and\
  \bibinfo {author} {\bibfnamefont {M.~J.}\ \bibnamefont {Savage}},\
  }\href@noop {} {\bibinfo {title} {Standard model physics and the digital
  quantum revolution: Thoughts about the interface}} (\bibinfo {year} {2021}),\
  \Eprint {https://arxiv.org/abs/2107.04769} {arXiv:2107.04769 [quant-ph]}
  \BibitemShut {NoStop}%
\bibitem [{\citenamefont {de~Jong}\ \emph {et~al.}(2021)\citenamefont
  {de~Jong}, \citenamefont {Lee}, \citenamefont {Mulligan}, \citenamefont
  {Płoskoń}, \citenamefont {Ringer},\ and\ \citenamefont
  {Yao}}]{dejong2021quantum}%
  \BibitemOpen
  \bibfield  {author} {\bibinfo {author} {\bibfnamefont {W.~A.}\ \bibnamefont
  {de~Jong}}, \bibinfo {author} {\bibfnamefont {K.}~\bibnamefont {Lee}},
  \bibinfo {author} {\bibfnamefont {J.}~\bibnamefont {Mulligan}}, \bibinfo
  {author} {\bibfnamefont {M.}~\bibnamefont {Płoskoń}}, \bibinfo {author}
  {\bibfnamefont {F.}~\bibnamefont {Ringer}},\ and\ \bibinfo {author}
  {\bibfnamefont {X.}~\bibnamefont {Yao}},\ }\href@noop {} {\bibinfo {title}
  {Quantum simulation of non-equilibrium dynamics and thermalization in the
  schwinger model}} (\bibinfo {year} {2021}),\ \Eprint
  {https://arxiv.org/abs/2106.08394} {arXiv:2106.08394 [quant-ph]} \BibitemShut
  {NoStop}%
\bibitem [{\citenamefont {Meurice}(2021)}]{meurice2021theoretical}%
  \BibitemOpen
  \bibfield  {author} {\bibinfo {author} {\bibfnamefont {Y.}~\bibnamefont
  {Meurice}},\ }\href@noop {} {\bibinfo {title} {Theoretical methods to design
  and test quantum simulators for the compact abelian higgs model}} (\bibinfo
  {year} {2021}),\ \Eprint {https://arxiv.org/abs/2107.11366} {arXiv:2107.11366
  [quant-ph]} \BibitemShut {NoStop}%
\bibitem [{\citenamefont {Zohar}(2021)}]{zohar2021quantum}%
  \BibitemOpen
  \bibfield  {author} {\bibinfo {author} {\bibfnamefont {E.}~\bibnamefont
  {Zohar}},\ }\href@noop {} {\bibinfo {title} {Quantum simulation of lattice
  gauge theories in more than one space dimension -- requirements, challenges,
  methods}} (\bibinfo {year} {2021}),\ \Eprint
  {https://arxiv.org/abs/2106.04609} {arXiv:2106.04609 [quant-ph]} \BibitemShut
  {NoStop}%
\bibitem [{\citenamefont {Armon}\ \emph {et~al.}(2021)\citenamefont {Armon},
  \citenamefont {Ashkenazi}, \citenamefont {García-Moreno}, \citenamefont
  {González-Tudela},\ and\ \citenamefont {Zohar}}]{armon2021photonmediated}%
  \BibitemOpen
  \bibfield  {author} {\bibinfo {author} {\bibfnamefont {T.}~\bibnamefont
  {Armon}}, \bibinfo {author} {\bibfnamefont {S.}~\bibnamefont {Ashkenazi}},
  \bibinfo {author} {\bibfnamefont {G.}~\bibnamefont {García-Moreno}},
  \bibinfo {author} {\bibfnamefont {A.}~\bibnamefont {González-Tudela}},\ and\
  \bibinfo {author} {\bibfnamefont {E.}~\bibnamefont {Zohar}},\ }\href@noop {}
  {\bibinfo {title} {Photon-mediated stroboscopic quantum simulation of a
  $\mathbb{Z}_{2}$ lattice gauge theory}} (\bibinfo {year} {2021}),\ \Eprint
  {https://arxiv.org/abs/2107.13024} {arXiv:2107.13024 [quant-ph]} \BibitemShut
  {NoStop}%
\bibitem [{\citenamefont {Andrade}\ \emph {et~al.}(2022)\citenamefont
  {Andrade}, \citenamefont {Davoudi}, \citenamefont {Gra{\ss}}, \citenamefont
  {Hafezi}, \citenamefont {Pagano},\ and\ \citenamefont
  {Seif}}]{andrade2021engineering}%
  \BibitemOpen
  \bibfield  {author} {\bibinfo {author} {\bibfnamefont {B.}~\bibnamefont
  {Andrade}}, \bibinfo {author} {\bibfnamefont {Z.}~\bibnamefont {Davoudi}},
  \bibinfo {author} {\bibfnamefont {T.}~\bibnamefont {Gra{\ss}}}, \bibinfo
  {author} {\bibfnamefont {M.}~\bibnamefont {Hafezi}}, \bibinfo {author}
  {\bibfnamefont {G.}~\bibnamefont {Pagano}},\ and\ \bibinfo {author}
  {\bibfnamefont {A.}~\bibnamefont {Seif}},\ }\bibfield  {title} {\bibinfo
  {title} {Engineering an effective three-spin hamiltonian in trapped-ion
  systems for applications in quantum simulation},\ }\href
  {https://doi.org/10.1088/2058-9565/ac5f5b} {\bibfield  {journal} {\bibinfo
  {journal} {Quantum Science and Technology}\ }\textbf {\bibinfo {volume}
  {7}},\ \bibinfo {pages} {034001} (\bibinfo {year} {2022})}\BibitemShut
  {NoStop}%
\bibitem [{\citenamefont {Kogut}\ and\ \citenamefont
  {Susskind}(1975)}]{PhysRevD.11.395}%
  \BibitemOpen
  \bibfield  {author} {\bibinfo {author} {\bibfnamefont {J.}~\bibnamefont
  {Kogut}}\ and\ \bibinfo {author} {\bibfnamefont {L.}~\bibnamefont
  {Susskind}},\ }\bibfield  {title} {\bibinfo {title} {Hamiltonian formulation
  of wilson's lattice gauge theories},\ }\href
  {https://doi.org/10.1103/PhysRevD.11.395} {\bibfield  {journal} {\bibinfo
  {journal} {Phys. Rev. D}\ }\textbf {\bibinfo {volume} {11}},\ \bibinfo
  {pages} {395} (\bibinfo {year} {1975})}\BibitemShut {NoStop}%
\bibitem [{\citenamefont {Kogut}(1979)}]{RevModPhys.51.659}%
  \BibitemOpen
  \bibfield  {author} {\bibinfo {author} {\bibfnamefont {J.~B.}\ \bibnamefont
  {Kogut}},\ }\bibfield  {title} {\bibinfo {title} {An introduction to lattice
  gauge theory and spin systems},\ }\href
  {https://doi.org/10.1103/RevModPhys.51.659} {\bibfield  {journal} {\bibinfo
  {journal} {Rev. Mod. Phys.}\ }\textbf {\bibinfo {volume} {51}},\ \bibinfo
  {pages} {659} (\bibinfo {year} {1979})}\BibitemShut {NoStop}%
\bibitem [{\citenamefont {Robson}\ and\ \citenamefont
  {Webber}(1980)}]{Robson:1980nt}%
  \BibitemOpen
  \bibfield  {author} {\bibinfo {author} {\bibfnamefont {D.}~\bibnamefont
  {Robson}}\ and\ \bibinfo {author} {\bibfnamefont {D.~M.}\ \bibnamefont
  {Webber}},\ }\bibfield  {title} {\bibinfo {title} {Gauge theories on a small
  lattice},\ }\href {https://doi.org/10.1007/BF01577321} {\bibfield  {journal}
  {\bibinfo  {journal} {Z. Phys. C}\ }\textbf {\bibinfo {volume} {7}},\
  \bibinfo {pages} {53} (\bibinfo {year} {1980})}\BibitemShut {NoStop}%
\bibitem [{\citenamefont {Ligterink}\ \emph {et~al.}(2000)\citenamefont
  {Ligterink}, \citenamefont {Walet},\ and\ \citenamefont
  {Bishop}}]{LIGTERINK2000983c}%
  \BibitemOpen
  \bibfield  {author} {\bibinfo {author} {\bibfnamefont {N.}~\bibnamefont
  {Ligterink}}, \bibinfo {author} {\bibfnamefont {N.}~\bibnamefont {Walet}},\
  and\ \bibinfo {author} {\bibfnamefont {R.}~\bibnamefont {Bishop}},\
  }\bibfield  {title} {\bibinfo {title} {A many-body treatment of hamiltonian
  lattice gauge theory},\ }\href
  {https://doi.org/https://doi.org/10.1016/S0375-9474(99)00748-4} {\bibfield
  {journal} {\bibinfo  {journal} {Nuclear Physics A}\ }\textbf {\bibinfo
  {volume} {663-664}},\ \bibinfo {pages} {983c} (\bibinfo {year}
  {2000})}\BibitemShut {NoStop}%
\bibitem [{\citenamefont {Bronzan}(1985)}]{PhysRevD.31.2020}%
  \BibitemOpen
  \bibfield  {author} {\bibinfo {author} {\bibfnamefont {J.~B.}\ \bibnamefont
  {Bronzan}},\ }\bibfield  {title} {\bibinfo {title} {Explicit hamiltonian for
  su(2) lattice gauge theory},\ }\href
  {https://doi.org/10.1103/PhysRevD.31.2020} {\bibfield  {journal} {\bibinfo
  {journal} {Phys. Rev. D}\ }\textbf {\bibinfo {volume} {31}},\ \bibinfo
  {pages} {2020} (\bibinfo {year} {1985})}\BibitemShut {NoStop}%
\bibitem [{\citenamefont {Peruzzo}\ \emph {et~al.}(2014)\citenamefont
  {Peruzzo}, \citenamefont {McClean}, \citenamefont {Shadbolt}, \citenamefont
  {Yung}, \citenamefont {Zhou}, \citenamefont {Love}, \citenamefont
  {Aspuru-Guzik},\ and\ \citenamefont {O’Brien}}]{Peruzzo_2014}%
  \BibitemOpen
  \bibfield  {author} {\bibinfo {author} {\bibfnamefont {A.}~\bibnamefont
  {Peruzzo}}, \bibinfo {author} {\bibfnamefont {J.}~\bibnamefont {McClean}},
  \bibinfo {author} {\bibfnamefont {P.}~\bibnamefont {Shadbolt}}, \bibinfo
  {author} {\bibfnamefont {M.-H.}\ \bibnamefont {Yung}}, \bibinfo {author}
  {\bibfnamefont {X.-Q.}\ \bibnamefont {Zhou}}, \bibinfo {author}
  {\bibfnamefont {P.~J.}\ \bibnamefont {Love}}, \bibinfo {author}
  {\bibfnamefont {A.}~\bibnamefont {Aspuru-Guzik}},\ and\ \bibinfo {author}
  {\bibfnamefont {J.~L.}\ \bibnamefont {O’Brien}},\ }\bibfield  {title}
  {\bibinfo {title} {A variational eigenvalue solver on a photonic quantum
  processor},\ }\bibfield  {journal} {\bibinfo  {journal} {Nature
  Communications}\ }\textbf {\bibinfo {volume} {5}},\ \href
  {https://doi.org/10.1038/ncomms5213} {10.1038/ncomms5213} (\bibinfo {year}
  {2014})\BibitemShut {NoStop}%
\bibitem [{\citenamefont {McClean}\ \emph {et~al.}(2016)\citenamefont
  {McClean}, \citenamefont {Romero}, \citenamefont {Babbush},\ and\
  \citenamefont {Aspuru-Guzik}}]{McClean_2016}%
  \BibitemOpen
  \bibfield  {author} {\bibinfo {author} {\bibfnamefont {J.~R.}\ \bibnamefont
  {McClean}}, \bibinfo {author} {\bibfnamefont {J.}~\bibnamefont {Romero}},
  \bibinfo {author} {\bibfnamefont {R.}~\bibnamefont {Babbush}},\ and\ \bibinfo
  {author} {\bibfnamefont {A.}~\bibnamefont {Aspuru-Guzik}},\ }\bibfield
  {title} {\bibinfo {title} {The theory of variational hybrid quantum-classical
  algorithms},\ }\href {https://doi.org/10.1088/1367-2630/18/2/023023}
  {\bibfield  {journal} {\bibinfo  {journal} {New Journal of Physics}\ }\textbf
  {\bibinfo {volume} {18}},\ \bibinfo {pages} {023023} (\bibinfo {year}
  {2016})}\BibitemShut {NoStop}%
\bibitem [{\citenamefont {O’Malley}\ \emph {et~al.}(2016)\citenamefont
  {O’Malley}, \citenamefont {Babbush}, \citenamefont {Kivlichan},
  \citenamefont {Romero}, \citenamefont {McClean}, \citenamefont {Barends},
  \citenamefont {Kelly}, \citenamefont {Roushan}, \citenamefont {Tranter},
  \citenamefont {Ding},\ and\ \citenamefont {et~al.}}]{O_Malley_2016}%
  \BibitemOpen
  \bibfield  {author} {\bibinfo {author} {\bibfnamefont {P.}~\bibnamefont
  {O’Malley}}, \bibinfo {author} {\bibfnamefont {R.}~\bibnamefont {Babbush}},
  \bibinfo {author} {\bibfnamefont {I.}~\bibnamefont {Kivlichan}}, \bibinfo
  {author} {\bibfnamefont {J.}~\bibnamefont {Romero}}, \bibinfo {author}
  {\bibfnamefont {J.}~\bibnamefont {McClean}}, \bibinfo {author} {\bibfnamefont
  {R.}~\bibnamefont {Barends}}, \bibinfo {author} {\bibfnamefont
  {J.}~\bibnamefont {Kelly}}, \bibinfo {author} {\bibfnamefont
  {P.}~\bibnamefont {Roushan}}, \bibinfo {author} {\bibfnamefont
  {A.}~\bibnamefont {Tranter}}, \bibinfo {author} {\bibfnamefont
  {N.}~\bibnamefont {Ding}},\ and\ \bibinfo {author} {\bibnamefont {et~al.}},\
  }\bibfield  {title} {\bibinfo {title} {Scalable quantum simulation of
  molecular energies},\ }\bibfield  {journal} {\bibinfo  {journal} {Physical
  Review X}\ }\textbf {\bibinfo {volume} {6}},\ \href
  {https://doi.org/10.1103/physrevx.6.031007} {10.1103/physrevx.6.031007}
  (\bibinfo {year} {2016})\BibitemShut {NoStop}%
\bibitem [{\citenamefont {Kandala}\ \emph {et~al.}(2017)\citenamefont
  {Kandala}, \citenamefont {Mezzacapo}, \citenamefont {Temme}, \citenamefont
  {Takita}, \citenamefont {Brink}, \citenamefont {Chow},\ and\ \citenamefont
  {Gambetta}}]{Kandala_2017}%
  \BibitemOpen
  \bibfield  {author} {\bibinfo {author} {\bibfnamefont {A.}~\bibnamefont
  {Kandala}}, \bibinfo {author} {\bibfnamefont {A.}~\bibnamefont {Mezzacapo}},
  \bibinfo {author} {\bibfnamefont {K.}~\bibnamefont {Temme}}, \bibinfo
  {author} {\bibfnamefont {M.}~\bibnamefont {Takita}}, \bibinfo {author}
  {\bibfnamefont {M.}~\bibnamefont {Brink}}, \bibinfo {author} {\bibfnamefont
  {J.~M.}\ \bibnamefont {Chow}},\ and\ \bibinfo {author} {\bibfnamefont
  {J.~M.}\ \bibnamefont {Gambetta}},\ }\bibfield  {title} {\bibinfo {title}
  {Hardware-efficient variational quantum eigensolver for small molecules and
  quantum magnets},\ }\href {https://doi.org/10.1038/nature23879} {\bibfield
  {journal} {\bibinfo  {journal} {Nature}\ }\textbf {\bibinfo {volume} {549}},\
  \bibinfo {pages} {242–246} (\bibinfo {year} {2017})}\BibitemShut {NoStop}%
\bibitem [{\citenamefont {Hempel}\ \emph {et~al.}(2018)\citenamefont {Hempel},
  \citenamefont {Maier}, \citenamefont {Romero}, \citenamefont {McClean},
  \citenamefont {Monz}, \citenamefont {Shen}, \citenamefont {Jurcevic},
  \citenamefont {Lanyon}, \citenamefont {Love}, \citenamefont {Babbush},
  \citenamefont {Aspuru-Guzik}, \citenamefont {Blatt},\ and\ \citenamefont
  {Roos}}]{PhysRevX.8.031022}%
  \BibitemOpen
  \bibfield  {author} {\bibinfo {author} {\bibfnamefont {C.}~\bibnamefont
  {Hempel}}, \bibinfo {author} {\bibfnamefont {C.}~\bibnamefont {Maier}},
  \bibinfo {author} {\bibfnamefont {J.}~\bibnamefont {Romero}}, \bibinfo
  {author} {\bibfnamefont {J.}~\bibnamefont {McClean}}, \bibinfo {author}
  {\bibfnamefont {T.}~\bibnamefont {Monz}}, \bibinfo {author} {\bibfnamefont
  {H.}~\bibnamefont {Shen}}, \bibinfo {author} {\bibfnamefont {P.}~\bibnamefont
  {Jurcevic}}, \bibinfo {author} {\bibfnamefont {B.~P.}\ \bibnamefont
  {Lanyon}}, \bibinfo {author} {\bibfnamefont {P.}~\bibnamefont {Love}},
  \bibinfo {author} {\bibfnamefont {R.}~\bibnamefont {Babbush}}, \bibinfo
  {author} {\bibfnamefont {A.}~\bibnamefont {Aspuru-Guzik}}, \bibinfo {author}
  {\bibfnamefont {R.}~\bibnamefont {Blatt}},\ and\ \bibinfo {author}
  {\bibfnamefont {C.~F.}\ \bibnamefont {Roos}},\ }\bibfield  {title} {\bibinfo
  {title} {Quantum chemistry calculations on a trapped-ion quantum simulator},\
  }\href {https://doi.org/10.1103/PhysRevX.8.031022} {\bibfield  {journal}
  {\bibinfo  {journal} {Phys. Rev. X}\ }\textbf {\bibinfo {volume} {8}},\
  \bibinfo {pages} {031022} (\bibinfo {year} {2018})}\BibitemShut {NoStop}%
\bibitem [{\citenamefont {Colless}\ \emph {et~al.}(2018)\citenamefont
  {Colless}, \citenamefont {Ramasesh}, \citenamefont {Dahlen}, \citenamefont
  {Blok}, \citenamefont {Kimchi-Schwartz}, \citenamefont {McClean},
  \citenamefont {Carter}, \citenamefont {de~Jong},\ and\ \citenamefont
  {Siddiqi}}]{PhysRevX.8.011021}%
  \BibitemOpen
  \bibfield  {author} {\bibinfo {author} {\bibfnamefont {J.~I.}\ \bibnamefont
  {Colless}}, \bibinfo {author} {\bibfnamefont {V.~V.}\ \bibnamefont
  {Ramasesh}}, \bibinfo {author} {\bibfnamefont {D.}~\bibnamefont {Dahlen}},
  \bibinfo {author} {\bibfnamefont {M.~S.}\ \bibnamefont {Blok}}, \bibinfo
  {author} {\bibfnamefont {M.~E.}\ \bibnamefont {Kimchi-Schwartz}}, \bibinfo
  {author} {\bibfnamefont {J.~R.}\ \bibnamefont {McClean}}, \bibinfo {author}
  {\bibfnamefont {J.}~\bibnamefont {Carter}}, \bibinfo {author} {\bibfnamefont
  {W.~A.}\ \bibnamefont {de~Jong}},\ and\ \bibinfo {author} {\bibfnamefont
  {I.}~\bibnamefont {Siddiqi}},\ }\bibfield  {title} {\bibinfo {title}
  {Computation of molecular spectra on a quantum processor with an
  error-resilient algorithm},\ }\href
  {https://doi.org/10.1103/PhysRevX.8.011021} {\bibfield  {journal} {\bibinfo
  {journal} {Phys. Rev. X}\ }\textbf {\bibinfo {volume} {8}},\ \bibinfo {pages}
  {011021} (\bibinfo {year} {2018})}\BibitemShut {NoStop}%
\bibitem [{\citenamefont {Smart}\ and\ \citenamefont
  {Mazziotti}(2019)}]{PhysRevA.100.022517}%
  \BibitemOpen
  \bibfield  {author} {\bibinfo {author} {\bibfnamefont {S.~E.}\ \bibnamefont
  {Smart}}\ and\ \bibinfo {author} {\bibfnamefont {D.~A.}\ \bibnamefont
  {Mazziotti}},\ }\bibfield  {title} {\bibinfo {title} {Quantum-classical
  hybrid algorithm using an error-mitigating $n$-representability condition to
  compute the mott metal-insulator transition},\ }\href
  {https://doi.org/10.1103/PhysRevA.100.022517} {\bibfield  {journal} {\bibinfo
   {journal} {Phys. Rev. A}\ }\textbf {\bibinfo {volume} {100}},\ \bibinfo
  {pages} {022517} (\bibinfo {year} {2019})}\BibitemShut {NoStop}%
\bibitem [{\citenamefont {Sagastizabal}\ \emph {et~al.}(2019)\citenamefont
  {Sagastizabal}, \citenamefont {Bonet-Monroig}, \citenamefont {Singh},
  \citenamefont {Rol}, \citenamefont {Bultink}, \citenamefont {Fu},
  \citenamefont {Price}, \citenamefont {Ostroukh}, \citenamefont
  {Muthusubramanian}, \citenamefont {Bruno}, \citenamefont {Beekman},
  \citenamefont {Haider}, \citenamefont {O'Brien},\ and\ \citenamefont
  {DiCarlo}}]{PhysRevA.100.010302}%
  \BibitemOpen
  \bibfield  {author} {\bibinfo {author} {\bibfnamefont {R.}~\bibnamefont
  {Sagastizabal}}, \bibinfo {author} {\bibfnamefont {X.}~\bibnamefont
  {Bonet-Monroig}}, \bibinfo {author} {\bibfnamefont {M.}~\bibnamefont
  {Singh}}, \bibinfo {author} {\bibfnamefont {M.~A.}\ \bibnamefont {Rol}},
  \bibinfo {author} {\bibfnamefont {C.~C.}\ \bibnamefont {Bultink}}, \bibinfo
  {author} {\bibfnamefont {X.}~\bibnamefont {Fu}}, \bibinfo {author}
  {\bibfnamefont {C.~H.}\ \bibnamefont {Price}}, \bibinfo {author}
  {\bibfnamefont {V.~P.}\ \bibnamefont {Ostroukh}}, \bibinfo {author}
  {\bibfnamefont {N.}~\bibnamefont {Muthusubramanian}}, \bibinfo {author}
  {\bibfnamefont {A.}~\bibnamefont {Bruno}}, \bibinfo {author} {\bibfnamefont
  {M.}~\bibnamefont {Beekman}}, \bibinfo {author} {\bibfnamefont
  {N.}~\bibnamefont {Haider}}, \bibinfo {author} {\bibfnamefont {T.~E.}\
  \bibnamefont {O'Brien}},\ and\ \bibinfo {author} {\bibfnamefont
  {L.}~\bibnamefont {DiCarlo}},\ }\bibfield  {title} {\bibinfo {title}
  {Experimental error mitigation via symmetry verification in a variational
  quantum eigensolver},\ }\href {https://doi.org/10.1103/PhysRevA.100.010302}
  {\bibfield  {journal} {\bibinfo  {journal} {Phys. Rev. A}\ }\textbf {\bibinfo
  {volume} {100}},\ \bibinfo {pages} {010302} (\bibinfo {year}
  {2019})}\BibitemShut {NoStop}%
\bibitem [{\citenamefont {Grimsley}\ \emph {et~al.}(2019)\citenamefont
  {Grimsley}, \citenamefont {Economou}, \citenamefont {Barnes},\ and\
  \citenamefont {Mayhall}}]{Grimsley_2019}%
  \BibitemOpen
  \bibfield  {author} {\bibinfo {author} {\bibfnamefont {H.~R.}\ \bibnamefont
  {Grimsley}}, \bibinfo {author} {\bibfnamefont {S.~E.}\ \bibnamefont
  {Economou}}, \bibinfo {author} {\bibfnamefont {E.}~\bibnamefont {Barnes}},\
  and\ \bibinfo {author} {\bibfnamefont {N.~J.}\ \bibnamefont {Mayhall}},\
  }\bibfield  {title} {\bibinfo {title} {An adaptive variational algorithm for
  exact molecular simulations on a quantum computer},\ }\bibfield  {journal}
  {\bibinfo  {journal} {Nature Communications}\ }\textbf {\bibinfo {volume}
  {10}},\ \href {https://doi.org/10.1038/s41467-019-10988-2}
  {10.1038/s41467-019-10988-2} (\bibinfo {year} {2019})\BibitemShut {NoStop}%
\bibitem [{\citenamefont {Kandala}\ \emph {et~al.}(2019)\citenamefont
  {Kandala}, \citenamefont {Temme}, \citenamefont {Córcoles}, \citenamefont
  {Mezzacapo}, \citenamefont {Chow},\ and\ \citenamefont
  {Gambetta}}]{Kandala_2019}%
  \BibitemOpen
  \bibfield  {author} {\bibinfo {author} {\bibfnamefont {A.}~\bibnamefont
  {Kandala}}, \bibinfo {author} {\bibfnamefont {K.}~\bibnamefont {Temme}},
  \bibinfo {author} {\bibfnamefont {A.~D.}\ \bibnamefont {Córcoles}}, \bibinfo
  {author} {\bibfnamefont {A.}~\bibnamefont {Mezzacapo}}, \bibinfo {author}
  {\bibfnamefont {J.~M.}\ \bibnamefont {Chow}},\ and\ \bibinfo {author}
  {\bibfnamefont {J.~M.}\ \bibnamefont {Gambetta}},\ }\bibfield  {title}
  {\bibinfo {title} {Error mitigation extends the computational reach of a
  noisy quantum processor},\ }\href {https://doi.org/10.1038/s41586-019-1040-7}
  {\bibfield  {journal} {\bibinfo  {journal} {Nature}\ }\textbf {\bibinfo
  {volume} {567}},\ \bibinfo {pages} {491–495} (\bibinfo {year}
  {2019})}\BibitemShut {NoStop}%
\bibitem [{\citenamefont {Arute}\ \emph {et~al.}(2020)\citenamefont {Arute},
  \citenamefont {Arya}, \citenamefont {Babbush}, \citenamefont {Bacon},
  \citenamefont {Bardin}, \citenamefont {Barends}, \citenamefont {Boixo},
  \citenamefont {Broughton}, \citenamefont {Buckley},\ and\ \citenamefont
  {et~al.}}]{2020}%
  \BibitemOpen
  \bibfield  {author} {\bibinfo {author} {\bibfnamefont {F.}~\bibnamefont
  {Arute}}, \bibinfo {author} {\bibfnamefont {K.}~\bibnamefont {Arya}},
  \bibinfo {author} {\bibfnamefont {R.}~\bibnamefont {Babbush}}, \bibinfo
  {author} {\bibfnamefont {D.}~\bibnamefont {Bacon}}, \bibinfo {author}
  {\bibfnamefont {J.~C.}\ \bibnamefont {Bardin}}, \bibinfo {author}
  {\bibfnamefont {R.}~\bibnamefont {Barends}}, \bibinfo {author} {\bibfnamefont
  {S.}~\bibnamefont {Boixo}}, \bibinfo {author} {\bibfnamefont
  {M.}~\bibnamefont {Broughton}}, \bibinfo {author} {\bibfnamefont {B.~B.}\
  \bibnamefont {Buckley}},\ and\ \bibinfo {author} {\bibnamefont {et~al.}},\
  }\bibfield  {title} {\bibinfo {title} {Hartree-fock on a superconducting
  qubit quantum computer},\ }\href {https://doi.org/10.1126/science.abb9811}
  {\bibfield  {journal} {\bibinfo  {journal} {Science}\ }\textbf {\bibinfo
  {volume} {369}},\ \bibinfo {pages} {1084–1089} (\bibinfo {year}
  {2020})}\BibitemShut {NoStop}%
\bibitem [{\citenamefont {Gyawali}\ and\ \citenamefont
  {Lawler}(2021)}]{gyawali2021insights}%
  \BibitemOpen
  \bibfield  {author} {\bibinfo {author} {\bibfnamefont {G.}~\bibnamefont
  {Gyawali}}\ and\ \bibinfo {author} {\bibfnamefont {M.~J.}\ \bibnamefont
  {Lawler}},\ }\href@noop {} {\bibinfo {title} {Insights from an adaptive
  variational wave function study of the fermi-hubbard model}} (\bibinfo {year}
  {2021}),\ \Eprint {https://arxiv.org/abs/2109.12126} {arXiv:2109.12126
  [quant-ph]} \BibitemShut {NoStop}%
\bibitem [{\citenamefont {Yamamoto}(2019)}]{yamamoto2019natural}%
  \BibitemOpen
  \bibfield  {author} {\bibinfo {author} {\bibfnamefont {N.}~\bibnamefont
  {Yamamoto}},\ }\href@noop {} {\bibinfo {title} {On the natural gradient for
  variational quantum eigensolver}} (\bibinfo {year} {2019}),\ \Eprint
  {https://arxiv.org/abs/1909.05074} {arXiv:1909.05074 [quant-ph]} \BibitemShut
  {NoStop}%
\bibitem [{\citenamefont {Stokes}\ \emph {et~al.}(2020)\citenamefont {Stokes},
  \citenamefont {Izaac}, \citenamefont {Killoran},\ and\ \citenamefont
  {Carleo}}]{2020QNG}%
  \BibitemOpen
  \bibfield  {author} {\bibinfo {author} {\bibfnamefont {J.}~\bibnamefont
  {Stokes}}, \bibinfo {author} {\bibfnamefont {J.}~\bibnamefont {Izaac}},
  \bibinfo {author} {\bibfnamefont {N.}~\bibnamefont {Killoran}},\ and\
  \bibinfo {author} {\bibfnamefont {G.}~\bibnamefont {Carleo}},\ }\bibfield
  {title} {\bibinfo {title} {Quantum natural gradient},\ }\href
  {https://doi.org/10.22331/q-2020-05-25-269} {\bibfield  {journal} {\bibinfo
  {journal} {Quantum}\ }\textbf {\bibinfo {volume} {4}},\ \bibinfo {pages}
  {269} (\bibinfo {year} {2020})}\BibitemShut {NoStop}%
\bibitem [{\citenamefont {Zhang}\ \emph {et~al.}(2021)\citenamefont {Zhang},
  \citenamefont {Ferguson}, \citenamefont {Kühn}, \citenamefont {Haase},
  \citenamefont {Wilson}, \citenamefont {Jansen},\ and\ \citenamefont
  {Muschik}}]{zhang2021simulating}%
  \BibitemOpen
  \bibfield  {author} {\bibinfo {author} {\bibfnamefont {J.}~\bibnamefont
  {Zhang}}, \bibinfo {author} {\bibfnamefont {R.}~\bibnamefont {Ferguson}},
  \bibinfo {author} {\bibfnamefont {S.}~\bibnamefont {Kühn}}, \bibinfo
  {author} {\bibfnamefont {J.~F.}\ \bibnamefont {Haase}}, \bibinfo {author}
  {\bibfnamefont {C.~M.}\ \bibnamefont {Wilson}}, \bibinfo {author}
  {\bibfnamefont {K.}~\bibnamefont {Jansen}},\ and\ \bibinfo {author}
  {\bibfnamefont {C.~A.}\ \bibnamefont {Muschik}},\ }\href@noop {} {\bibinfo
  {title} {Simulating gauge theories with variational quantum eigensolvers in
  superconducting microwave cavities}} (\bibinfo {year} {2021}),\ \Eprint
  {https://arxiv.org/abs/2108.08248} {arXiv:2108.08248 [quant-ph]} \BibitemShut
  {NoStop}%
\bibitem [{\citenamefont {Ferguson}\ \emph {et~al.}(2021)\citenamefont
  {Ferguson}, \citenamefont {Dellantonio}, \citenamefont {Balushi},
  \citenamefont {Jansen}, \citenamefont {D\"ur},\ and\ \citenamefont
  {Muschik}}]{PhysRevLett.126.220501}%
  \BibitemOpen
  \bibfield  {author} {\bibinfo {author} {\bibfnamefont {R.~R.}\ \bibnamefont
  {Ferguson}}, \bibinfo {author} {\bibfnamefont {L.}~\bibnamefont
  {Dellantonio}}, \bibinfo {author} {\bibfnamefont {A.~A.}\ \bibnamefont
  {Balushi}}, \bibinfo {author} {\bibfnamefont {K.}~\bibnamefont {Jansen}},
  \bibinfo {author} {\bibfnamefont {W.}~\bibnamefont {D\"ur}},\ and\ \bibinfo
  {author} {\bibfnamefont {C.~A.}\ \bibnamefont {Muschik}},\ }\bibfield
  {title} {\bibinfo {title} {Measurement-based variational quantum
  eigensolver},\ }\href {https://doi.org/10.1103/PhysRevLett.126.220501}
  {\bibfield  {journal} {\bibinfo  {journal} {Phys. Rev. Lett.}\ }\textbf
  {\bibinfo {volume} {126}},\ \bibinfo {pages} {220501} (\bibinfo {year}
  {2021})}\BibitemShut {NoStop}%
\bibitem [{\citenamefont {Ba\~nuls}\ \emph {et~al.}(2017)\citenamefont
  {Ba\~nuls}, \citenamefont {Cichy}, \citenamefont {Cirac}, \citenamefont
  {Jansen},\ and\ \citenamefont {K\"uhn}}]{PhysRevX.7.041046}%
  \BibitemOpen
  \bibfield  {author} {\bibinfo {author} {\bibfnamefont {M.~C.}\ \bibnamefont
  {Ba\~nuls}}, \bibinfo {author} {\bibfnamefont {K.}~\bibnamefont {Cichy}},
  \bibinfo {author} {\bibfnamefont {J.~I.}\ \bibnamefont {Cirac}}, \bibinfo
  {author} {\bibfnamefont {K.}~\bibnamefont {Jansen}},\ and\ \bibinfo {author}
  {\bibfnamefont {S.}~\bibnamefont {K\"uhn}},\ }\bibfield  {title} {\bibinfo
  {title} {Efficient basis formulation for ($1+1$)-dimensional su(2) lattice
  gauge theory: Spectral calculations with matrix product states},\ }\href
  {https://doi.org/10.1103/PhysRevX.7.041046} {\bibfield  {journal} {\bibinfo
  {journal} {Phys. Rev. X}\ }\textbf {\bibinfo {volume} {7}},\ \bibinfo {pages}
  {041046} (\bibinfo {year} {2017})}\BibitemShut {NoStop}%
\bibitem [{\citenamefont {Klco}\ and\ \citenamefont
  {Savage}(2020{\natexlab{a}})}]{Klco_2020F}%
  \BibitemOpen
  \bibfield  {author} {\bibinfo {author} {\bibfnamefont {N.}~\bibnamefont
  {Klco}}\ and\ \bibinfo {author} {\bibfnamefont {M.~J.}\ \bibnamefont
  {Savage}},\ }\bibfield  {title} {\bibinfo {title} {Fixed-point quantum
  circuits for quantum field theories},\ }\bibfield  {journal} {\bibinfo
  {journal} {Physical Review A}\ }\textbf {\bibinfo {volume} {102}},\ \href
  {https://doi.org/10.1103/physreva.102.052422} {10.1103/physreva.102.052422}
  (\bibinfo {year} {2020}{\natexlab{a}})\BibitemShut {NoStop}%
\bibitem [{\citenamefont {Lanczos}(1950)}]{Lanczos:1950zz}%
  \BibitemOpen
  \bibfield  {author} {\bibinfo {author} {\bibfnamefont {C.}~\bibnamefont
  {Lanczos}},\ }\bibfield  {title} {\bibinfo {title} {An iteration method for
  the solution of the eigenvalue problem of linear differential and integral
  operators},\ }\href {https://doi.org/10.6028/jres.045.026} {\bibfield
  {journal} {\bibinfo  {journal} {J. Res. Natl. Bur. Stand. B}\ }\textbf
  {\bibinfo {volume} {45}},\ \bibinfo {pages} {255} (\bibinfo {year}
  {1950})}\BibitemShut {NoStop}%
\bibitem [{\citenamefont {Motta}\ \emph {et~al.}(2019)\citenamefont {Motta},
  \citenamefont {Sun}, \citenamefont {Tan}, \citenamefont {O’Rourke},
  \citenamefont {Ye}, \citenamefont {Minnich}, \citenamefont {Brandão},\ and\
  \citenamefont {Chan}}]{Motta_2019}%
  \BibitemOpen
  \bibfield  {author} {\bibinfo {author} {\bibfnamefont {M.}~\bibnamefont
  {Motta}}, \bibinfo {author} {\bibfnamefont {C.}~\bibnamefont {Sun}}, \bibinfo
  {author} {\bibfnamefont {A.~T.~K.}\ \bibnamefont {Tan}}, \bibinfo {author}
  {\bibfnamefont {M.~J.}\ \bibnamefont {O’Rourke}}, \bibinfo {author}
  {\bibfnamefont {E.}~\bibnamefont {Ye}}, \bibinfo {author} {\bibfnamefont
  {A.~J.}\ \bibnamefont {Minnich}}, \bibinfo {author} {\bibfnamefont {F.~G.
  S.~L.}\ \bibnamefont {Brandão}},\ and\ \bibinfo {author} {\bibfnamefont
  {G.~K.-L.}\ \bibnamefont {Chan}},\ }\bibfield  {title} {\bibinfo {title}
  {Determining eigenstates and thermal states on a quantum computer using
  quantum imaginary time evolution},\ }\href
  {https://doi.org/10.1038/s41567-019-0704-4} {\bibfield  {journal} {\bibinfo
  {journal} {Nature Physics}\ }\textbf {\bibinfo {volume} {16}},\ \bibinfo
  {pages} {205–210} (\bibinfo {year} {2019})}\BibitemShut {NoStop}%
\bibitem [{\citenamefont {Klco}\ and\ \citenamefont
  {Savage}(2020{\natexlab{b}})}]{icons2020}%
  \BibitemOpen
  \bibfield  {author} {\bibinfo {author} {\bibfnamefont {N.}~\bibnamefont
  {Klco}}\ and\ \bibinfo {author} {\bibfnamefont {M.~J.}\ \bibnamefont
  {Savage}},\ }\bibfield  {title} {\bibinfo {title} {Minimally entangled state
  preparation of localized wave functions on quantum computers},\ }\bibfield
  {journal} {\bibinfo  {journal} {Physical Review A}\ }\textbf {\bibinfo
  {volume} {102}},\ \href {https://doi.org/10.1103/physreva.102.012612}
  {10.1103/physreva.102.012612} (\bibinfo {year}
  {2020}{\natexlab{b}})\BibitemShut {NoStop}%
\bibitem [{\citenamefont {Zache}\ \emph {et~al.}(2021)\citenamefont {Zache},
  \citenamefont {Damme}, \citenamefont {Halimeh}, \citenamefont {Hauke},\ and\
  \citenamefont {Banerjee}}]{zache2021achieving}%
  \BibitemOpen
  \bibfield  {author} {\bibinfo {author} {\bibfnamefont {T.~V.}\ \bibnamefont
  {Zache}}, \bibinfo {author} {\bibfnamefont {M.~V.}\ \bibnamefont {Damme}},
  \bibinfo {author} {\bibfnamefont {J.~C.}\ \bibnamefont {Halimeh}}, \bibinfo
  {author} {\bibfnamefont {P.}~\bibnamefont {Hauke}},\ and\ \bibinfo {author}
  {\bibfnamefont {D.}~\bibnamefont {Banerjee}},\ }\href@noop {} {\bibinfo
  {title} {Achieving the continuum limit of quantum link lattice gauge theories
  on quantum devices}} (\bibinfo {year} {2021}),\ \Eprint
  {https://arxiv.org/abs/2104.00025} {arXiv:2104.00025 [hep-lat]} \BibitemShut
  {NoStop}%
\bibitem [{\citenamefont {Tong}\ \emph {et~al.}(2021)\citenamefont {Tong},
  \citenamefont {Albert}, \citenamefont {McClean}, \citenamefont {Preskill},\
  and\ \citenamefont {Su}}]{tong2021provably}%
  \BibitemOpen
  \bibfield  {author} {\bibinfo {author} {\bibfnamefont {Y.}~\bibnamefont
  {Tong}}, \bibinfo {author} {\bibfnamefont {V.~V.}\ \bibnamefont {Albert}},
  \bibinfo {author} {\bibfnamefont {J.~R.}\ \bibnamefont {McClean}}, \bibinfo
  {author} {\bibfnamefont {J.}~\bibnamefont {Preskill}},\ and\ \bibinfo
  {author} {\bibfnamefont {Y.}~\bibnamefont {Su}},\ }\href@noop {} {\bibinfo
  {title} {Provably accurate simulation of gauge theories and bosonic systems}}
  (\bibinfo {year} {2021}),\ \Eprint {https://arxiv.org/abs/2110.06942}
  {arXiv:2110.06942 [quant-ph]} \BibitemShut {NoStop}%
\bibitem [{\citenamefont {Nielsen}\ and\ \citenamefont
  {Chuang}(2011)}]{Nielsen:2011:QCQ:1972505}%
  \BibitemOpen
  \bibfield  {author} {\bibinfo {author} {\bibfnamefont {M.~A.}\ \bibnamefont
  {Nielsen}}\ and\ \bibinfo {author} {\bibfnamefont {I.~L.}\ \bibnamefont
  {Chuang}},\ }\href@noop {} {\emph {\bibinfo {title} {Quantum Computation and
  Quantum Information: 10th Anniversary Edition}}},\ \bibinfo {edition} {10th}\
  ed.\ (\bibinfo  {publisher} {Cambridge University Press},\ \bibinfo {address}
  {New York, NY, USA},\ \bibinfo {year} {2011})\BibitemShut {NoStop}%
\bibitem [{\citenamefont {Frazier}(2018)}]{frazier2018tutorial}%
  \BibitemOpen
  \bibfield  {author} {\bibinfo {author} {\bibfnamefont {P.~I.}\ \bibnamefont
  {Frazier}},\ }\bibfield  {title} {\bibinfo {title} {A tutorial on bayesian
  optimization},\ }\href@noop {} {\bibfield  {journal} {\bibinfo  {journal}
  {arXiv preprint arXiv:1807.02811}\ } (\bibinfo {year} {2018})}\BibitemShut
  {NoStop}%
\bibitem [{mit()}]{mittikhonovnotes}%
  \BibitemOpen
  \href {https://www.mit.edu/~9.520/scribe-notes/cl7.pdf} {\bibinfo {title}
  {Tikhonov regularization and erm}}\BibitemShut {NoStop}%
\bibitem [{\citenamefont {Truong}(2019)}]{truong2019convergence}%
  \BibitemOpen
  \bibfield  {author} {\bibinfo {author} {\bibfnamefont {T.~T.}\ \bibnamefont
  {Truong}},\ }\bibfield  {title} {\bibinfo {title} {Convergence to minima for
  the continuous version of backtracking gradient descent},\ }\href@noop {}
  {\bibfield  {journal} {\bibinfo  {journal} {arXiv preprint arXiv:1911.04221}\
  } (\bibinfo {year} {2019})}\BibitemShut {NoStop}%
\bibitem [{\citenamefont {Tamiya}\ and\ \citenamefont
  {Yamasaki}(2021)}]{tamiya2021stochastic}%
  \BibitemOpen
  \bibfield  {author} {\bibinfo {author} {\bibfnamefont {S.}~\bibnamefont
  {Tamiya}}\ and\ \bibinfo {author} {\bibfnamefont {H.}~\bibnamefont
  {Yamasaki}},\ }\href@noop {} {\bibinfo {title} {Stochastic gradient line
  bayesian optimization: Reducing measurement shots in optimizing parameterized
  quantum circuits}} (\bibinfo {year} {2021}),\ \Eprint
  {https://arxiv.org/abs/2111.07952} {arXiv:2111.07952 [quant-ph]} \BibitemShut
  {NoStop}%
\bibitem [{\citenamefont {Schuld}\ \emph {et~al.}(2020)\citenamefont {Schuld},
  \citenamefont {Bocharov}, \citenamefont {Svore},\ and\ \citenamefont
  {Wiebe}}]{PhysRevA.101.032308}%
  \BibitemOpen
  \bibfield  {author} {\bibinfo {author} {\bibfnamefont {M.}~\bibnamefont
  {Schuld}}, \bibinfo {author} {\bibfnamefont {A.}~\bibnamefont {Bocharov}},
  \bibinfo {author} {\bibfnamefont {K.~M.}\ \bibnamefont {Svore}},\ and\
  \bibinfo {author} {\bibfnamefont {N.}~\bibnamefont {Wiebe}},\ }\bibfield
  {title} {\bibinfo {title} {Circuit-centric quantum classifiers},\ }\href
  {https://doi.org/10.1103/PhysRevA.101.032308} {\bibfield  {journal} {\bibinfo
   {journal} {Phys. Rev. A}\ }\textbf {\bibinfo {volume} {101}},\ \bibinfo
  {pages} {032308} (\bibinfo {year} {2020})}\BibitemShut {NoStop}%
\bibitem [{\citenamefont {{IBM Quantum Experience}}(2021)}]{ibmManila}%
  \BibitemOpen
  \bibfield  {author} {\bibinfo {author} {\bibnamefont {{IBM Quantum
  Experience}}},\ }\href@noop {} {\bibinfo {title} {ibmq\_manila v1.0.18}},\
  \bibinfo {howpublished} {\url{https://quantum-computing.ibm.com}} (\bibinfo
  {year} {2021})\BibitemShut {NoStop}%
\bibitem [{\citenamefont {Lüscher}(2004)}]{LUSCHER2004209}%
  \BibitemOpen
  \bibfield  {author} {\bibinfo {author} {\bibfnamefont {M.}~\bibnamefont
  {Lüscher}},\ }\bibfield  {title} {\bibinfo {title} {Solution of the dirac
  equation in lattice qcd using a domain decomposition method},\ }\href
  {https://doi.org/https://doi.org/10.1016/S0010-4655(03)00486-7} {\bibfield
  {journal} {\bibinfo  {journal} {Computer Physics Communications}\ }\textbf
  {\bibinfo {volume} {156}},\ \bibinfo {pages} {209} (\bibinfo {year}
  {2004})}\BibitemShut {NoStop}%
\bibitem [{\citenamefont {Frommer}\ \emph {et~al.}(2014)\citenamefont
  {Frommer}, \citenamefont {Kahl}, \citenamefont {Krieg}, \citenamefont
  {Leder},\ and\ \citenamefont {Rottmann}}]{frommer2014adaptive}%
  \BibitemOpen
  \bibfield  {author} {\bibinfo {author} {\bibfnamefont {A.}~\bibnamefont
  {Frommer}}, \bibinfo {author} {\bibfnamefont {K.}~\bibnamefont {Kahl}},
  \bibinfo {author} {\bibfnamefont {S.}~\bibnamefont {Krieg}}, \bibinfo
  {author} {\bibfnamefont {B.}~\bibnamefont {Leder}},\ and\ \bibinfo {author}
  {\bibfnamefont {M.}~\bibnamefont {Rottmann}},\ }\href@noop {} {\bibinfo
  {title} {Adaptive aggregation based domain decomposition multigrid for the
  lattice wilson dirac operator}} (\bibinfo {year} {2014}),\ \Eprint
  {https://arxiv.org/abs/1303.1377} {arXiv:1303.1377 [hep-lat]} \BibitemShut
  {NoStop}%
\bibitem [{\citenamefont {Heybrock}\ \emph {et~al.}(2014)\citenamefont
  {Heybrock}, \citenamefont {Joo}, \citenamefont {Kalamkar}, \citenamefont
  {Smelyanskiy}, \citenamefont {Vaidyanathan}, \citenamefont {Wettig},\ and\
  \citenamefont {Dubey}}]{Heybrock_2014}%
  \BibitemOpen
  \bibfield  {author} {\bibinfo {author} {\bibfnamefont {S.}~\bibnamefont
  {Heybrock}}, \bibinfo {author} {\bibfnamefont {B.}~\bibnamefont {Joo}},
  \bibinfo {author} {\bibfnamefont {D.~D.}\ \bibnamefont {Kalamkar}}, \bibinfo
  {author} {\bibfnamefont {M.}~\bibnamefont {Smelyanskiy}}, \bibinfo {author}
  {\bibfnamefont {K.}~\bibnamefont {Vaidyanathan}}, \bibinfo {author}
  {\bibfnamefont {T.}~\bibnamefont {Wettig}},\ and\ \bibinfo {author}
  {\bibfnamefont {P.}~\bibnamefont {Dubey}},\ }\bibfield  {title} {\bibinfo
  {title} {Lattice qcd with domain decomposition on intel xeon phi
  co-processors},\ }\bibfield  {journal} {\bibinfo  {journal} {SC14:
  International Conference for High Performance Computing, Networking, Storage
  and Analysis}\ }\href {https://doi.org/10.1109/sc.2014.11}
  {10.1109/sc.2014.11} (\bibinfo {year} {2014})\BibitemShut {NoStop}%
\bibitem [{\citenamefont {White}(1992)}]{PhysRevLett.69.2863}%
  \BibitemOpen
  \bibfield  {author} {\bibinfo {author} {\bibfnamefont {S.~R.}\ \bibnamefont
  {White}},\ }\bibfield  {title} {\bibinfo {title} {Density matrix formulation
  for quantum renormalization groups},\ }\href
  {https://doi.org/10.1103/PhysRevLett.69.2863} {\bibfield  {journal} {\bibinfo
   {journal} {Phys. Rev. Lett.}\ }\textbf {\bibinfo {volume} {69}},\ \bibinfo
  {pages} {2863} (\bibinfo {year} {1992})}\BibitemShut {NoStop}%
\bibitem [{\citenamefont {Experience}(2020)}]{ibmmeaserror}%
  \BibitemOpen
  \bibfield  {author} {\bibinfo {author} {\bibfnamefont {I.~Q.}\ \bibnamefont
  {Experience}},\ }\href@noop {} {\bibinfo {title} {ibmq\_measurement error}},\
  \bibinfo {howpublished}
  {\url{https://qiskit.org/documentation/tutorials/noise/3_measurement_error_mitigation.html}}
  (\bibinfo {year} {2020})\BibitemShut {NoStop}%
\bibitem [{\citenamefont {Li}\ and\ \citenamefont
  {Benjamin}(2017)}]{PhysRevX.7.021050}%
  \BibitemOpen
  \bibfield  {author} {\bibinfo {author} {\bibfnamefont {Y.}~\bibnamefont
  {Li}}\ and\ \bibinfo {author} {\bibfnamefont {S.~C.}\ \bibnamefont
  {Benjamin}},\ }\bibfield  {title} {\bibinfo {title} {Efficient variational
  quantum simulator incorporating active error minimization},\ }\href
  {https://doi.org/10.1103/PhysRevX.7.021050} {\bibfield  {journal} {\bibinfo
  {journal} {Phys. Rev. X}\ }\textbf {\bibinfo {volume} {7}},\ \bibinfo {pages}
  {021050} (\bibinfo {year} {2017})}\BibitemShut {NoStop}%
\bibitem [{\citenamefont {Temme}\ \emph {et~al.}(2017)\citenamefont {Temme},
  \citenamefont {Bravyi},\ and\ \citenamefont
  {Gambetta}}]{PhysRevLett.119.180509}%
  \BibitemOpen
  \bibfield  {author} {\bibinfo {author} {\bibfnamefont {K.}~\bibnamefont
  {Temme}}, \bibinfo {author} {\bibfnamefont {S.}~\bibnamefont {Bravyi}},\ and\
  \bibinfo {author} {\bibfnamefont {J.~M.}\ \bibnamefont {Gambetta}},\
  }\bibfield  {title} {\bibinfo {title} {Error mitigation for short-depth
  quantum circuits},\ }\href {https://doi.org/10.1103/PhysRevLett.119.180509}
  {\bibfield  {journal} {\bibinfo  {journal} {Phys. Rev. Lett.}\ }\textbf
  {\bibinfo {volume} {119}},\ \bibinfo {pages} {180509} (\bibinfo {year}
  {2017})}\BibitemShut {NoStop}%
\bibitem [{\citenamefont {Jin}(2018)}]{bayesopt_presentation}%
  \BibitemOpen
  \bibfield  {author} {\bibinfo {author} {\bibfnamefont {H.}~\bibnamefont
  {Jin}},\ }\href {https://www.cs.uic.edu/~hjin/files/bayesian_opt.pdf}
  {\bibinfo {title} {Bayesian optimization}} (\bibinfo {year}
  {2018})\BibitemShut {NoStop}%
\bibitem [{\citenamefont {Do}\ and\ \citenamefont
  {Lee}(2008)}]{gprocess_presentation}%
  \BibitemOpen
  \bibfield  {author} {\bibinfo {author} {\bibfnamefont {C.~B.}\ \bibnamefont
  {Do}}\ and\ \bibinfo {author} {\bibfnamefont {H.}~\bibnamefont {Lee}},\
  }\href {http://cs229.stanford.edu/section/cs229-gaussian_processes.pdf}
  {\bibinfo {title} {Gaussian processes}} (\bibinfo {year} {2008})\BibitemShut
  {NoStop}%
\bibitem [{\citenamefont {Cressie}(1990)}]{Cressie1990}%
  \BibitemOpen
  \bibfield  {author} {\bibinfo {author} {\bibfnamefont {N.}~\bibnamefont
  {Cressie}},\ }\bibfield  {title} {\bibinfo {title} {The origins of kriging},\
  }\bibfield  {journal} {\bibinfo  {journal} {Math. Geol.}\ }\textbf {\bibinfo
  {volume} {22}},\ \href {https://doi.org/10.1007/BF00889887}
  {10.1007/BF00889887} (\bibinfo {year} {1990})\BibitemShut {NoStop}%
\bibitem [{\citenamefont {Vidal}(2003)}]{2003}%
  \BibitemOpen
  \bibfield  {author} {\bibinfo {author} {\bibfnamefont {G.}~\bibnamefont
  {Vidal}},\ }\bibfield  {title} {\bibinfo {title} {Efficient classical
  simulation of slightly entangled quantum computations},\ }\bibfield
  {journal} {\bibinfo  {journal} {Physical Review Letters}\ }\textbf {\bibinfo
  {volume} {91}},\ \href {https://doi.org/10.1103/physrevlett.91.147902}
  {10.1103/physrevlett.91.147902} (\bibinfo {year} {2003})\BibitemShut
  {NoStop}%
\bibitem [{\citenamefont {Vidal}(2004)}]{2004Eff}%
  \BibitemOpen
  \bibfield  {author} {\bibinfo {author} {\bibfnamefont {G.}~\bibnamefont
  {Vidal}},\ }\bibfield  {title} {\bibinfo {title} {Efficient simulation of
  one-dimensional quantum many-body systems},\ }\bibfield  {journal} {\bibinfo
  {journal} {Physical Review Letters}\ }\textbf {\bibinfo {volume} {93}},\
  \href {https://doi.org/10.1103/physrevlett.93.040502}
  {10.1103/physrevlett.93.040502} (\bibinfo {year} {2004})\BibitemShut
  {NoStop}%
\bibitem [{\citenamefont {Verstraete}\ \emph {et~al.}(2004)\citenamefont
  {Verstraete}, \citenamefont {García-Ripoll},\ and\ \citenamefont
  {Cirac}}]{2004MPS}%
  \BibitemOpen
  \bibfield  {author} {\bibinfo {author} {\bibfnamefont {F.}~\bibnamefont
  {Verstraete}}, \bibinfo {author} {\bibfnamefont {J.~J.}\ \bibnamefont
  {García-Ripoll}},\ and\ \bibinfo {author} {\bibfnamefont {J.~I.}\
  \bibnamefont {Cirac}},\ }\bibfield  {title} {\bibinfo {title} {Matrix product
  density operators: Simulation of finite-temperature and dissipative
  systems},\ }\bibfield  {journal} {\bibinfo  {journal} {Physical Review
  Letters}\ }\textbf {\bibinfo {volume} {93}},\ \href
  {https://doi.org/10.1103/physrevlett.93.207204}
  {10.1103/physrevlett.93.207204} (\bibinfo {year} {2004})\BibitemShut
  {NoStop}%
\end{thebibliography}%

\end{document}